\documentclass[12pt]{article}

\usepackage{amsfonts}
\usepackage{graphicx}
\begin{document}

\title{{\bf Lecture notes on \lq\lq Three\\ Supersymmetric/Topological systems \\ in Quantum Field
Theory\rq\rq}}
\author{J. Mateos Guilarte
\\ {\normalsize Departamento de Fisica},
{\rm University of Salamanca, SPAIN} \\ }

\date{}
\maketitle
{\bf X Autumn Colloquium on Geometry and Physics} 

{\bf Miraflores de la Sierra, SPAIN, September 2001}

\begin{abstract}
(1+1)-dimensional ${\cal N}=1$ super-symmetric field theory and
(3+1)-dimensional ${\cal N}=2$ super-symmetric gauge theory are
discussed in a, more or less, unified way, designed to identify
the quantum BPS states in both systems. Euclidean 4-dimensional
gauge theory with ${\cal N}=2$ twisted super-symmetry is also
analized. ${\bf C}^\infty$-topological invariants are identified
as certain n-point correlation functions in this QFT framework. 
The twist of the effective dual Abelian gauge theory
is briefly described, both from mathematical and physical viewpoints.
The physical nature of the topological defects arising in these systems, kinks, BPS and Dirac monopoles,
BPST instantons, Liouville and Abrikosov-Nielsen-Olesen selfdual vortices, etcetera, is analyzed, 

The thread of the story connecting the QFT systems treated respectively in Sections \S.3 and \S.4 is the process of TWIST that leads from a conventional extended Supersymetric Gauge Theory to the topological ${\cal N}=2$ SUSY Donaldson QFT. Within Section \S.3
the $SL(2,\mathbb{Z}$-dualities establish a link between the weak coupling regime of the original gauge theory and the 
Wilsonian (abelian) effective gauge theory arising at low energies. We shall also look after the reminiscences of these dualities between
the twisted TQFT systems of Section \S.4. 

\end{abstract}

\tableofcontents

%\clearpage
\section{Introduction}
The MacPherson Manifesto on the IAS/Princeton Quantum Field Theory
Program ends with this sentence: \lq\lq the goal is to develop the
sort of intuition common among physicists for those who are used
to thought processes stemming from geometry and algebra", see
\cite{Del}.

These Lecture notes are, somehow, conceived in this spirit. To
contribute to this goal I have tried to make explicit some
concepts which are second nature to physicists whereas are crucial
in the realm of the Mathematical Physics, so productive in the
last twenty years. In particular, I shall discuss the ubiquitous
phenomena of quantum BPS states and anomalies in two
supersymmetric field theory models. Quantum BPS states and
anomalies in the central charge of the SUSY algebra play an
important r$\hat{{\rm o}}$le in both (1+1)dimensional ${\cal N}=1$
supersymmetric field theory and (3+1)-dimensional ${\cal N}=2$
supersymmetric gauge theory. I shall explain, first, these
concepts in the (1+1)-D baby model because the complexities that
plague the higher dimensional systems are minimized.

The count of the BPS states and the derivation of the anomaly is
much more difficult in ${\cal N}=2$ SUSY Yang-Mills, but I will
present this topic in a way as close to the (1+1)-D case as
possible. The electric-magnetic duality transformation is behind
the astonishing discovery of the Wilson effective action for any
energy scale. The Seiberg-Witten proposal for the Wilson action
allows to identify the anomaly in the central charge and elucidate
the properties of the BPS states in terms of beautiful mathematics
involving elliptic curves, holomorphic differentials and modular
forms. The remarkable fact is that the whole construction is based
on a few highly plausible physical facts.

Similar structures appear in: (a) supersymmetric sigma models
where BPS states are related to hyper-Kahler metrics in
4n-dimensional target manifolds. (b) twisted ${\cal N}=2$
supersymmetric gauge theory leading to Donaldson and
Seiberg-Witten invariants of differentiable 4-manifolds. (c)
dimensional reduction of supersymmetric string theory. Calabi-Yau
and mirror symmetry of 6-dimensional manifolds arise using the
same circle of ideas. (d) dimensional reduction of M theory. Short
multiplets give rise to 7-dimensional manifolds with either ${\rm
Spin}_7$ or $G_2$ holonomy. No wonder on the interest in the magic
of these supersymmetric phenomena.

We organize the material in three, non-homogeneous neither in size nor in difficulty, Chapters. In the first Chapter we describe the
${\cal N}=1$ supersymmetric extension of $(1+1)$-dimensional scalar field theory as a warmup. Our construction is standard but we apply
the general framework to a particular system with two scalar fields and a rich variety of BPS kinks. In particular we offer a computation 
of the central charge arising in the Supersymmetric algebra as an anomaly in the $(1+1)$-dimensional QFT system. The second Chapter is devoted to the development of the Seiberg-Witten formulation of ${\cal N}=2$ Supersymmetric Yang-Mills Theory in $(3+1)$ dimensions.
The treatment is standard following, besides Seiberg-Witten original paper \cite{SeiWi}, the comprehensive reviews of Bilal \cite{Bilal} 
and Alvarez-Gaum\'e/Hassan \cite{Alvarez}. Particular attention is concentrated on the definition of the effective theory that appears at
low energies and the description of its structure and topological properties. The last Chapter addresses the construction of topological gauge theories in Euclidean four manifolds. Starting from supersymmetric Yang-Mills systems we end in Euclidean ${\cal N}=2$ SUSY through
the process of twisting. We discuss both the high energy regime, where one meets with Donaldson invariants of four manifolds, and the low energy effective theory touching with the Seiberg-Witten equations.

\clearpage
\section{${\cal N}=1$ (1+1)-dimensional Supersymmetric Field
Theory}

\subsection{The classical theory }

Points $(x^\mu,\theta)$ in the physical ${\mathbb R}^{2|2}$
superspace , see \cite{Del} , are characterized in terms of local
coordinates $x^\mu\in {\mathbb R}^{1,1}$ in Minkowski space-time and
$\displaystyle \theta=\left( \begin{array}{c} \theta_1\\ \theta_2
\end{array} \right)$ Grassman Majorana spinors:
\[
\theta_{\alpha}^*=\theta_{\alpha}, \qquad \alpha=1,2 \qquad ;
\qquad \theta_1^2=\theta_2^2=\theta_1\theta_2+\theta_2\theta_1=0
\]
We choose the metric tensor
$x^{\mu}x_{\mu}=g^{\mu\nu}x_{\mu}x_{\nu}$ as $g={\rm diag} (1,-1)$
and specify the Clifford algebra
$\{\gamma^\mu,\gamma^\nu\}=g^{\mu\nu}$ in the Majorana
representation in terms of the Pauli matrices:
\[
\gamma^0=\sigma^2\ ,\qquad \gamma^1=i\sigma^1\ ;\qquad
\gamma^5=\gamma^0\, \gamma^1=\sigma^3
\]
The Dirac adjoint is $\bar{\theta}=\theta^t \gamma^0$ and
$\theta\bar{\theta}=-2i\theta_1\theta_2$.

If $\varepsilon $ is a Grassman Majorana spinor parameter, the
\lq\lq vector field" $Q=\frac{\partial }{\partial \bar{\theta}} \,
+i\, \gamma^{\mu} \theta \partial_{\mu}$ is the generator of
infinitesimal supertranslations in ${\mathbb R}^{2|2}$:
\[
\theta \rightarrow \theta+i\varepsilon \quad \cong \quad i \bar{Q}
\varepsilon \theta=i \varepsilon  \qquad \qquad x^{\mu}
\rightarrow x^{\mu} + \bar{\theta} \gamma^{\mu} \varepsilon \quad
\cong \quad i\bar{Q}\varepsilon x^{\mu} = \bar{\theta}
\gamma^{\mu} \varepsilon
\]
The components $Q_\alpha$ of the super-charge $Q$ satisfy the
anti-commutation relations:
\begin{eqnarray}
\left\{ Q_{\alpha} , Q_{\beta}\right\}&=&
Q_{\alpha}Q_{\beta}+Q_{\beta}Q_{\alpha}=2\left( \gamma^{\mu}
C\right)_{\alpha\beta}\, P_{\mu} \label{eq:susy}\\&=& -2 i \left(
\begin{array}{cc}\partial_0-\partial_1 & 0 \\ 0 &
\partial_0+\partial_1 \end{array} \right)_{\alpha\beta} \qquad ;
\qquad C=-\gamma^0 \nonumber
\end{eqnarray}

Our aim is to build the most general theory for the super-field
\[
\vec{\Phi}(x^{\mu},\theta)=\sum_{a=1}^N \Phi_a (x^{\mu},\theta) \,
\vec{e}_a\, : {\mathbb R}^{2|2}\to {\mathbb R}^N \quad ;\quad
\vec{e}_a\cdot \vec{e}_b \, = \, \delta_{ab}
\]
invariant with respect to the super-Poincare algebra, i.e.
symmetry under the Poincare and the (\ref{eq:susy}) SUSY
transformations, the super-Poincare algebra, is required.
$\vec{\Phi}(x^{\mu},\theta)$ is a map from the super-space to the
${\mathbb R}^N$ \lq\lq internal" space - ${\vec e}_a , a=1,2, \cdots
,N$ are ortho-normal vectors in ${\mathbb R}^N$ - that , through the
power series expansion
\[
\displaystyle \vec{\Phi}(x^\mu,
\theta)=\vec{\phi}(x^{\mu})+\bar{\theta} \,
\vec{\psi}(x^{\mu})-\frac{1}{2} \bar{\theta}\theta \,
\vec{F}(x^{\mu}),
\]
can be seen as including $N$ scalar fields, $\displaystyle
\vec{\phi}(x^{\mu}) = \sum_{a=1}^N \phi_a (x^{\mu}) \, \vec{e}_a$
, N Majorana spinor fields , $\displaystyle \vec{\psi}(x^\mu)
=\sum_{a=1}^N \psi_a(x^{\mu}) \, \vec{e}_a$ , and N auxiliary
scalar fields $\vec{F}(x^{\mu})=\sum_{a=1}^N F_a(x^{\mu})\,
\vec{e}_a$.

\subsubsection{The supersymmetric action}
Bearing in mind that the action of the SUSY charges on the
superfield is
\[
\delta \vec{\Phi} =i \bar{\varepsilon}\, Q \, \vec{\Phi}
\Rightarrow \, \left\{ \begin{array}{l} \delta \vec{\phi}=i
\bar{\varepsilon} \vec{\psi}\\ \delta \vec{\psi} =\left(
\gamma^\mu \partial_{\mu} \vec{\phi}+\vec{F}\right) \varepsilon\\
\delta \vec{F}=i\bar{\varepsilon}\gamma^{\mu} \partial_{\mu}
\vec{\psi}\end{array} \right.
\]
and realizing that the covariant derivative
$D=\frac{\partial}{\partial \bar{\theta}}-i \gamma^{\mu} \theta
\partial_\mu$ anti-commutes with $Q$, it is not difficult to
achieve this goal. The dynamics of the ${\cal N}=1$ supersymmetric
field theory on ${\mathbb R}^{1,1}$ with no interactions is governed
by the action:
\begin{eqnarray*}
S_0[\vec{\Phi}]&=&-\frac{1}{2} \int d^2x d^2\theta \, \bar{D}
\vec{\Phi} \, D \vec{\Phi}\\ &=& \frac{1}{2} \int d^2x\left\{
\partial_\mu \vec{\phi}\partial^\mu \vec{\phi} + i
\vec{\bar{\psi}}\gamma^\mu \partial_\mu \vec{\psi}-\frac{1}{2}
\vec{F}\vec{F}\right\}
\end{eqnarray*}
The last identity is obtained through Berezin integration along
the odd (Grasmann) variables in the superspace. To switch on
supersymmetric interactions the recipe is well known: one adds
\[
S_I[\vec{\Phi}]=2 \int d^2xd^2\theta \, W[\vec{\Phi}]\quad ,
\]
where $W[\vec{\Phi}]$ is a unspecified \lq\lq superpotential", to
the free action $S_0$.

Expanding $W[\vec{\Phi}]$ in a power series in the Grassman
variables

\[
W[\vec{\Phi}]=W[\vec{\phi}]+\vec{\nabla} W\cdot \bar{\theta}
\vec{\psi}-\frac{1}{2} \bar{\theta}\theta \left( i \vec{\nabla}W
\cdot \vec{F}+\frac{1}{2} \vec{\bar{\psi}}\cdot
\vec{\vec{\Delta}}W \cdot \vec{\psi}\right)
\]
\[
\vec{\nabla}W=\sum_{a=1}^{N}\frac{\partial W}{\partial\phi_a}\cdot
\vec{e}_a; \quad \vec{\vec{\Delta}}W=\vec{\nabla}\otimes
\vec{\nabla}W=\sum_{a=1}^N\sum_{b=1}^N \vec{e}_a\otimes \vec{e}_b
\frac{\partial^2W}{\partial\phi_a \partial \phi_b}
\]
\[
\vec{\bar{\psi}}\cdot \vec{\vec{\Delta}}W\cdot
\vec{\psi}=\sum_{a=1}^N\sum_{b=1}^N \bar{\psi}_a \cdot
\frac{\partial^2W}{\partial\phi_a \partial \phi_b}\cdot \psi_b
\]
and performing the Berezin integration, we find
\[
S_I[\vec{\Phi}]=-\int d^2x \, \left( i \vec{\nabla}W\cdot
\vec{F}+\frac{1}{2} \vec{\bar{\psi}}\cdot \vec{\vec{\Delta}}W\cdot
\vec{\psi}\right)
\]
as the interacting piece of the ${\cal N}=1$ supersymmetric action
$S=S_0+S_I$.

Solving for the auxiliary fields in the constraint equations
$\vec{F}=-i \vec{\nabla}W$ one checks that the \lq\lq on shell"
action reads
\[
S=\frac{1}{2} \int d^2x \, \left\{ \partial_{\mu}\vec{\phi}
\partial^{\mu} \vec{\phi}+i \vec{\bar{\psi}} \gamma^{\mu}
\partial_\mu \vec{\psi}-\vec{\nabla} W\cdot
\vec{\nabla}W-\vec{\bar{\psi}}\cdot \vec{\vec{\Delta}}\cdot
\vec{\psi}\right\}
\]
so that the field equations are:
\begin{equation}
\square \vec{\phi}+\vec{\nabla}W\cdot
\vec{\vec{\Delta}}W+\vec{\bar{\psi}}\cdot
\vec{\vec{\vec{\Delta}}}W\cdot \vec{\psi}=\vec{0}\label{eq:scel}
\end{equation}
\begin{equation}
i\, \partial_\mu\vec{\bar{\psi}}\gamma^\mu+\vec{\bar{\psi}}\cdot
\vec{\vec{\Delta}}W=\vec{0}\label{eq:spel}
\end{equation}
The N\"other theorem provides us with the Hamiltonian functions
associated to the vector fields $Q_1$ and $Q_2$:
\[
\tilde{Q}=\int dx \, \left\{ \gamma^\mu \gamma^0 \vec{\psi}
\partial_\mu \vec{\phi}+i \gamma^0 \vec{\psi}
\vec{\nabla}W\right\}
\]
$\tilde{Q}_1$ and $\tilde{Q_2}$ induce respectively the flows
associated to the super-translations $\theta^1+i\varepsilon^1$ and
$\theta^2+i\varepsilon^2$ in the co-tangent bundle to the
configuration space, the space of initial conditions for the PDE
system (\ref{eq:scel})-(\ref{eq:spel}).

\subsubsection{BPS and non-BPS super-solutions}

We shall not discuss the super-solutions to the field equations
(\ref{eq:scel})-(\ref{eq:spel}) in full generality, see instead
\cite{Del}. We describe in this sub-section several types of
specially significant super-waves.

A. Homogeneous.

Let us choose choose $W$ such that exist space-time independent
scalar field configurations $\vec{\phi}_c$ for which
$\left.\vec{\nabla}W\right|_{\vec{\phi}_c}=\vec{0}$. Then,
\[
\vec{\phi}(x^\mu)=\vec{\phi}_c \qquad  , \qquad
\vec{\psi}(x^\mu)=\vec{\psi}_c=\vec{0}
\]
are the homogeneous super-solutions to
(\ref{eq:scel})-(\ref{eq:spel}).

B. Plane super-waves.

Considering small fluctuations of the scalar and spinor fields
around one homogeneous solution,
\[
\vec{\phi}(x,t)=\vec{\phi}_c+\delta\vec{\phi}(x,t)\qquad , \qquad
\vec{\psi}(x,t)=\vec{\psi}_c+\delta\vec{\psi}(x,t) \quad ,
\]
such that second order effects are negligible - ${\cal O}(\delta
\Phi )^2$ -, one checks that
\begin{eqnarray}
\delta\vec{\phi}(x,t)&=&\sum_{j=1}^N\sum_{k_j\in{\mathbb Z}}(
\vec{a}(k_j) \, e^{i\omega_jt-ik_jx}+\vec{a}^*(k_j) \,
e^{-i\omega_jt+ik_jx})\label{eq:planB}\\
\delta\vec{\psi}(x,t)&=&\sum_{j=1}^N\sum_{k_j\in{\mathbb Z}
}(\vec{u}(k_j)\,e^{i\omega_j
t-ik_jx}+\vec{u}^*(k_j)\,e^{-i\omega_j t+ik_jx})\label{eq:planF}
\end{eqnarray}
are the plane super-wave solutions on a large but finite interval
of (\ref{eq:scel})-(\ref{eq:spel}) if and only if:
\[
\omega_j^2=k_j^2+\lambda_j^2 \quad , \quad \lambda_j^2\in {\rm
Spec}\left( \left.\vec{\vec{\Delta}}W\right|_{\vec{\phi}_c}.
\left.\vec{\vec{\Delta}}W\right|_{\vec{\phi}_c}\right) \quad .
\]

C. Super-solitons.

For time-independent super fields, $\vec{\Phi}[x,\theta]\neq
\vec{f}(t)$, the field equations reduce to the ODE system:

\begin{equation}
\frac{d^2\vec{\phi}}{dx^2}=\vec{\nabla}W\cdot
\vec{\vec{\Delta}}W+\vec{\bar{\psi}} \vec{\nabla}
\vec{\vec{\Delta}}W\cdot \vec{\psi} \label{eq:sscel}
\end{equation}
\begin{equation}
i\, \frac{d\vec{\bar{\psi}}}{dx} \cdot
\gamma^1=\vec{\bar{\psi}}\vec{\vec{\Delta}}W \label{eq:sspel}
\end{equation}
The \lq\lq on shell" energy is,
\[
E=\frac{1}{2} \int dx \left( \frac{d\vec{\phi}}{dx}\cdot
\frac{d\vec{\phi}}{dx} +\vec{\nabla}W \cdot \vec{\nabla}W+i
\vec{\bar{\psi}}\gamma^1 \frac{d\vec{\psi}}{dx}+\vec{\bar{\psi}}
\vec{\vec{\Delta}} W \vec{\psi}\right)
\]
and the bosonic contribution can be arranged \'a la Bogomolny.
\[
E_B=\frac{1}{2} \int dx \, \left( \frac{d\vec{\phi}}{dx}\pm
\vec{\nabla}W\right)\left( \frac{d\vec{\phi}}{dx}\pm
\vec{\nabla}W\right)\mp \int dx \frac{d\vec{\phi}}{dx} \cdot
\vec{\nabla}W
\]
The Bogomolny bound $\displaystyle E_{{\rm BPS}}=\left| \int
dW\right|$ is saturated by the solutions of
\begin{equation}
\frac{d\vec{\phi}}{dx}=\mp \vec{\nabla}W \label{eq:fsscel}
\end{equation}
and
\begin{equation}
\frac{d\psi^a_\pm}{dx}= -\sum_{b=1}^N
\frac{\partial^2W}{\partial\phi^a\partial\phi^b}\cdot\psi^b_\pm
\quad ,\label{eq:dsspel}
\end{equation}
where $\vec{\psi}_\pm=\frac{1}{\sqrt{2}} \left( \vec{\psi}_1\pm
\vec{\psi}_2\right)$ are chiral (Majorana-Weyl) spinors.

Note that the solutions of (\ref{eq:fsscel}) solve also
(\ref{eq:sscel}) whereas the system (\ref{eq:dsspel}) is nothing
but (\ref{eq:sspel}) diagonalized. Super-solitons are (bosonic)
flow lines of $\pm {\rm grad}W$ - solutions of (\ref{eq:fsscel})-
and their fermionic partners, the solutions of (\ref{eq:dsspel}),
which do not contribute to the energy.

A central element in supersymmetric/topological field theory and
its application to Geometry and Topology is the concept of BPS
state. Why are super-solitons so distinguished? The answer is
because they are annihilated by some combination of the
super-symmetry generators. Acting on shell static super-fields;,
\[
\vec{\Phi}[x,\theta]=\vec{\phi}(x)+\bar{\theta}
\vec{\psi}(x)-\frac{1}{2} \bar{\theta}\theta\vec{\nabla}W
\]
the effective super-charges are:  $Q=\frac{\partial}{\partial
\bar{\theta}}-i\gamma^1\theta \frac{d}{dx}$. Therefore,

\[
Q\vec{\Phi}=-i \gamma^1 \theta
\frac{d\vec{\phi}}{dx}+\vec{\psi}-\theta \vec{\nabla}W \qquad \,
\Leftrightarrow \qquad \left\{ \begin{array} {c} \left(Q
\vec{\Phi}\right)_1=\theta_2\frac{d\vec{\phi}}{dx}
+\vec{\psi}_1-\theta_1 \vec{\nabla}W  \\  \left(Q
\vec{\Phi}\right)_2=\theta_1\frac{d\vec{\phi}}{dx}
+\vec{\psi}_2-\theta_2 \vec{\nabla}W \end{array} \right .
\]
The classical BPS states are those for which $\left( Q
\vec{\Phi}\right)_1\pm \left( Q \vec{\Phi}\right)_2=\vec{0}$ but,
\[
\left( Q \vec{\Phi}\right)_1\pm \left( Q
\vec{\Phi}\right)_2=(\theta_2\pm \theta_1)
\frac{d\vec{\phi}}{dx}+( \vec{\psi}_1\pm
\vec{\psi}_2)-(\theta_1\pm \theta_2) \vec{\nabla}W \quad .
\]
Thus, the super-solitons which satisfy
\begin{equation}
\frac{d\vec{\phi}}{dx}\pm \vec{\nabla}W=\vec{0}\qquad , \quad
\vec{\psi}_{\pm}=\vec{0} \qquad {\rm and} \quad
\frac{d\vec{\psi}_{\mp}}{dx}=-\left.
\vec{\vec{\Delta}}W\right|_{\vec{\phi}_0}\cdot \vec{\psi}_{\mp}
\label{eq:ssol}
\end{equation}
are classical BPS states. Witten's list \cite{Wit1} -gradient flow
lines, holomorphic curves, gauge theory instantons, monopoles ,
Seiberg-Witten solutions, hyper-Kahler structures, Calabi-Yau
metrics, metrics of $G_2$ and Spin$_7$ holonomy- of this
aristocracy shows the importance, both in Physics and Mathematics,
of the BPS states

\subsection{A rapid look at the quantum theory}
The standard canonical quantization procedure promotes the field
configurations to field operators by decreeing that the Poisson
super-brackets of the classical theory ought be replaced by
super-commutators; if $\vec{\pi}(x)=\dot{\vec{\phi}}(x)$ is the
momentum,
\begin{equation}
\left[ \hat{\vec{\phi}}^a(x),\hat{\vec{\pi}}^b(y) \right]=i
\delta^{ab} \delta(x-y) \qquad , \qquad \left\{
\hat{\vec{\psi}}_\alpha^a(x),\hat{\vec{\psi}}_\beta^b(y) \right\}=
\delta^{ab}\delta_{\alpha\beta} \delta(x-y) \label{eq:comm}
\end{equation}
in the natural system of units $\hbar=c=1$.

The quantum SUSY charges
\[
\hat{Q}=\int dx \left\{ \gamma^\mu\gamma^0\hat{\vec{\psi}}
\partial_\mu\hat{\vec{\phi}}+i
\gamma^0\hat{\vec{\psi}}\vec{\nabla}\hat{W}\right\}
\]
becomes also operators, and the quantum SUSY algebra
\begin{equation}
\left\{ \hat{Q}_\alpha,\hat{Q}_\beta\right\} = 2
(\gamma^\mu\gamma^0)_{\alpha\beta} \hat{P}_\mu-2
\gamma_{\alpha\beta}^1\hat{T} \label{eq:qalg}
\end{equation}
is given not only in terms of the energy-momentum N\"other
invariants,

\[
\hat{H}=\hat{P}_0=\frac{1}{2} \int dx \left\{
\hat{\vec{\pi}}\hat{\vec{\pi}}+\frac{d\hat{\vec{\phi}}}{dx}\frac{d\hat{\vec{\phi}}}{dx}+
i\hat{\vec{\psi}}^t\gamma^5
\frac{d\hat{\vec{\psi}}}{dx}+\vec{\nabla}\hat{W}
\vec{\nabla}\hat{W}+\hat{\bar{\vec{\psi}}}
\vec{\vec{\Delta}}\hat{W} \hat{\vec{\psi}}\right\}
\]
\[
\hat{P}_1=\int dx \, \left\{ \hat{\vec{\pi}}\cdot
\frac{d\hat{\vec{\phi}}}{dx}+\frac{i}{2}
\hat{\vec{\psi}}\frac{d\hat{\vec{\psi}}}{dx}\right\}
\]
but also includes the central charge:
\[
\hat{T}=\int dx \, \frac{d\hat{\vec{\phi}}}{dx} \cdot
\vec{\nabla}\hat{W} \quad ,
\]
and old friend in the disguise of Bogomolny bound.

One can ask a natural question: what is the r\"ole of the
classical super-solitons in the quantum theory ? In perturbation
theory $\footnote{For a lucid analysis of the mathematical
meaning of the Fock space and perturbation theory see
\cite{Rab} }$ one expands $H$ around $\vec{\Phi}_c$ and splits the
Hamiltonian in \lq\lq free" and \lq\lq interaction parts:
$\hat{H}=\hat{H}_0+\hat{H}_{\rm int}$,
\begin{equation}
\hat{H}_0=\frac{1}{2}\int dx \left\{ \left[
\hat{\vec{\pi}}\hat{\vec{\pi}}+\frac{d\hat{\vec{\phi}}}{dx}\frac{d\hat{\vec{\phi}}}{dx}
+\hat{\vec{\phi}}\left( \left.
\vec{\vec{\Delta}}W\right|_{\vec{\phi}_c}\right)^2
\hat{\vec{\phi}}\right]\right. \left. +\left[ i \hat{\vec{\psi}}^t
\gamma^5 \frac{\hat{\vec{\psi}}}{dx}+ \hat{\vec{\bar{\psi}}}
\left. \vec{\vec{\Delta}}W\right|_{\vec{\phi}_c}
\hat{\vec{\psi}}\right] \right\}
\end{equation}
The plane super-waves (\ref{eq:planB})-(\ref{eq:planF}) are
compatible with the canonical quantization rules (\ref{eq:comm})
if the Fourier coefficients become non-commuting operators that
satisfy:
\begin{equation}
\left[ \hat{a}_a^\dagger(k_j),\hat{a}_b(q_l)\right]
=\delta_{ab}\delta_{k_jq_l} \qquad , \qquad \left\{
\hat{u}^\dagger_{a\alpha}(k_j),\hat{u}_{b\beta}(q_l)\right\}
=\delta_{ab} \delta_{\alpha\beta}\delta_{k_jq_l} \label{eq:fcomm}
\end{equation}
From (\ref{eq:fcomm}) one easily derives the spectrum of the
operators $\hat{N}^B_a(k_j)=\hat{a}^\dagger_a(k_j)\hat{a}_a(k_j)$
and
$\hat{N}^F_{a\alpha}(k_j)=\hat{u}\dagger_{a\alpha}(k_j)\hat{u}_{a\alpha}(k_j)$:
\[
\hat{N}^B_a(k_j)|n^B_a(k_j)\rangle=n^B_a(k_j)|n^B_a(k_j)\rangle
\quad , \quad
\hat{N}^F_{a\alpha}(k_j)|n^F_{a\alpha}(k_j)\rangle=n^F_{a\alpha}(k_j)|n^F_{a\alpha}(k_j)\rangle
\]
where $n^B_a(k_j)$ is a natural number and
$n^F_{a\alpha}(k_j)=0,1$. Because ( after normal-ordering )
\[
\hat{H}_0=\sum_{j=1}^N
\omega_j\sum_{a,k_j}\left(\hat{N}^B_a(k_j)+\hat{N}^F_{a\alpha}(k_j)\right)\quad
,
\]
the ground state of the free theory -the eigen-state with smallest
eigenvalue of $\hat{H}_0$- is the vacuum state:
\begin{eqnarray*}| 0 \rangle &=& \otimes{\atop j,a,k_j}|n^B_a(k_j)=0\rangle \otimes{\atop \alpha}
|n^F_{a\alpha}(k_j)=0\rangle\\ \hat{a}_a(k_j)\, |0\rangle &=&
\hat{u}_{a\alpha}(k_j)\, |0\rangle =0,\quad \forall j,a,\alpha,k_j
\end{eqnarray*}
Thus, the expectation value of $\hat{\vec{\phi}}(x^\mu)$ at the
vacuum is
\[
\langle 0|\hat{\vec{\phi}}(x^\mu)|0\rangle =\vec{\phi}_c
\]
and there is a one-to-one correspondence between the ground states
of the quantum free theory and the homogeneous super-solutions of
the classical theory.

The two types of one-particle states
\[
\hat{a}_a^\dagger (k_j) |0\rangle = |n^B_a(k_j)=1 \rangle \quad ,
\quad \hat{u}^\dagger_{a\alpha}(k_j) |0\rangle = |
n^F_{a\alpha}(k_j)=1 \rangle
\]
are degenerated in energy, which is $\omega_j$. There is a
one-to-one correspondence between plane super-waves in the
classical theory and the one-particle states with definite
momentum in the free quantum theory. Multi-particle states show
the statistics of each type of particle: bosonic, the state
remains the same under the exchange of two particles or fermionic,
the state changes sign under such an exchange.

Before of trying to identify the states in the quantum theory
related to the BPS super-solitons there is the need of addressing
a very delicate point: we are dealing in fact not with
operator-valued functions but with operator-valued distributions.
Note the delta-functions in (\ref{eq:comm}). The na\"if SUSY
algebra relations (\ref{eq:qalg}) are therefore non-sense because
undefined products of distributions at the same points of
space-time are involved. Fortunately, there is a unique method for
calculating equal time (super)commutators from the (known) Green's
functions of renormalized perturbation theory: the
Bjorken-Johnson-Low limit, \cite{Jac}.

Consider the Fourier transform of the matrix elements of the
\lq\lq chronological " product of two supercharge operators
between any two states $|S_1\rangle , |S_2\rangle$ in the Fock
space:
\begin{eqnarray}
T(q^\mu)&=&\int d^2 x e^{i q^\mu x_\mu}\left< S_1 \right| T
\hat{Q}_\alpha (x^\mu) \hat{Q}_\beta(0) |S_2\rangle \\
T\hat{Q}_\alpha(x^\mu)\hat{Q}_\beta(0)&=&
\hat{Q}_\alpha(x^\mu)\hat{Q}_\beta(0)\quad , \quad x^0>0 \nonumber
\\&=&
\hat{Q}_\beta(0)\hat{Q}_\alpha(x^\mu)\quad , \quad x^0<0 \nonumber
\end{eqnarray}
The BJL definition of such matrix elements is:
\begin{equation}
\lim_{q^0 \rightarrow \infty} q^0 T(q^\mu)
\underbrace{\equiv}_{{\rm def}} i \int dx e^{-i q x} \left< S_1
\right| \{ \hat{Q}_\alpha (0,x), \hat{Q}_\beta(0,0) \} |S_2\rangle
\label{eq:BJL}
\end{equation}
We also need the Wick's theorem. If,
\[
\left< 0 \right| T \hat{\phi}_a(x_1^\mu) \hat{\phi}_a (x_2^\mu)
|0\rangle= \Delta_{aa}^B(x_1^\mu-x_2^\mu) \qquad , \qquad \left< 0
\right| T \hat{\psi}^t_a(x_1^\mu) \hat{\psi}_a (x_2^\mu)
|0\rangle= \Delta_{aa}^F(x_1^\mu-x_2^\mu)
\]
are the one-particle Green's functions, the chronological product
of a string of n field operators is:
\begin{eqnarray}
&& TA_1(x_1^\mu)... A_n(x^\mu_n) = : A_1(x_1^\mu)... A_n(x^\mu_n):
+\\ && + \Delta(x_1^\mu-x_2^\mu) :
A_3(x_3^\mu)...A_n(x_n^\mu):...+ ...+\\&&+ \Delta(x_1^\mu-x_2^\mu)
\Delta(x_3^\mu-x_4^\mu) ... \Delta(x_{n-1}^\mu-x_n^\mu) \qquad ,
\end{eqnarray}
where $:A_1(x_1^\mu) \cdots A_n(x_n^\mu):$ is the normal-ordered
product - all the creation operators are put to the left of all
the annihilation operators -. Here, by $A(x^\mu)$ we mean either
$\hat{\phi}_a(x^\mu)$ or $\hat{\psi}_{a\alpha}(x^\mu)$ and by
$\Delta(x^\mu-y^\mu)$ we denote either
$\Delta_{aa}^B(x^\mu-y^\mu)$ or $\Delta_{aa}^F(x^\mu-y^\mu)$.

The procedure for finding the quantum (super)commutator
(\ref{eq:qalg}) is now clear:
\begin{itemize}
\item Wick's theorem applied to $T(q^\mu)$ tells what states
$|S_1\rangle , |S_2\rangle $ give a non-zero answer.

\item Choose $|S_1\rangle$ and  $|S_2\rangle$ accordingly, and
compute $\lim_{q^0 \rightarrow \infty} q^0 T(q^\mu)$.

\item After a long calculation one finds a result compatible with
the new relation:
\begin{equation}
\{\hat{Q}_\alpha, \hat{Q}_\beta\} = 2 (\gamma^\mu
\gamma^0)_{\alpha \beta} \hat{P}_\mu - 2 \gamma_{\alpha \beta}^1
\hat{T}_R \label{eq:qqalg}
\end{equation}
where $\hat{T}_R=\hat{T}+\frac{1}{4 \pi} \left| \int dx \frac{d
}{dx}(\Delta \hat{W}) \right|,  \Delta W= \vec\nabla \vec\nabla
W$.
\end{itemize}
The quantum SUSY algebra is $\underline{{\rm anomalous}}$: the
classical and quantum central charges differ in something
proportional to the difference between the values of the Laplacian
of the super-potential at $x=\pm\infty$. Only if the
super-potential is harmonic, the central charge does not receive
quantum corrections; but a harmonic super-potential is the
necessary condition for the existence of ${\cal N}=2$
super-symmetry.

From(\ref{eq:qqalg}) one writes the Hamiltonian operator in the
form:
\[
\hat{P}_0=(\hat{Q}_1\pm \hat{Q}_2)^2+ \left| \hat{T}_R \right|
\]
Thus, the expectation value of the energy operator at any state
$|S\rangle$ satisfies the inequality:
\[
\left< S \right| \hat{P}_0 |S \rangle \geq \left< S \right|
\,|\hat{T}_R|\, |S \rangle
\]
Equality is attained by  the quantum BPS states -$\left< BPS
\right|(\hat{Q}_1\pm \hat{Q}_2)^2  |BPS \rangle =0 $ -

\begin{eqnarray*}
\left< BPS \right| \hat{P}_0 |BPS \rangle &=& \left<
BPS\right||\hat{T}_R ||BPS\rangle =
\\ = \left| W\left( < \hat{\vec{\phi}}
>|_\infty \right) - W\left( < \hat{\vec{\phi}}
>|_{-\infty} \right)\right|&+&\frac{1}{4 \pi}\left|  \Delta
W\left( < \hat{\vec{\phi}}
>|_\infty \right) - \Delta W\left( < \hat{\vec{\phi}}
>|_{-\infty} \right)\right|
\end{eqnarray*}
The quantum BPS states are coherent states such that
$\left<BPS\right|\hat{\vec{\phi}}(x)|BPS\rangle ,
\left<BPS\right|\hat{\vec{\psi}}(x)|BPS\rangle $ are the classical
super-solitons which solve (\ref{eq:ssol}).

\subsection{The super-symmetric BNRT model}
Let us study a $N=2$ ${\cal N}=1$ super-symmetric system where the
super-potential is.
\[
W[\vec{\Phi}]=\sqrt{\lambda} \vec{\Phi} \cdot \vec{e}_1 \left[
\frac{4}{3} (\vec{\Phi}\cdot \vec{e}_1)^2-\frac{m^2}{\lambda}+2
\sigma^2 (\vec{\Phi}\cdot \vec{e}_2)^2 \right] \qquad ,
\]
and $\lambda$ , $m$ are coupling constants with dimensions of
inverse length whereas $\sigma^2$ is a non-dimensional parameter.
For more details on the so-called BNRT model, see \cite{AMJ1}.

It is convenient to use non-dimensional super-space and field
variables:
\[
x^\mu \rightarrow \frac{1}{m} x^\mu, \hspace{1cm} \theta_\alpha
\rightarrow \frac{1}{\sqrt{m}} \theta_\alpha, \hspace{1cm} d
\theta_\alpha \rightarrow \sqrt{m} d \theta_\alpha
\]
\[
\vec{\Phi} \longrightarrow  \frac{m}{\sqrt{\lambda}}
\vec{\Phi}\qquad \Leftrightarrow \qquad\{ \vec{\phi} \rightarrow
\frac{m}{\sqrt{\lambda}} \vec{\phi} \quad , \quad \vec{\psi}
\rightarrow m \sqrt{\frac{m}{\lambda}} \vec{\psi} \quad , \quad
\vec{F} \rightarrow \frac{m^2}{\sqrt{\lambda}} \vec{F}\}
\]
Thus, $ D\vec{\Phi} \rightarrow m \sqrt\frac{m}{\lambda}
D\vec{\Phi}$ , $ W[\vec{\Phi}] \rightarrow \frac{m^3}{\lambda}
W[\vec{\Phi}]$ and the ${\cal N}=1$ SUSY action reads:
\[
S[\vec{\Phi}]= - \frac{m^2}{\lambda} \left\{ \int d^2 x d^2 \theta
\left( \frac{1}{2} \bar{D}\vec{\Phi} D \vec{\Phi}+ 2
\vec{\Phi}\vec{e}_1[\frac{4}{3}(\vec{\Phi}\vec{e}_1)^2-1+2\sigma^2(\vec{\Phi}\vec{e}_2)^2]
\right) \right \}
\]
From $\vec\nabla W= [(4 \phi_1^2+2 \sigma^2 \phi_2^2-1)
\vec{e}_1+4 \sigma^2 \phi_1 \phi_2 \vec{e}_2]$, and
\[
\vec{\vec{\Delta}} W= 4 [2 \phi_1 \vec{e}_1 \otimes \vec{e}_1+
\sigma^2 \phi_2 (\vec{e}_1 \otimes \vec{e}_2+ \vec{e}_2 \otimes
\vec{e}_1)+\sigma^2 \phi_1 \vec{e}_2 \otimes \vec{e}_2]\quad ,
\]
we obtain the on shell SUSY action:
\[
S_B[\vec{\phi}]= \frac{m^2}{2 \lambda} \int d^2 x \left\{
\partial_\mu \vec\phi \partial^\mu \vec\phi - (4 \phi_1^2+2 \sigma^2 \phi_2^2-1)^2-16 \sigma^4
\phi_1^2 \phi_2^2 \right\} ;  S_F[\vec{\psi}]= \frac{m^2}{2
\lambda} \int d^2 x \left\{i \vec{\bar{\psi}} \gamma^\mu
\partial_\mu \vec\psi  \right\}
\]
\[
S_{BF}[\vec{\phi},\vec{\psi}]= \frac{m^2}{2 \lambda} \int d^2 x
\left\{ 8 \bar\psi_1 \phi_1 \psi_1+4 \sigma^2(\bar\psi_1 \phi_2
\psi_2+\bar\psi_2 \phi_2 \psi_1)+\bar\psi_2 \phi_1 \psi_2 \right\}
\]
Observe that $\Delta W=8[2+\sigma^2] \phi_1$  and the
super-potential is harmonic only for $\sigma=\pm i \sqrt{2}$,
where one finds the celebrated Wess-Zumino model, see \cite{AMJ2}.

Besides the super-Poincar\`e symmetry the system is invariant
under the discrete group ${\mathbb G}={\mathbb Z}_2\times {\mathbb Z}_2$
generated by the internal reflections
$\pi_1\vec{\Phi}=-\Phi_1\vec{e}_1+\Phi_2\vec{e}_2$ and
$\pi_2\vec{\Phi}=\Phi_1\vec{e}_1-\Phi_2\vec{e}_2$. The homogeneous
super-solutions are the critical points of $W$,
\[
\vec{\Phi}_{v_{\pm}}^{(1)}=\pm \frac{1}{2} \vec{e}_1
 \hspace{2cm} \vec{\Phi}_{v_{\pm}}^{(2)}=\pm \frac{1}{\sigma \sqrt{2}}
 \vec{e}_2
\]
and the vacuum manifold ${\cal V}$ is the union of two orbits of
${\mathbb G}$. Thus, the moduli space of vacua ${\cal M}=\frac{{\cal
V}}{{\mathbb G}}=(\vec{\Phi}^{(1)}_v,\vec{\Phi}^{(2)}_v)$ contains
two points.

The plane super-waves transmute to the fundamental Bosons and
Fermions in the quantum world. Because
\[
\vec{\vec{\Delta}} W \left(\vec{\phi}_{v_{\pm}}^{(1)} \right)= \pm
[4 \vec{e}_1 \otimes \vec{e}_1+ 2 \sigma^2  \vec{e}_2 \otimes
\vec{e}_2]
\]
we identify  $\vec{\Phi}_{v_{+}}^{(1)}$ and
$\vec{\Phi}_{v_{-}}^{(1)}$ respectively as minimum and maximum of
$W$.
\begin{figure}[ht]
\centerline{ \includegraphics[height=5cm]{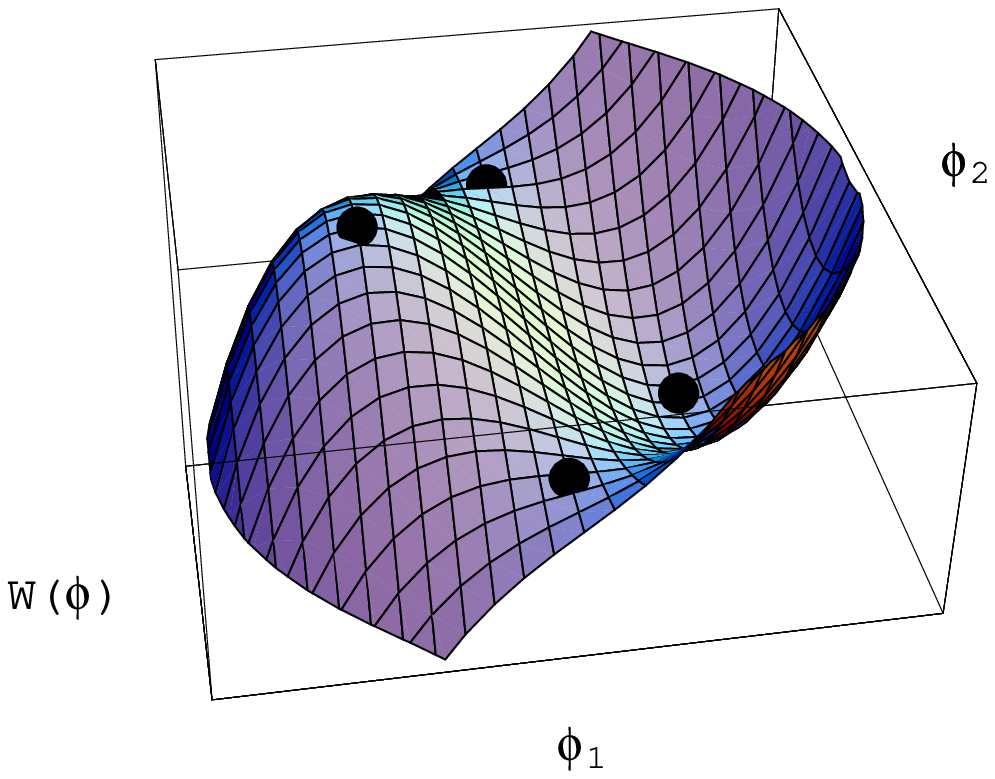} \hspace{3cm}
\includegraphics[height=5cm]{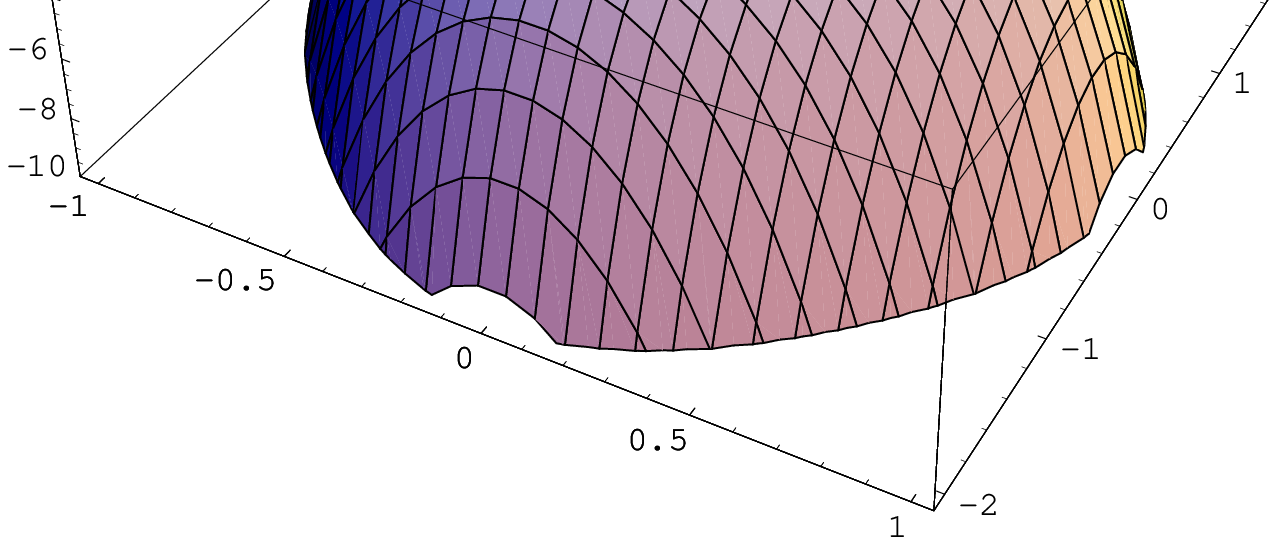}} \caption{Plots of $W$ and $U$
as functions of $\phi_1$ and $\phi_2$.}
\end{figure}

As dictated by super-symmetry, we see that there are two
Boson/Fermion branches - degenerated in mass - on the
$\vec{\Phi}^{(1)}_v$ point in ${\cal M}$:
\[
\begin{array}{|cc|c|cc|} \hline &&&& \\[-0.3cm]  & &
  \omega_1(k_1=0) & & \omega_2(k_2=0) \\ \hline &&&& \\
  [-0.3cm]
 {\rm Bosons} & & 4 m & & 2 \sigma^2 m \\ \hline &&&& \\[-0.3cm]
 {\rm Fermions} & & 4 m & & 2 \sigma^2 m \\ \hline
\end{array}
\]

We could be tempted also to put the minus sign in the Fermi
masses, but the Dirac sea paradigm and the \lq\lq particle "
\lq\lq hole " identity in the Majorana representation liberate us
of such a worry.

Simili modo,
\[
\vec{\vec{\Delta}} W \left(\vec{\phi}_{v_{\pm}}^{(2)} \right)= \pm
2 \sqrt{2} \sigma [\vec{e}_1 \otimes \vec{e}_2+ \vec{e}_2 \otimes
\vec{e}_1]
\]
tells us that both $\vec{\Phi}_{v_{+}}^{(2)}$ and
$\vec{\Phi}_{v_{-}}^{(2)}$ are saddle points of $W$. A
$\frac{\pi}{4}$ rotation in ${\mathbb R}^2$ to
$2\vec{\Phi}=\Phi_+(\vec{e}_1+\vec{e}_2)+\Phi_-(\vec{e}_1-\vec{e}_2)$
allows us to identify the two Boson/Fermion branches on the
$\vec{\Phi}_v^{(2)}$ point in ${\cal M}$:
\[
\begin{array}{|cc|c|cc|} \hline &&&& \\[-0.3cm]  & &
  \omega_+(k_+=0) & & \omega_-(k_-=0) \\ \hline &&&& \\
  [-0.3cm]
 {\rm Bosons} & & 2 \sqrt{2}\sigma m & & 2 \sqrt{2} \sigma^2 m \\ \hline &&&& \\[-0.3cm]
 {\rm Fermions} & & 2 \sqrt{2} \sigma m & & 2 \sqrt{2} \sigma^2 m \\ \hline
\end{array}
\]

What about the super-solitons that gives rise to extended BPS
states in the quantum theory? In this model, the bosonic ODE
system (\ref{eq:ssol}) becomes
\begin{equation}
\frac{d \phi_1}{dx}= \mp 4 \phi_1^2+ 2 \sigma^2 \phi_2^2-1 \qquad
, \qquad \frac{d \phi_2}{dx}=\mp 4 \sigma^2 \phi_1 \phi_2
\end{equation}
The flows lines of $\pm{\rm grad}W$ are the solutions of the ODE:
\begin{equation}
\frac{d \phi_1}{d \phi_2}=\frac{4 \phi_1^2+ 2 \sigma^2 \phi_2^2-1
}{4 \sigma^2 \phi_1 \phi_2} \label{eq:flow}
\end{equation}
which admits the integrating factor
$|\phi_2|^{-\frac{2}{\sigma}}\phi_2^{-1}$, if $\sigma\neq 1$ and
$\sigma\neq 0$, thereby allowing us to find all the flow-lines as
the family of curves
\begin{equation}
\phi_1^2 + \frac{\sigma}{2 (1-\sigma)} \phi_2^2 = \frac{1}{4}+
\frac{c}{2 \sigma} |\phi_2|^{\frac{2}{\sigma}} \label{eq:traba1}
\end{equation}
parametrized by the real integration constant $c$. There is a
critical value
\[
c^S=\frac{1}{4} \frac{\sigma}{1-\sigma} \left(
2\sigma\right)^{\frac{\sigma+1}{\sigma}}
\]
and the behaviour of a particular curve in the (\ref{eq:traba1})
family is described in the following items:
\begin{itemize}
\item For $c \in (-\infty,c^S)$, formula (\ref{eq:traba1}) describes
closed curves in the internal space ${\mathbb R}^2$ that connect the
vacua $\vec{\Phi}_{v_+}^{(1)}$ and $\vec{\Phi}_{v_-}^{(1)}$, see
Figure 2. Thus, they provide a kink family in the topological
sector {\it (1;1}. Henceforth, we refer to these kinks as {\bf
TK2}$^{(1;1)}(c)$. A fixed value of $c$ determines four members in
the kink variety related amongst one another by spatial parity and
internal reflections. The kink moduli space is defined as the
quotient of the kink variety by the action of the symmetry group:
\[
{\cal M}_{\rm K}=\frac{{\cal V}_{\rm K}}{{\mathbb P}\times{\mathbb
G}}=(-\infty , c^S ),
\]
the real open half-line parametrized by $c$.

\item In the range $c\in (c^S,\infty)$, equation (\ref{eq:traba1})
describes open curves and no vacua are connected. These ${\rm
grad}\,W$ flow-lines are infinite energy solutions that do not
belong to the configuration space ${\cal C}$, see Figure 2.

\item At the other point of the boundary of ${\cal M}_{\rm K}$, $c=c^S$,
we find the separatrices between bounded and unbounded motion and
the envelop of all kink orbits in the {\it (1;1)} topological
sector, see Figure 2.
\end{itemize}

We briefly discuss the $\sigma=1$ case. The $\sigma=0$ case is not
interesting because the $\phi_2$ dependence disappears in the
potential: it is a \lq\lq direct sum" of an $N=1$ $\phi^4$ model
and an $N=1$ free model. Integration of $(\ref{eq:flow})$ when
$\sigma=1$ gives
\begin{equation}
\phi_1^2-\phi_2^2 \left( \frac{c}{2}+\log |
\phi_2|\right)=\frac{1}{4} \label{eq:traba2}
\end{equation}
where the kink trajectories now appear in the $c\in (-\infty,
c^S]$ range, with $c^S=-1+\ln 2$. The description of the kink
orbits is analogous to the description for $\sigma\neq 1$ above.
\begin{figure}[ht] \centerline{\includegraphics[height=4.cm]{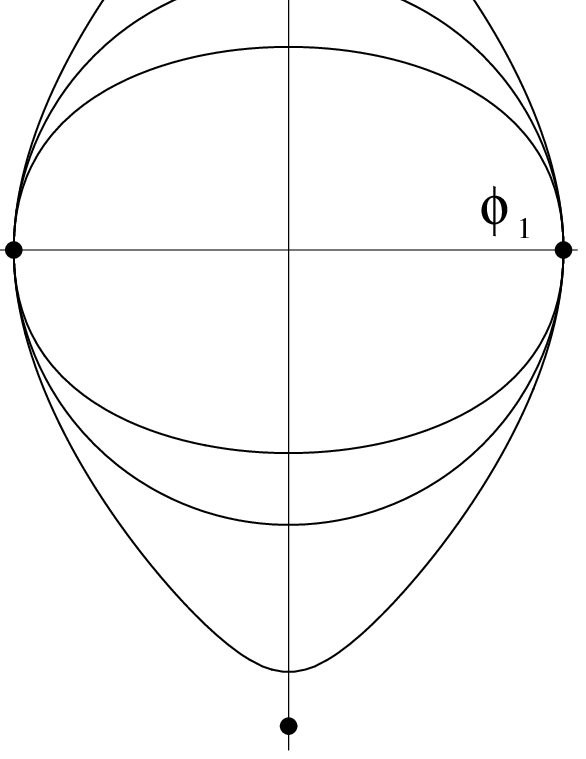} \hspace{1.5cm} \includegraphics[height=4.cm]{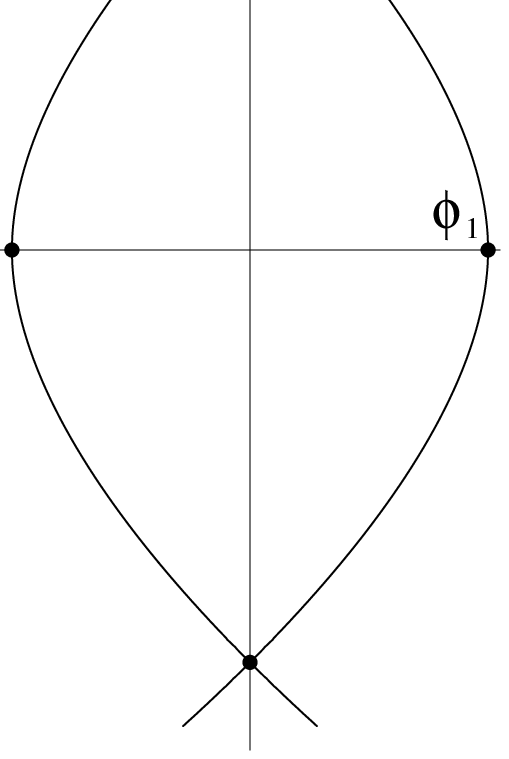}
\hspace{1.5cm}\includegraphics[height=4.cm]{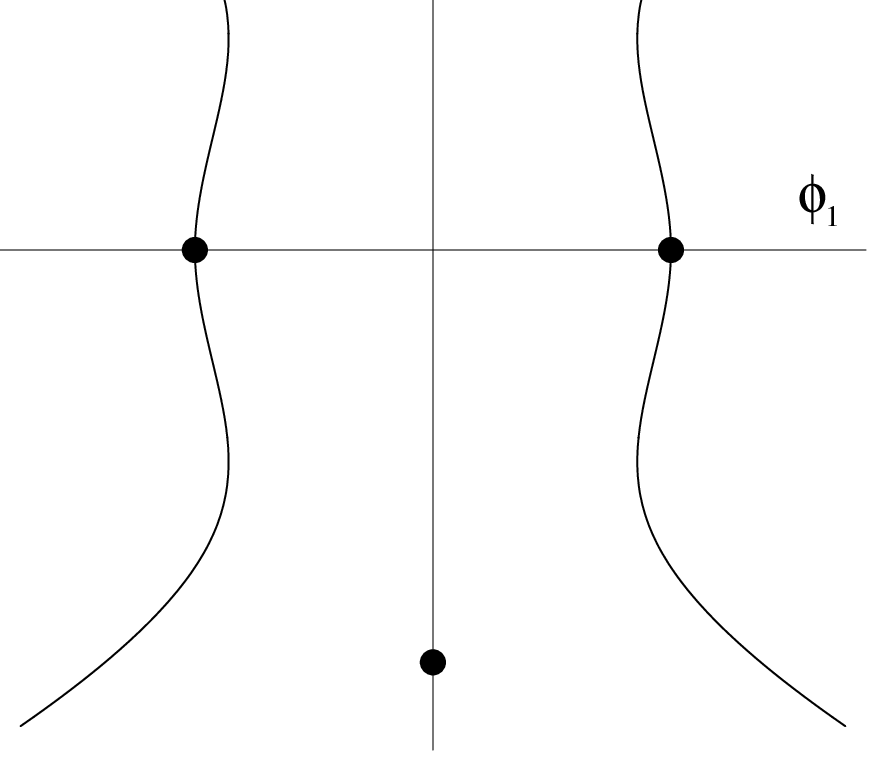}} \caption{\small
\it Flow-lines given by (\ref{eq:traba1}): for $c \in
(-\infty,c^S)$ (left), $c=c^S$ (middle), and $c \in (c^S ,\infty)$
(right).}
\end{figure}

With the exception of the one-component topological kinks with
$\phi_1=0$, there are no other flow lines connecting the vacua
$\vec{\Phi}_{v_+}^{(2)}$ and $\vec{\Phi}_{v_-}^{(2)}$, for generic
$\sigma^2$. If $\sigma^2<2$, the spectrum of the second order
fluctuation operator around the ${\rm TK}1^{(1;1)}$ kinks shows
some negative eigen-value and these kinks are unstable -non-BPS-.
For the critical values $\sigma^2=\frac{4}{l(l+1)-4}$,
$l=3,4,\cdots$, there are Jacobi fields in the spectrum announcing
the existence of a whole continuous family of these non-BPS
solitons and, probably, the availability of a non-continuously
differentiable second superpotential.

 If $\sigma^2=2$, there exists a \lq\lq honest" superpotential,
 where the r$\hat{\rm o}$les of $\Phi_1$ and $\Phi_2$ are
 exchanged; the \lq\lq vertical" $(1;1)$ topological sector is
 plenty of BPS solitons.
\clearpage
\section{${\cal N}=2$ Supersymmetric Gauge Theory}

\subsection{${\cal N}$=1 SUSY Yang-Mills theory}
\subsubsection{${\mathbb R}^{4|4}$ super-space and ${\cal N}=1$ SUSY
algebra} We now move to physical ${\mathbb R}^{1,3}$ Minkowski
space-time with the following choices of metric tensor and
associated Clifford algebra:
\[
g^{\mu\nu}={\rm diag} (1,-1,-1,-1)
\]
\[
\{\gamma^\mu,\gamma^\nu\}=2g^{\mu\nu} ,\qquad \gamma^\mu=\left(
\begin{array}{cc}0 & \sigma^\mu \\ \bar{\sigma}^\mu &
0 \end{array} \right) ;\qquad \gamma^5=\gamma^0\, \gamma^1\,
\gamma^2\, \gamma^3
\]
\[
\sigma^\mu=({\bf 1}_2,\sigma^1,\sigma^2,\sigma^3)\qquad ,\qquad
\bar{\sigma}^\mu=({\bf 1}_2,-\sigma^1,-\sigma^2,-\sigma^3)
\]
In this representation of the Clifford algebra the Dirac spinors
are the direct sum of Weyl spinors $\psi_\alpha \in
(\frac{1}{2},0)$, $\bar{\chi}^{\dot{\alpha}} \in (0,\frac{1}{2})$
that belong to the fundamental representations of ${\rm SO}(3,1)$,
the connected piece to the identity of the Lorentz group:
\[
\Psi_D(x^\mu)=\left ( \begin{array}{c}\psi_\alpha(x^\mu) \\
\bar{\chi}^{\dot{\alpha}}(x^\mu)
\end{array}\right)\in (\frac{1}{2},0)\oplus (0,\frac{1}{2})
\]
Spinor indices are raised and lowered with the anti-symmetric
$\varepsilon$-tensor:
$\psi_\alpha=\varepsilon_{\alpha\beta}\psi^\beta ,\quad
\bar{\chi}^{\dot{\alpha}}=\varepsilon^{\dot{\alpha}\dot{\beta}}\bar{\chi}_{\dot{\beta}}$.
Thus, $\psi\eta=\psi^\alpha\eta_\alpha ,\quad
\bar{\chi}\bar{\xi}=\bar{\chi}_{\dot{\alpha}}\bar{\xi}^{\dot{\alpha}}$
are scalar products, whereas $\psi\sigma^\mu\bar{\chi},\quad
 \bar{\xi}\bar{\sigma}^\mu\eta$ are vector combinations.

Odd coordinates in the ${\mathbb R}^{4|4}$ super-space  are Grassman
Weyl spinors, $\theta_\alpha $ , $\bar{\theta}^{\dot{\alpha}}$,
which satisfy:
\[
 \theta^2 \equiv\theta \theta=\theta^\alpha
\theta_\alpha=-2\theta_1\theta_2 ,\    \qquad,\qquad
\theta_{\alpha_1}\theta_{\alpha_2}+\theta_{\alpha_2}\theta_{\alpha_1}=\bar{\theta}^{\dot{\alpha}_1}\bar{\theta}^{\dot{\alpha}_2}+\bar{\theta}^{\dot{\alpha}_2}\bar{\theta}^{\dot{\alpha}_1}=0
\]
If $\varepsilon$ and $\bar{\varepsilon}$ are Grassman Weyl spinor
parameters, the vector fields
\[
Q_\alpha=\frac{\partial}{\partial\theta^\alpha}-i\sigma_{\alpha\dot{\alpha}}^\mu\bar{\theta}^{\dot{\alpha}}\partial_\mu
\qquad , \qquad
\bar{Q}_{\dot{\alpha}}=-\frac{\partial}{\partial\bar{\theta}^{\dot{\alpha}}}+i\theta^\alpha\sigma_{\alpha\dot{\alpha}}^\mu\partial_\mu
\]
are spinor generators of the super-translation group:
$P(x,\varepsilon,\bar{\varepsilon})=e^{{\displaystyle i[-x^\mu
P_\mu+\varepsilon Q+\bar{\varepsilon}\bar{Q}]}}$. Infinitesimally,
\[
\theta_\alpha \to \theta_\alpha + i\varepsilon_\alpha \qquad ,
\qquad  \bar{\theta}^{\dot{\alpha}} \to
\bar{\theta}^{\dot{\alpha}}
-i\bar{\varepsilon}^{\dot{\alpha}}\qquad , \qquad x^\mu\to
x^\mu+\varepsilon^\alpha\sigma^\mu_{\alpha\dot{\alpha}}\bar{\theta}^{\dot{\alpha}}+
\theta^\alpha\bar{\sigma}^\mu_{\alpha\dot{\alpha}}\bar{\varepsilon}^{\dot{\alpha}}
\]
The super-charges satisfy the anti-commutation relations:
\[
\{Q_\alpha,\bar{Q}_{\dot{\alpha}}\}=2 i
\sigma_{\alpha\dot{\alpha}}^\mu\partial_\mu \qquad ; \qquad
\{Q_\alpha,Q_\beta\}=\{\bar{Q}_{\dot{\alpha}},\bar{Q}_{\dot{\beta}}\}=0
.
\]
The covariant derivatives:
\[
D_\alpha=\frac{\partial}{\partial\theta^\alpha}+i\sigma_{\alpha\dot{\alpha}}^\mu\bar{\theta}^{\dot{\alpha}}\partial_\mu
\qquad , \qquad
\bar{D}_{\dot{\alpha}}=-\frac{\partial}{\partial\bar{\theta}^{\dot{\alpha}}}-i\theta^\alpha\sigma^\mu_{\alpha\dot{\alpha}}\partial_\mu
\]
which satisfy,
\[
\{D_\alpha,Q_\beta\}=\{D_\alpha,\bar{Q}_{\dot{\beta}}\}=\{\bar{D}_{\dot{\alpha}},Q_\beta\}=\{\bar{D}_{\dot{\alpha}},\bar{Q}_{\dot{\beta}}\}=0
\]
and
\[ \{D_\alpha,\bar{D}_{\dot{\alpha}}\}=-2 i
\sigma_{\alpha,\dot{\alpha}}^\mu\partial_\mu \qquad ; \qquad
\{D_\alpha,D_\beta\}=\{\bar{D}_{\dot{\alpha}},\bar{D}_{\dot{\beta}}\}=0
\quad ,
\]
will also play an important r$\hat{{\rm o}}$le in the formulation
of the theory.

We build ${\cal N}=1$ super-symmetric Yang-Mills theory ( without
matter hyper-multiplets ) out of two kinds of super-fields.

\subsubsection{The ${\cal N}=1$ super-symmetric action for chiral
super-fields} The chiral (anti-chiral) super-fields,
\[
\Phi(x^\mu,\theta,\bar{\theta})\, :\; {\mathbb R}^{4|4}
\longrightarrow {\rm adj}_{\mathbb C}\; {\rm SU(N)}
\]
are maps from the ${\cal N}=1$ super-space to the adjoint
representation of the $SU(N)$ group which are constrained by the
$\bar{D}_{\dot{\alpha}}\Phi=0$ chiral ( $D_\alpha\Phi^\dagger=0$
anti-chiral ) condition.

The chiral constraint is easily solved in terms of
$y^\mu=x^\mu+i\theta\sigma^\mu\bar{\theta}\ $ and $\theta$.
Because
$\bar{D}_{\dot{\alpha}}y^\mu=\bar{D}_{\dot{\alpha}}\theta=0$ one
finds
\[
\Phi (y,\theta)=\phi(y)+\sqrt{2}\theta\psi(y)+\theta^2F(y)
\]
and a chiral field contains a complex scalar field, $\phi(\equiv
\phi^a\frac{\tau^a}{2})$, a Weyl spinor field,
$\psi_\alpha\equiv\psi_\alpha^a \frac{\tau^a}{2}$, and a complex
auxiliary field, $F(\equiv F^a\frac{\tau^a}{2})$, all of them in
the adjoint representation of the $SU(N)$ gauge group - $\tau^a$
are the  generators of ${\rm Lie} SU(N)$-. The action of the
super-charges on the chiral superfields is:
\[
 \, \Phi \Rightarrow \, \left\{
\begin{array}{l} \delta \phi=i\sqrt{2}\varepsilon
\psi\\ \delta \psi = -\sqrt{2}\sigma^\mu\bar{\varepsilon}
\partial_{\mu} \phi+i\sqrt{2}\varepsilon F\\ \delta
F=-\sqrt{2}\bar{\varepsilon}\bar{\sigma}^{\mu} \partial_{\mu}
\psi\end{array} \right.
\]
Because $\delta \Phi=i \varepsilon\, Q$ and $\delta \Phi^\dagger=i
\bar{\varepsilon}\,\bar{Q}$ are super-derivations, the
infinitesimal action of $P(a^\mu,\varepsilon,\bar{\varepsilon})$
on the super-action functional
\[
S_0=\frac{1}{4}\int d^4x\; d^2\theta\; d^2\bar{\theta}\; {\rm
tr}\Phi^\dagger \Phi \quad ,
\]
is a total super-derivative and, thus, $S_0$ is ${\cal N}=1$ SUSY
invariant. The Berezin integral - ($\int d^2\theta\;
d^2\bar{\theta}\;\theta^2 \bar{\theta}^2=4$)- and the power
``series" expansion of the chiral super-field:
\begin{equation}
\Phi(y,
\theta)=\phi(x)+i\theta\sigma^\mu\bar{\theta}\partial_\mu\phi(x)-
\frac{1}{4}\theta^2\bar{\theta}^2\Box \phi(x)
+\sqrt{2}\theta\psi(x)-\frac{i}{\sqrt{2}}\theta^2\partial_\mu\psi(x)\sigma^\mu\bar{\theta}+\theta^2F(x)
\end{equation}
allow us to write $S_0$ in terms of the super-field components:
\begin{equation}
S_0 =\int d^4x\; {\rm tr}
(\partial_\mu\phi^\dagger\partial^\mu\phi-i\bar{\psi}\bar{\sigma}^\mu\partial_\mu\psi+F^\dagger
F) \quad .
\end{equation}
$S_0$ determines a free dynamics but one can switch on
interactions compatibles with ${\cal N}=1$ super-symmetry using a
chiral super-potential:
\[
S_1=\int d^4x[\int d^2\theta\;{\cal W}(\Phi)+\int
d^2\bar{\theta}\;\bar{{\cal W}}(\Phi^\dagger)] .
\]

\subsubsection{${\cal N}=1$ supersymmetric action for vector
super-fields}
The vector super-fields are also maps from the
super-space to the adjoint representation of $SU(N)$
\[
V(x^\mu,\theta,\bar{\theta})\, :\; {\mathbb R}^{4|4} \longrightarrow
{\rm adj}\; {\rm SU(N)}
\]
which satisfy a reality condition: $V=V^\dagger$. In the
Wess-Zumino gauge, the power expansion in the odd variables reads:
\[
V(x,\theta,\bar{\theta})=-\theta\sigma^\mu\bar{\theta}A_\mu(x)+i(\theta^2\bar{\theta}\bar{\lambda}(x)-\bar{\theta}^2\theta\lambda(x))+\frac{1}{2}\theta^2\bar{\theta}^2D(x)
\]
The super-connection $V$ includes the connection $A_\mu\equiv
A_\mu^a\frac{\tau^a}{2}$ on a $SU(N)$ bundle over ${\mathbb
R}^{1,3}$, its SUSY partner, the Weyl spinor
$\lambda\equiv\lambda^a \frac{\tau^a}{2}$, and the real auxiliary
field $D$. Chiral fields $\Lambda$ give rise to super gauge
transformations on $V$, $V'=V+\Lambda+\Lambda^\dagger$, and the
spinor field strength,
\[
W_\alpha=\frac{1}{8g}\bar{D}^2(e^{2gV}D_\alpha e^{-2gV}) \, \,  ,
\]
where $g$ is the coupling constant, transform under super-gauge
transformations as: 
\[
W'_\alpha=e^{-i2g\Lambda}W_\alpha \,
e^{i2g\Lambda} \, \, . 
\]
Note that $\bar{D}_{\dot{\alpha}}W_\alpha=0$ and
$W$ is a chiral spinor field. In the appropriate coordinates the
Grassman power expansion of $W$ reads,
\[
W_\alpha(y,\theta)=(-i\lambda_\alpha+\theta_\alpha D-i\sigma^{\mu \nu}\theta_\alpha F_{\mu
\nu}+\theta^2\sigma^\mu\nabla_\mu\bar{\lambda}_\alpha)(y)
\]
where the following tensors
\[
\sigma^{\mu\nu}=\frac{1}{4}(\sigma^\mu\bar{\sigma}^\nu-\sigma^\nu\bar{\sigma}^\mu)\quad;\quad
F_{\mu\nu}=\partial_\mu A_\nu-\partial_\nu
A_\mu+g[A_\mu,A_\nu]\quad ; \quad
\nabla_\mu\lambda_\alpha=\partial_\mu\lambda_\alpha+g[A_\mu, \lambda_\alpha]
\]
enter.

One sees easily that
\begin{eqnarray*}
S_2=-\frac{1}{2}\int d^4x\; d^2\theta\; {\rm tr}W^\alpha W_\alpha
=\int d^4x\;{\rm
tr}(-\frac{1}{4}F_{\mu\nu}F^{\mu\nu}+\frac{i}{4}F_{\mu\nu}\tilde{F}^{\mu\nu}-
i\lambda\sigma^\mu\nabla_\mu\bar{\lambda}+\frac{D^2}{2})\nonumber
\end{eqnarray*}
is a ${\cal N}$=1 SUSY and super-gauge invariant action. Moreover,
re-scaling $g V\to V$ and defining the complex gauge coupling
$\tau=\frac{\Theta}{2\pi}+i\frac{4\pi}{g^2}$ in terms of the
instanton angle $\Theta$, one writes $S_2$ in the form:
\begin{eqnarray*}
S_2&=&\frac{{\rm Im}\tau}{16 \pi}\int d^4x\; d^2\theta\; {\rm tr}
W^\alpha W_\alpha \\ &=&\frac{1}{g^2}\int d^4x\; {\rm
tr}(-\frac{1}{4}F_{\mu\nu}F^{\mu\nu}-i\lambda\sigma^\mu\nabla_\mu\bar{\lambda}+\frac{D^2}{2})+\frac{\Theta}{32\pi^2}\int
d^4x\;F_{\mu\nu}\tilde{F}^{\mu\nu}
\end{eqnarray*}

The minimal coupling principle -which dictates the gauge
interactions between charged fields by changing derivatives by
covariant derivatives- is generalized to the super-symmetric world
by replacing $S_0$ by
\begin{eqnarray*}
S_3&=&\frac{1}{4g^2}\int d^4x\; d^2\theta\; d^2\bar{\theta}\;{\rm
tr}\;\Phi^\dagger e^{-2V}\Phi  =\int d^4x \;{\rm
tr}(|\nabla_\mu\phi|^2-i\bar{\psi}\bar{\sigma}^\mu\nabla_\mu\psi+F^\dagger
F)\\ &-&\int d^4x\;{\rm tr}(\phi^\dagger
[D,\phi]-\sqrt{2}i(\phi^\dagger\{\lambda,\psi\}-\bar{\psi}[\bar{\lambda},\phi])
\end{eqnarray*}
and adding $S_3$ to the $S_1$ and $S_2$ actions. Thus, the action
of the so called ${\cal N}=1$ SUSY minimal gauge theory is:
\begin{eqnarray}
S_3+S_2+S_1&=&\frac{1}{4g^2}\int d^4x\; d^2\theta\;
d^2\bar{\theta}\;{\rm tr}\;\Phi^\dagger e^{-2V}\Phi +\frac{{\rm
Im}\tau}{16 \pi}\int d^4x\; d^2\theta\; {\rm tr}\; W^\alpha
W_\alpha \nonumber \\&+&\int d^4x\;[\int d^2\theta {\cal
W}(\Phi)+\int d^2\bar{\theta}\bar{{\cal W}}(\Phi^\dagger)]
\end{eqnarray}

\subsection{${\cal N}$=2 supersymmetric Yang-Mills theory}

\subsubsection{${\cal N}=2$ supersymmetry and holomorphy}
Four dimensional ${\cal N}=2$ supersymmetric theories are based on
the extended super-space ${\mathbb R}^{4|(4,4)}$  where the odd
dimensions are parametrized by two sets of Weyl spinor pairs: $
\theta_\alpha^I,\ \bar{\theta}^{\dot{\alpha}}_I,\quad I=1,2$.
Sometimes we shall also use the notation:
$\theta_\alpha^1=\theta_\alpha$ ,
$\theta_\alpha^2=\tilde{\theta}_\alpha$ ,
$\bar{\theta}^{\dot{\alpha}}_1=\bar{\theta}^{\dot{\alpha}}$ ,
$\bar{\theta}^{\dot{\alpha}}_2=\bar{\tilde{\theta}}^{\dot{\alpha}}$
, in order to compare the ${\cal N}=2$ with the ${\cal N}=1$
theory.

There are accordingly, two pairs of ${\cal N}=2$ super-charges,
vector fields that generate super-translations in ${\mathbb
R}^{4|(4,4)}$,
\[
Q_\alpha^I=\frac{\partial}{\partial\theta^\alpha_I}-i\sigma_{\alpha\dot{\alpha}}^\mu\bar{\theta}^{\dot{\alpha}I}\partial_\mu
\qquad , \qquad \bar{Q}_{\dot{\alpha}
I}=-\frac{\partial}{\partial\bar{\theta}^{\dot{\alpha}I}}+
i\theta^\alpha_I\sigma_{\alpha\dot{\alpha}}\partial_\mu ,
\]
and close the super-algebra:
\[
\{Q_\alpha^I,\bar{Q}_{\dot{\alpha}J}\}=2\sigma^\mu_{\alpha\dot{\alpha}}\hat{P}_\mu\delta_J^I
\]
\[
\{Q_\alpha^I,Q_\beta^J\}=0,\qquad
\{\bar{Q}_{I\dot{\alpha}},\bar{Q}_{J \dot{\beta}}\}=0 \quad .
\]

An ${\cal N}=2$ chiral super-field $\Psi
(x,\theta^I,\bar{\theta}_I$ is an ${\cal N}=2$ super-field which
satisfies the constraints:
$\bar{D}_I^{\dot{\alpha}}\Psi(x,\theta^I,\bar{\theta}_I)=0$.
Because $\bar{D}_I^{\dot{\alpha}}\tilde{y}^\mu=0$, one obtains
easily the general solution of the constraints as a function of
$\tilde{y}^\mu=y^\mu+i\tilde{\theta}\sigma^\mu\bar{\tilde{\theta}}$:
\[
\Psi(\tilde{y},\theta)=\Phi(\tilde{y},\theta)+\sqrt{2}\tilde{\theta}^\alpha
W_\alpha(\tilde{y},\theta)+\tilde{\theta}^\alpha\tilde{\theta}_\alpha
G(\tilde{y},\theta)
\]
Here, $\Phi(\tilde{y},\theta)$ and $G(\tilde{y},\theta)$ are
${\cal N}=1$ chiral super-fields whereas
$W_\alpha(\tilde{y},\theta)$ is an ${\cal N}=1$ chiral spinor
super-field. We choose for $\Phi$ and $W_\alpha$ the ${\cal N}=1$
chiral super-fields of SUSY Yang-Mills theory discussed in the
previous subsection. Moreover, we impose the extra reality
constraints
$D^{I\alpha}D^J_\alpha\Psi=\bar{D}_I^{\dot{\alpha}}\bar{D}_{J\dot{\alpha}}\Psi^\dagger
$ , which allow to solve for the auxiliary field $G$ in terms of
$\Phi$ and $V$:
\[
G(\tilde{y},\theta)=-\frac{1}{2}\int
d^2\bar{\theta}\;[\Phi(x^\mu+i\tilde{\theta}\sigma^\mu\bar{\tilde{\theta}},\theta,\bar{\theta})]^\dagger
\exp[-2
V(x^\mu+i\tilde{\theta}\sigma^\mu\bar{\tilde{\theta}},\theta,\bar{\theta})]\quad
\]
Thus, all the basic field of minimal SUSY Yang-Mills are
encompassed in the ${\cal N}=2$ chiral super-field $\Psi$ and
shown - together with their spins- in the next array:
\[
\begin{array}{lccc}
 &\quad s=1\quad &\quad s=1/2\quad&\quad s=0\quad\\
W_\alpha&A_\mu&\lambda_\alpha& \\ \Phi& &\psi_\alpha &\phi
\end{array} \quad .
\]

Note that the action
\begin{eqnarray*}
S&=&{\rm Im}[\frac{\tau}{16 \pi}\int d^4x\; d^2\theta\;
d^2\tilde{\theta}\;\frac{1}{2}{\rm tr}\Psi^2]\\ &=&{\rm
Im}\frac{\tau}{16 \pi}\int d^4x\;{\rm tr}[\int d^2\theta\;
W^\alpha W_\alpha +\int d^2\theta\; d^2\bar{\theta}\;\Phi^\dagger
e^{-2V}\Phi ]
\end{eqnarray*}
is invariant under the ${\cal N}=2$ super-translations. Therefore,
the choice of equal coefficients for the kinetic terms of the
$s=\frac{1}{2}$ fields,
$\bar{\lambda}\bar{\sigma}^\mu\nabla_\mu\lambda\quad\&\quad
\bar{\psi}\bar{\sigma}^\mu\nabla_\mu\psi$, and the prohibition of
superpotentials in the ${\cal N}=1$ minimal SUSY Yang-Mills action
leads to the admission of extended  ${\cal N}=2$ supersymmetry.

Forgetting about renormalizability we can write the most general
action, invariant under ${\cal N}=2$ super-translations, for a
chiral ${\cal N}$=2 super-field in the form:
\begin{eqnarray}
S&=&\frac{1}{16 \pi}{\rm Im}\int d^4x\; d^2\theta\;
d^2\tilde{\theta}\; {\cal F}(\Psi)\nonumber\\ &=&\frac{1}{16
\pi}{\rm Im}\int d^4x\; [\int d^2\theta {\cal F}_{ab}(\Phi)
W^{\alpha a} W_\alpha^b +\int d^2\theta
d^2\bar{\theta}(\Phi^\dagger e^{-2 V})^a{\cal
F}_a(\Phi)]\label{eq:hen}
\end{eqnarray}
The holomorphic functional  ${\cal F}(\Psi)$ is called the
prepotential and we have a ${\cal N}=2$ gauged supersymmetric
sigma model where ${\cal F}_a\Phi^a=\frac{\partial {\cal
F}}{\partial \Phi^a}\Phi^a$ is the Kahler potential and ${\cal
F}_{ab}=\frac{\partial^2 {\cal F}}{\partial \Phi^a
\Phi^b}W^{\alpha a} W_\alpha^b$ is the gauge kinetic function.
Thus, $g_{ab}={\rm Im}\frac{\partial^2 {\cal F}}{\partial \Phi^a
\Phi^b}$ can be interpreted as a metric -of special Kahler type-
in the ${\rm adj SU}(2)$ target space.

Obviously,
\[
{\cal F}_{YM}(\Psi)=\frac{1}{2}\;{\rm tr}\; \tau\Psi^2\qquad ({\rm
classically})
\]
is the right prepotential in the weak g-coupling limit, the
classical limit for the basic fields above. Seiberg and Witten in
an spectacular paper, \cite{SeiWi}, posed the following question:
what is ${\cal F}_{YM}^{quantum}(\Psi)$? How looks the
prepotential for stronger values of g (low energies) where quantum
effects on the basic fields are important? Holomorphy/${\cal N}=2$
supersymmetry and electric-magnetic duality guided to these
authors to provide a most beautiful answer.

\subsubsection{Classical vacuum moduli space: super-symmetric Higss
mechanism}

The search for super-solutions begins by integrating out the
auxiliary fields:
\begin{equation}
S_{aux}=\frac{1}{g^2}\int d^4x\;{\rm tr}(\frac{D^2}{2}-
\phi^\dagger [D,\phi]+F^\dagger F)=-\int d^4x\;\frac{1}{2g^2}{\rm
tr}[\phi^\dagger,\phi]^2 \quad .
\end{equation}
The homogeneous super-solutions are super-fields without soul - no
dependence on Grassman variables- which are zeroes of the Higgs
potential: $ U(\phi)=\frac{1}{2}{\rm tr}[\phi^\dagger,\phi]^2$. We
shall restrict ourselves to the gauge group ${\rm G=SU(2)}$ case.
Then, it is obvious that
\[
\phi_0=\frac{1}{2}\tau^3 a\qquad,\ \ a\in {\mathbb C}
\]
is a continuous set of zeroes of $U$. The \lq\lq classical vacuum
orbit" ${\cal V}$ -the manifold of all the zeroes of $U$- is:
\begin{eqnarray*}
{\cal V}\equiv ad\;G(\phi_0)\times {\mathbb C}^*&=& {\rm
\frac{SU(2)}{U(1)}(\simeq S^2)}\times {\mathbb C}^*\quad , \quad {\rm
if}\quad a\in{\mathbb C}^*={\mathbb C}-\{0\}\\ {\cal V}&\simeq &{\rm
point}\qquad , \qquad {\rm if}\quad a=0.
\end{eqnarray*}
The classical vacuum moduli space ${\cal M}$ is the quotient of
${\cal V}$ by the action of ${\rm adG}$: ${\cal M}\equiv
\frac{{\cal V}}{{\rm adG}}={\mathbb C}$.

We look now for the plane super-wave solutions becoming
fundamental quanta in the quantum perspective. The vacuum
degeneracy induces a super-symmetric Higgs mechanism which is
read from the quadratic terms in the expansion of the action
around a point in ${\cal M}$ for the basic fields $A_\mu , \lambda_\alpha
, \psi_\alpha$, and $\phi$. Small fluctuations of the scalar field of the
form $\phi(x)=\phi_0+\eta(x)$ give three types of quadratic terms
in the expansion of the Lagrangian:
\begin{itemize}
\item From $\frac{1}{g^2}[A_\mu,\phi^\dagger][A^\mu,\phi]\cong
\frac{1}{g^2}[A_\mu,\phi_0^\dagger][A^\mu,\phi_0]+o(\eta)$, we see
that the vector field plane wave solutions are of two types: there
is no quadratic term for $A_\mu^3$ and the corresponding plane
waves propagate at the speed of light: $m_3^2=0$. The plane waves
of the linear combination of vector fields $W_\mu^\pm=A_\mu^1\pm
iA_\mu^2$ carry energy $ m_{W^\pm}^2=|a|^2g^2$ - after re-scaling
back $\Psi\to g\Psi$- in the zero momentum limit.

\item $\frac{1}{2g^2}{\rm tr}[\phi^\dagger,\phi]^2\simeq
\frac{1}{2}[\phi_0^\dagger,\eta][\eta^\dagger,\phi_0]+o(\eta^3)$,
and the situation is identical for the scalar field. There are two
\lq\lq massive" field combinations, $\eta^\pm=\eta^1\pm i\eta^2$
with masses $ m_{\eta_\pm}^2=|a|^2g^2$, and one \lq\lq massless"
scalar field $\eta^3$: $m_{\eta^3}^2=0$.

\item
The mass terms for the fermions are more difficult to elucidate,
but from
\[
\frac{i\sqrt{2}}{g^2}{\rm tr}\left
[\phi^\dagger\{\psi,\lambda\}-\{\bar{\psi},\bar{\lambda}\}\phi
\right]\cong \frac{i\sqrt{2}|a|}{2g^2}\left
[e^{-i\omega}(\psi^1\lambda^2+\lambda^1\psi^2)-e^{i\omega}(\bar{\psi}^1\bar{\lambda}^2+
\bar{\lambda}^1\bar{\psi^2})\right ]
\]
it is not difficult to see that the field combinations
$\lambda_\pm=\frac{1}{2}(\psi^2\pm e^{-i\omega}\lambda^1)$ and
$\psi_\pm=\frac{1}{2}(\psi^1\pm e^{-i\omega}\lambda^2)$ have
masses $m_{\lambda_\pm}=m_{\psi_\pm}=|a|g$, whereas $\lambda^3$
and $\psi^3$ are massless: $m_{\lambda^3}=m_{\psi^3}=0$.
\end{itemize}

The particle spectrum arising from this analysis is summarized in
the following arrays, which show the basic fields and the
associated quanta in the physics folklore nomenclature.

\noindent A. Massless ${\cal N}=2$ super-multiplet:
\[
\begin{array}{ccc}
s=1&s=1/2&s=0\\ A_\mu^3\ ({\rm photon})& \lambda^3_\alpha\ ({\rm
photino})& \\
 &\psi^3_\alpha\  ({\rm Higgsino})&\eta^3\ ({\rm complex\ Higgs\ boson})
\end{array}
\]

\noindent B. Two massive (but short) ${\cal N}=2$ supermultiplets:
\[
\begin{array}{ccc}
s=1&s=1/2&s=0\\ W_\mu^\pm\ ({\rm massive\ vector\ bosons})&
\lambda^\pm_\alpha\ ({\rm gauginos})& \\
 &\psi^\pm_\alpha\  ({\rm goldstino})&\eta^\pm\ ({\rm complex\ Goldstone\ bosons})
\end{array}
\]
It is amazing to realize that the would be massive Higgs particle
in the usual (bosonic) mechanism remains massles in the
super-symmetric version of spontaneous gauge symmetry breaking.
Also, the Goldstone (massless ) bosons, hidden in the gauge fields
grown massive, are themselves massive in the super-symmetric
theory.

The choice of a vacuum to quantize leads to a particle spectrum
organized in ${\cal N}=2$ super-multiplets and ${\cal N}=2$
super-symmetry is unbroken. The degeneracy in the particle
spectrum with respect to the gauge $G=SU(2)$ group is lost after
the vacuum's choice and this symmetry is spontaneously broken to
$H=U(1)$. There is an important CAVEAT: singularities arise at
$a=0$ and many more particles are massless.

\subsubsection{Monopoles and dyons}
The search for super-solitons is identical to the search performed
in ${\cal N}=1$ supersymmetric (1+1)-dimensional field theory,
albeit analytically more involved. The energy for static purely
bosonic configurations can also be arranged a la Bogomolny:
\begin{eqnarray*}
E_B&=&\int d^3x \;[\frac{1}{2}{\rm
tr}\vec{B}\vec{B}+\frac{1}{2}{\rm
tr}\vec{D}\phi^\dagger\vec{D}\phi+\frac{1}{2}{\rm
tr}[\phi,\phi^\dagger]]\\ &\geq&\int d^3x\; \frac{1}{2}[{\rm
tr}\vec{B}\vec{B}+{\rm tr}\vec{D}\phi^\dagger\vec{D}\phi]
=\frac{1}{2}\int d^3x\;{\rm tr}\{
(\vec{B}-\vec{D}\phi^\dagger)(\vec{B}-\vec{D}\phi)\}+\frac{a}{2}(Q_m+Q_m^\dagger)
\end{eqnarray*}
where
\[
Q_m=\int_{S^2_\infty}\vec{b}\cdot d
\vec{S}=\frac{1}{a}\int_{S^2_\infty}{\rm tr}(\phi \vec{B})\cdot
d\vec{S} =\frac{1}{a}\int d^3x\; {\rm tr}(\vec{B}\cdot
\vec{D}\phi)
\]
is the field configuration magnetic charge.

If $[\phi,\phi^\dagger]=0$ the Bogomolny bound is saturated by
field configurations which solve the PDE system:
\begin{equation}
\vec{B}=\vec{D}\phi \qquad . \label{eq:mon}
\end{equation}
There is a fairly complete knowledge of the moduli space of
solutions of (\ref{eq:mon}), see \cite{AtHi}-\cite{Col}. We simply
collect three important results:
\begin{enumerate}
\item Every finite energy solution of (\ref{eq:mon}) belong to a
topological sector characterized by the winding number of the map
from the boundary of space to the vacuum orbit, provided by the
behaviour of the Higss field at infinity:
$\vec{\phi}|_{\infty}:S^2\simeq
\partial {\mathbb R}^3\longrightarrow {\cal V}\equiv S^2$.

\item The finite energy solutions are solitons and for topological
charge equal to 1 the BPS magnetic monopole is found:
\[
\phi_1^a(\vec{x})\;\frac{\tau^a}{2}=a_1
f(r)\frac{x^a}{r}\frac{\tau^a}{2} , \quad
\phi_2^a(\vec{x})\;\frac{\tau^a}{2}=a_2
f(r)\frac{x^a}{r}\frac{\tau^a}{2} , \quad
A_i^a(\vec{x})=\frac{1}{g}A(r)\varepsilon_{ij}^a\frac{x^j}{r^2}\frac{\tau^a}{2}
\]
\[
f(r)=\frac{{\rm cosh}\;r}{{\rm sinh}\;r}-\frac{1}{r}\qquad;\qquad
A(r)=1-\frac{r}{{\rm sinh}\;r} \, \, .
\]
$\phi_1^a$ and $\phi_2^a$ are respectively the real and imaginary
part of $\phi^a$, $a=a_1+ia_2$ sets the Higgs field vacuum value, and this solution shows a magnetic
monopole centered at the origin. Its orbit by the action of the
translation group and the unbroken $U(1)$ subgroup of the gauge
group is the moduli space of magnetic monopoles with topological
charge equal to 1.

\item In general, for any finite energy solution of
(\ref{eq:mon}), the energy in the center of mass is:
\[
E_M=M_M=|aQ_m|=\frac{4\pi}{g}|n_m a|
\]
$n_m\in{\mathbb Z}$ is the winding number of the $S^2_\infty\to S^2$
map and, thus, the magnetic charge is quantized.
\end{enumerate}
In fact, it is possible invoke electric-magnetic duality to
understand (\ref{eq:mon}) as the $\alpha =0$ member of the PDE
family:
\begin{equation}
\vec{B}={\rm cos}\alpha\vec{D}\phi\qquad \vec{E}={\rm
sin}\alpha\vec{D}\phi \label{eq:dyo}.
\end{equation}
The solutions of (\ref{eq:dyo}) have also electric charge,
although it is not quantized at this level. The dyon mass is
\[
E_D=M_D=|a|(|Q_e|^2+|Q_m|^2)^{\frac{1}{2}}
\]
and thus, it is given in terms of the magnetic and electric
charges:
\[
Q_e=\int_{S^2_\infty}\vec{e}\cdot
d\vec{S}=\frac{1}{a}\int_{S^2_\infty}{\rm tr}(\phi\vec{E})\cdot
d\vec{S}
\]

\subsubsection{The lattice of classical BPS charges}

In the framework of ${\cal N}=2$ super-symmetry we notice that the
Hamiltonian functions associated to the vector fields $Q^I$ are,
\begin{equation}
\tilde{Q}^I=-\frac{i}{g^2}\int d^3x \{\vec{\sigma}\sigma_2{\rm
tr}\psi^{\dagger I}(i\vec{E}+\vec{B})+\sqrt{2}\varepsilon^{IJ}{\rm
tr}\;\psi^J\nabla_0\phi^\dagger +\sqrt{2}\vec{\sigma}\;{\rm
tr}\;\vec{\nabla}\phi^\dagger\varepsilon^{IJ}\psi^J+i\sigma_2\;{\rm
tr}\;\psi^{\dagger I}[\phi^\dagger,\phi]\}
\end{equation}
where we denote by $\psi=\psi^1,\lambda=\psi^2$ the two spinor
fields, explicitly showing that $\psi$ and $\lambda$ form a
$SU(2)_I$-doublet. The super-Poisson algebra
\[
\{\tilde{Q}_\alpha^I,\bar{\tilde{Q}}_{\dot{\alpha}J}\}=2\sigma^\mu_{\alpha\dot{\alpha}}
\tilde{P}_\mu\delta_J^I \quad , \quad
\{\tilde{Q}_\alpha^I,\tilde{Q}_\beta^J\}=\varepsilon_{\alpha\beta}Z^{IJ},\qquad
\{\bar{\tilde{Q}}_{I\dot{\alpha}},\bar{\tilde{Q}}_{J
\dot{\beta}}\}=\varepsilon_{\dot{\alpha}\dot{\beta}}Z_{IJ}^*
\]
admits a complex central charge: $Z^{IJ}=2\; \varepsilon^{IJ}
Z=2\; \varepsilon^{IJ} a(Q_e+iQ_m)$.

Define now
\[
a_\alpha=\frac{1}{\sqrt{2}}\;[\tilde{Q}_\alpha^1+\varepsilon_{\alpha\beta}
(\tilde{Q}^2_\beta)^\dagger]\quad , \qquad
b_\alpha=\frac{1}{\sqrt{2}}\;[\tilde{Q}_\alpha^1-\varepsilon_{\alpha\beta}(\tilde{Q}^2_\beta)^\dagger]
\]
In the center of mass reference system ${\tilde P}_0=M$ and the
non-null anti-commutators among the $a_\alpha, a_\alpha^\dagger ,
b_\alpha, b_\alpha^\dagger, $ variables are:
\[
\{a_\alpha,a_\beta^\dagger\}=2\;(M+|Z|)\;\delta_{\alpha\beta}\quad
, \quad \{b_\alpha,b_\beta^\dagger\}=2(M-|Z|)\delta_{\alpha\beta}
\]
Therefore, the classical BPS states, satisfying the relation
$M=|Z|$, are organized in short multiplets of the super-symmetry
algebra.

The classical BPS multiplets are, thus, uniquely characterized by
their electric and magnetic charges. The magnetic charge is
quantized because of topological reasons. The electric charge
should comply with the Dirac-Schwinger quantization condition, see
e.g. \cite{Oliv}, in a consistent quantum theory with electric and
magnetic charges. The electric charge is accordingly also
quantized and the effect of the $U(1)$ anomaly produced by the $F\wedge F$- term
on a magnetic charge is a net shift of the electric charge by an
amount proportional to the instanton angle and the magnetic charge
itself, \cite{Wi2}:
\[
Q_e=g(n_e-\frac{\Theta}{2\pi}n_m)\qquad , \qquad n_e\in{\mathbb
Z}\quad .
\]
Therefore, there is a lattice of dyonic classical BPS states with
masses given by the beautiful formula:
\[
\begin{array}{|c|}\hline \\[-0.3cm]
  M_D=g|a||n_e+i\tau n_m| \\\hline
\end{array}
\]
\qquad .

\subsection{Low energy effective theory}

\subsubsection{Wilson effective action: quantum moduli space of
vacua} ${\cal N}=2$ SUSY Yang-Mills theory is renormalizable and
asymptotically free. This means that the latter theory is the
microscopic theory which controls the weak coupling/classical/high
energy behaviour. The challenge is to elucidate the nature of the
low energy effective theory.

There are two types of effective actions in quantum field theory.
One is the standard generating functional $\Gamma[\phi]$ of
one-particle irreducible Feynman diagrams. Momentum integration in
loop-diagrams are from zero up to a UV-cutoff which is taken to
infinity at the end. $\Gamma[\phi]\equiv \Gamma[\mu ,\phi]$ also
depends on the scale $\mu$ used to define the renormalized vertex
functions. A quite different object is the Wilsonian effective
action $S_W[\mu ,\phi]$. It is defined as $\Gamma[\mu ,\phi]$
except that all the loop-momenta are integrated down to $\mu$
which serves as infrared cutoff.

Thus, if we choose $\mu^2\leq g^2|a|^2$ the Wilson effective
action $S_W[\mu ,\phi]$ depends only on the Abelian fields which
emerge from the super-symmetric Higgs mechanism as massless.
Denoting these fields as $ \phi=\phi^3, \psi=\psi^3, F=F^3,
A_\mu=A_\mu^3, \lambda=\lambda^3, D=D^3, f_{\mu\nu}=\partial_\mu
A_\nu-\partial_\nu A_\mu $, we collect them in either (two )
${\cal N}$=1,
\[
\varphi (y,\theta)=\phi(y)+\sqrt{2}\theta\psi(y)+\theta^2F(y)
\]
\[
v(x,\theta,\bar{\theta})=-\theta\sigma^\mu\bar{\theta}A_\mu+
i(\theta^2\bar{\theta}\bar{\lambda}-
\bar{\theta}^2\theta\lambda)+\frac{1}{2}\theta^2\bar{\theta}^2D;
\hspace{0.2cm} w(y,\theta)=(-i\lambda+\theta D-i\sigma^{\mu
\nu}\theta f_{\mu
\nu}+\theta^2\bar{\sigma}^\mu\partial_\mu\bar{\lambda})(y)
\]
or (one ) ${\cal N}$=2, Abelian super-multiplets:
\[
\chi(\tilde{y},\tilde{\theta})=\varphi(\tilde{y},\theta)+\sqrt{2}\;\tilde{\theta}^\alpha
w_\alpha(\tilde{y},\theta)+\tilde{\theta}^\alpha\tilde{\theta}_\alpha
G(\tilde{y},\theta);\hspace{0.1cm}
G(\tilde{y},\theta)=-\frac{1}{2}\int d^2\bar{\theta}[\varphi
(\tilde{y},\theta,\bar{\theta}) ]^\dagger \exp[-2
v(\tilde{y},\theta,\bar{\theta}) ].
\]
Moreover, the scale fixed as infrared cutoff is naturally provided
by $\langle\phi\rangle =a$. There is a subtlety here: the action
of the $SU(2)$ Weyl group sends a to -a which are thus the same
point in ${\cal M}$. Therefore a good coordinate in ${\cal M}$ is
$\langle\phi^2\rangle =a^2$. In any case, $S_W$ depends on which
point of the vacuum moduli space we choose to quantize.

The abelianization process leading to the Wilson effective action
do not spoil ${\cal N}=2$ super-symmetry. $S_W$ is of the general
form:
\[
S_W=\frac{1}{16 \pi}{\rm Im}\int d^4x\; d^2\theta\;
d^2\tilde{\theta}\; {\cal F}(\chi)=\frac{1}{16 \pi}{\rm Im}\int
d^4x [\int d^2\theta {\cal F}^{\prime\prime}(\varphi) w^{\alpha}
w_\alpha +\int d^2\theta d^2\bar{\theta}(\varphi^\dagger {\cal
F}^\prime(\varphi)]
\]
and the Seiberg-Witten problem is the determination of the
prepotential ${\cal F}$. One property will be crucial in achieving
this goal: ${\cal F}$ is holomorphic in the ${\cal N}=2$ chiral
super-field $\chi $ and, defining the parameters in terms of the
expectation value $\langle\chi\rangle$ in the vacuum of this
field, is also holomorphic in the parameters of the theory.

The mathematical/physical meaning of ${\cal F}$ is unveiled by
expanding $S_W$ in component fields:
\[
S_W^I[\mu,\varphi,w_\alpha]=\frac{{\rm Im}}{4\pi}\int d^4x[{\cal
F}^{\prime\prime}(\phi)(-\frac{1}{4})f_{\mu\nu}(f^{\mu\nu}-i\tilde{f}^{\mu\nu})-i{\cal
F}^{\prime\prime}(\phi)\bar{\lambda}\sigma^\mu\partial_\mu\lambda+\cdots]
\]
\[
S_W^{II}[\mu,\varphi,w_\alpha]=\frac{{\rm Im}}{4\pi}\int
d^4x[{\cal F}^{\prime\prime}(\phi)
\partial_\mu\phi\partial^\mu\phi-i{\cal
F}^{\prime\prime}(\phi)\psi\sigma^\mu\partial_\mu\bar{\psi}+\cdots]
\]
These formulas show that the effective action is nothing but the
action for a (3+1)D gauged non-linear ${\cal N}=2$ SUSY sigma
model where the target manifold is the vacuum moduli space: ${\cal
M}={\mathbb C}$. ${\rm Im}{\cal F}^{\prime\prime}(\phi)$ determines
the kinetic terms and plays the role of a metric in configuration
space. For constant configurations, it defines the metric in the
vacuum moduli space: $ds^2={\rm Im}{\cal F}^{\prime\prime}(a)da
d\bar{a}={\rm Im}\tau (a)da d\bar{a}$.

Thus, we can understand
\[
\tau(a)=\frac{\Theta_{eff}(a)}{2\pi}+i\frac{4\pi}{g^2_{eff}(a)}={\cal
F}^{\prime\prime}(a)
\]
as the running coupling constant at the $\mu^2\propto
<|\phi|^2>=|a|^2$ squared energy scale. In the classical/high
energy regime we have the classical vacuum moduli space
\[
{\cal M}={\mathbb C}\qquad ds^2={\rm Im}\;\tau_{cl}\;da d\bar{a}=
{\rm Im}\;(\frac{\Theta}{2\pi}+i\frac{4\pi}{g^2})\;da d\bar{a}
\]
whereas,
\[
{\cal M}_{quantum}={\mathbb C}\qquad ds^2={\rm Im}\;{\cal
F}^{\prime\prime}(a)da d\bar{a}
\]
is the quantum moduli space when quantum fluctuations become
important.

\subsubsection{Weak coupling regime}
In a neighborhood of the infinity point in ${\cal M}$, the
$a\to\infty$ limit, the fluctuation modes with large momentum $p$
dominate $S_W$. Asymptotic freedom tells us that large momentum
amounts to weak coupling and we can rely on perturbation theory in
this regime. The evaluation of the perturbation theory
contribution to the prepotential can be performed through the
analysis of the classical global symmetry of the theory:
$U(2)\simeq U(1)_C\otimes SU(2)_I$. The $U(1)_C$ sub-group is a
$R$-transformation with $R$-character equal to 2:
\newline
$U(1)_C\,:\,\Phi\longrightarrow \,
e^{2i\alpha}\Phi(e^{-i\alpha}\theta)$. $\theta , \tilde{\theta}$
form a doublet under $SU(2)_I$ -and so do $\psi=\psi^1 ,
\lambda=\psi^2$- but the $U(1)_I$ sub-group has $R$-character
equal to 0: $U(1)_I\,:\,\Phi\longrightarrow
\Phi(e^{-i\alpha}\theta).$ In perturbation theory  of the
microscopic Yang-Mills theory the fermionic triangle diagrams show
that $\partial_\mu J_5^\mu=-\frac{1}{4\pi}{\rm
Tr}F_{\mu\nu}\tilde{F}^{\mu\nu}$ and the $U(1)_C$ symmetry is
anomalous.

Integration of the fast quantum fluctuations lead to the effective
Abelian theory that, accordingly, should include a term of the
form :$\delta{\cal
L}_{eff}=-\frac{\alpha}{4\pi^2}f_{\mu\nu}\tilde{f}^{\mu\nu}\cdots$\,
. Thus,
\begin{equation}
{\cal L}^I_W=\frac{1}{16\pi}{\rm Im}[{\cal
F}^{\prime\prime}(\varphi)(-f_{\mu\nu}f^{\mu\nu}+if_{\mu\nu}\tilde{f}^{\mu\nu})]
-\frac{\alpha}{4\pi^2}f_{\mu\nu}\tilde{f}^{\mu\nu}+\cdots
\label{eq:anom}
\end{equation}
where $\alpha$ is the parameter of the $U(1)_C$ transformation. Is
this term encoded in the prepotential? The action of a $U(1)_C$
transformation on ${\cal L}_W^I$ reads:
\begin{equation}
{\cal L}_W^I(e^{2i\alpha}\varphi)=\frac{1}{16\pi}{\rm Im}[{\cal
F}^{\prime\prime}(e^{2i\alpha}\varphi)(-f_{\mu\nu}f^{\mu\nu}+
if_{\mu\nu}\tilde{f}^{\mu\nu})]+\cdots \label{eq:Rtra}
\end{equation}
Comparing (\ref{eq:anom}) and (\ref{eq:Rtra}), we find:
\[
{\cal F}^{\prime\prime}(e^{2i\alpha}\varphi)={\cal
F}^{\prime\prime}(\varphi)-\frac{4\alpha}{\pi}\quad\equiv\quad{\cal
F}^{\prime\prime\prime}(\varphi)=\frac{2i}{\pi\varphi}\quad .
\]
The integration of the last expression is elementary: in terms of
the integration constant $\Lambda^2$ we obtain ${\cal
F}_{pert}(\varphi)=\frac{i}{2\pi}\varphi^2\ln\frac{\varphi^2}{\Lambda^2}$
. Therefore, near the infinity point in ${\cal M}$ the
prepotential including one-loop effects is:
\[
{\cal
F}_{pert}(\chi)=\frac{i}{2\pi}\chi^2\ln\frac{\chi^2}{\Lambda^2}
\quad .
\]
From ${\cal F}(a)=\frac{i}{2\pi}a^2\ln\frac{a^2}{\Lambda^2}$ one
easily calculates the effective coupling constant
$\tau(a)=\frac{i}{\pi}(\ln\frac{a^2}{\Lambda^2}+3)$. Then,
\begin{equation}
\frac{4\pi^2}{g^2(a)}=\ln\frac{|a|^2}{\Lambda^2}+3 \label{eq:bet}
\end{equation}
and the effective beta function -$\frac{dg(a)}{d\ln|a|}=\beta(g)$-
is: $\beta(g)=-\frac{1}{4\pi^2}g^3$. $\Lambda^2$ is nothing but
the dynamically generated by quantum effects
renormalization-invariant scale: defining the renormalization
point as $\mu^2=e^3|a|^2$ we have:
$\Lambda^2=\mu^2\exp\{-\frac{4\pi^2}{g^2(\mu)}\}$. $|a|^2=\infty$
is the weak coupling renormalization scale, $\mu^2=\Lambda^2$ is
the strong coupling scale: $g(\Lambda^2)=\infty$.

There are anti-self-dual gauge connections in the Euclidean
version of the theory. Solutions of the self-duality equations
$F^E_{\mu\nu}=-\tilde{F}^E_{\mu\nu}$ give rise to instantons in
the parent Minkowskian theory, which interpolate between pure
gauge configurations with different winding number at $t=-\infty$
and $t=\infty$. Standard instanton physics assigns an Euclidean
action of  :
\begin{equation}
\exp(-\frac{8\pi^2
k}{g^2(a)})=e^{-6k}(\frac{\Lambda}{a})^{4k}\label{eq:ins}
\end{equation}
to a self-dual connection with Chern number k. In (\ref{eq:ins})
we have used the (complexified) equation (\ref{eq:bet}) to write
the instanton action in terms of $\Lambda$ and $a$, and followed
Seiberg, \cite{Sei}, in assigning an $R$-charge of 2 to $\Lambda
$.

Weyl-Dirac operators acting on sections of second Chern class k=1
spinor bundles have 8 zero modes according to the Atiyah-Singer
index theorem. Thus, Berezin integration in Euclidean fermionic
correlation functions in the instanton background provides a
non-zero answer only for 4 $\lambda$- and 4 $\psi$-spinors. Such a
correlator $G$ changes to $e^{i8\alpha}G$ under $U(1)_C$
transformation. This argument survives analytic continuation to
Minkowskian correlations and the $U(1)_C$ symmetry is broken to a
${\mathbb Z}_8$ sub-group by tunneling via one instanton. Therefore,
the prepotential taking into account this effect must be of the
form $\,\,{\cal F}_{instanton}={\cal
F}_1(\frac{\Lambda}{\varphi})^4\varphi^2$ -only the arbitrary
constant ${\cal F}_1$ coefficient is left to fix-, because, then,
${\cal F}_{instanton}={\cal
F}_1(\frac{\Lambda}{\varphi})^4\varphi^2e^{4i\alpha}=e^{4i\alpha}{\cal
F}_{instanton}(\varphi)$.

One can check that the ${\mathbb Z}_8$ sub-group is also a quantum
symmetry of ${\cal F}_{pert}$:
\[
{\cal F}_{pert}\rightarrow
e^{4i\alpha}(\frac{i}{2\pi}\chi^2\ln\frac{\chi^2}{\Lambda^2}-\frac{2\alpha}{\pi}\chi^2)\quad
.
\]
Then,
\[
\frac{2\alpha}{\pi}\frac{1}{16}{\rm Im}\int d^4xd^2\theta{\rm
Tr}W^\alpha W_\alpha=-\frac{\alpha}{4\pi^2}\int d^4x {\rm
Tr}F_{\mu\nu}\tilde{F}^{\mu\nu}=8\alpha k
\]
and the transformation do not alter the quantum action if
$\alpha=\frac{2\pi n}{8}, n\in {\mathbb Z}$. In the weak coupling
regime the super-potential collects the one-loop perturbative and
instanton corrections:
\begin{equation}
{\cal
F}(\chi)=\frac{i}{2\pi}\chi^2\ln\frac{\chi^2}{\Lambda^2}+\sum_{n=1}^\infty
{\cal F}_n(\frac{\Lambda}{\chi})^{4n}\chi^2 \label{eq:pre}\quad .
\end{equation}
It is worthwhile to mention that ${\cal N}=2$ super-symmetry
forbids any further corrections to the perturbative
$\beta$-function.

One might wonder about whether or not the description developed
for large values of a -in terms of the $\Phi ,W$ and ${\cal F}$ is
also appropriate in other regions of ${\cal M}$, e.g. around the
origin. Because ${\cal F}(a)$ is holomorphic, ${\rm Im}{\cal
F}^{\prime\prime}(a)={\rm Im}\tau(a)$ is harmonic:
$\partial\bar{\partial}\, {\rm Im}\tau(a)=0$. Then, $ {\rm
Im}\tau(a)>0$ $\forall a\in {\cal M}\equiv{\mathbb C}\cup\{\infty\}$
if and only if ${\rm Im}\; \tau(a)={\rm constant}$ (as in the
classical case). But $\tau(a)$ is non constant, at least near
$a=\infty$.

The way out is to allow for different local descriptions: the
coordinates $a,\bar{a}$ and ${\cal F}$ are appropriate only in a
certain region of ${\cal M}$. From $ {\rm
Im}\tau(a)=\frac{1}{\pi}(\ln\frac{|a|^2}{\Lambda^2}+3)$, we see
that there seem to be a priori three points in ${\cal M}$ where
another system of coordinates $\hat{a},\bar{\hat{a}}$ could be
necessary:
\begin{itemize}
\item ${\rm Im}\tau(a)\to 0$ and the effective action become
non-sense if $a=\pm e^{-3/2}\Lambda$. The strong coupling limit
provides two singular points.
\item
$\partial\bar\partial\ln\frac{|a|^2}{\Lambda^2}=\delta^{(2)}(a) $
and ${\rm Im}\tau(a)$ fails to be harmonic at the origin.
\end{itemize}

\subsubsection{Electric-magnetic duality}

Duality will provide a different set of (dual) fields $\Phi_D$ and
$W_D^\alpha$ that supply an appropriate description in other
region of the vacuum moduli space. How to define the duality
transformation ? First, how to define a good metric in all ${\cal
M}$?

One can write the metric in the form
\[
 ds^2={\rm Im}\;da_Dda=-\frac{i}{2}(da_D d\bar{a}-da d\bar{a}_D)
\]
where the new coordinate is: $a_D\equiv\frac{\partial {\cal
F}}{\partial a}$. Before of implementing this idea on the
super-fields, it is convenient to pause and explain mathematically
the situation. Let us introduce a complex space ${\mathbb
X}\cong{\mathbb C}^2$ with coordinates $a,a_D$ and endow ${\mathbb X}$
with the type $(1,1)$ symplectic form $w=\frac{i}{2}(da\wedge
d\bar{a}_D-da_D\wedge d\bar{a})$ and also with the holomorphic two
form $\omega_h=da\wedge da_D$. We describe a map $f:{\cal
M}\rightarrow {\mathbb X}$ by functions $a(u),a_D(u)$ such that
$f^*(\omega_h)=0$. This ensures that locally, if we pick $u=a$,
$a_D=\frac{\partial {\cal F}}{\partial a}$ with some holomorphic function
${\cal F}$. The metric in ${\cal M}$ is the one whose Kahler form
is $f^*(\omega)$. We must, however, recall that a good physical
parameter is provided by the choice $u={\rm
Tr}\langle\phi^2\rangle$ in the high energy regime.

The parallel duality transformation on the ${\cal N}=1$ chiral
super-field is easy to write:
\[
\varphi_D\equiv {\cal F}^\prime (\varphi)=\frac{d{\cal
F}}{d\varphi}\qquad ; \qquad {\cal F}^\prime_D(\varphi_D)\equiv
-\varphi \quad .
\]
One immediately realizes the invariance of the abelian effective
theory
\[
{\rm Im}\int d^4x\; d^2\theta\;
d^2\bar{\theta}\;\varphi^\dagger{\cal F}^\prime(\varphi) ={\rm
Im}\int d^4x\; d^2\theta\; d^2\bar{\theta}\;(-{\cal
F}_D^\prime(\varphi_D))^\dagger\varphi_D ={\rm Im}\int d^4x\;
d^2\theta\; d^2\bar{\theta}\;\varphi^\dagger_D{\cal
F}_D^\prime(\varphi_D)
\]
against this (``complex" ) canonical transformation:
\[
{\cal F}_D(\varphi_D)={\cal F}(\varphi)-\varphi\varphi_D
\]

The duality transformation on $W_\alpha$ is \underline{non-local}.
The Bianchi identity $df=0$ for the abelian gauge field
$f_{\mu\nu}$ is tantamount in the super-symmetric framework to the
constraint: ${\rm Im}(D_\alpha w^\alpha)=0$. We can trade
integration on the $v$ super-field by integration on the $w$
super-field in the quantum action provided that the constraint is
enforced by a Lagrange multiplier super-field $v_D$:
\begin{eqnarray*}
& &\int {\cal D}v\exp[\frac{i}{16 \pi}{\rm Im}\int d^4x\;
d^2\theta {\cal F}^{\prime\prime}(\varphi) w^\alpha w_\alpha ]\\
&\simeq&\int {\cal D}w{\cal D}v_D\exp[\frac{i}{16 \pi}{\rm Im}\int
d^4x \;d^2\theta {\cal F}^{\prime\prime}(\varphi) w^{\alpha}
w_\alpha +\frac{i}{32\pi}{\rm Im}\int d^4x\;d^2\theta\;
d^2\bar{\theta}\;v_DD_\alpha w^\alpha]
\end{eqnarray*}
Observe that
\begin{eqnarray*}
\int d^2\theta\; d^2\bar{\theta}v_DD_\alpha w^\alpha&=&-\int
d^2\theta\;d^2\bar{\theta} D_\alpha v_D w^\alpha+ \int d^2\theta\;
\bar{D}^2(D_\alpha v_D w^\alpha)\\ &=&\int
d^2\theta(\bar{D}^2D_\alpha v_D)w^\alpha=-4\int d^2\theta
w_{D\alpha}w^\alpha
\end{eqnarray*}
where we used $\bar{D}_{\dot{\beta}}w^\alpha=0$ and where the dual
$w_D$ is defined from $v_D$ as $w_{D\alpha}\equiv
-\frac{1}{4}\bar{D}^2D_\alpha v_D$. Then one can do the functional
integral over $w_\alpha$- a Gaussian- and one obtains:
\[
\int {\cal D} v_D\;{\rm exp}[\frac{i}{16 \pi}{\rm Im}\int d^4x\;
d^2\theta(-\frac{1}{{\cal F}^{\prime\prime}}(\varphi)w_D^\alpha
w_{D\alpha})]
\]
This formula re-expresses the ${\cal N}=1$ super-symmetric abelian
effective gauge theory in terms of the dual field $w^\alpha_D$
with the effective coupling $\tau(a)={\cal F}^{\prime\prime}(a)$
replaced by $-\frac{1}{\tau(a)}$. Moreover, because ${\cal
F}_D^{\prime\prime}=-\frac{d\varphi}{d\varphi_D}=-\frac{1}{{\cal
F}^{\prime\prime}(\varphi)}$, we can define
$\tau_D(a_D)\underbrace{\equiv}_{{\rm def}}{\cal
F}^{\prime\prime}(a_D)=-\frac{1}{\tau(a)}$, and write the
effective action in terms of the dual fields in a completely
symmetric way:
\begin{equation}
\frac{1}{16 \pi}{\rm Im}\int d^4x\; d^2\theta\;
d^2\tilde{\theta}\;{\cal F}_D(\chi_D)=\frac{1}{16 \pi}{\rm Im}\int
d^4x [\int d^2\theta {\cal F}_D^{\prime\prime}(\varphi_D)
w_D^{\alpha} w_{D\alpha}+\int d^2\theta
d^2\bar{\theta}(\varphi_D^\dagger {\cal
F}^\prime(\varphi_D)]\label{eq:dual})
\end{equation}

From the Berezin integration of the constraint term
\[
\frac{i}{32\pi}{\rm Im}\int
dx^4d\theta^2d\bar{\theta}^2v_DD_{\alpha}
w^\alpha=\frac{1}{8\pi}\int dx^4
v_D^\mu\partial^\nu\tilde{f}_{\mu\nu}=\frac{1}{8\pi}\int dx^4
\tilde{f}_D^{\mu\nu}f_{\mu\nu}
\]
one realizes that $v_D$ couples to magnetic -rather than electric-
charge -for a magnetic monopole $\partial_\mu
\tilde{f}^{0\mu}=8\pi\delta^{(3)}(x)$-. Therefore, $v_D$ is the
connection associated to the $U(1)_D$ sub-group of the $SU(2)_D$
dual Lie group to the $SU(2)$ gauge group. The duality
transformation is an electric-magnetic transformation analogous to
the duality symmetry -$f_{\mu\nu}\to\tilde{f}_{\mu\nu}$- of the
Maxwell equations in the vacuum. To keep EM duality at work when
sources are added to the Maxwell equations requires also the
transformations $q_e\to -q_m$, $q_m\to q_e$ among electric and
magnetic charges. In our case $\tau(a)\to\tau_D(a_D)$ does the job
and there is a parallel weak coupling/strong coupling
transformation. It seems thus, highly plausible that the right
effective action in the strong coupling regions of the moduli
space is (\ref{eq:dual}). A warning: unlike electromagnetism, this
statement means that there is no symmetry of the abelian effective
theories with respect to EM duality. We will come back to this
point later.

\subsubsection{$SL(2,{\mathbb Z})$ invariance: the lattice of quantum
BPS charges}

The exploration of such a possibility calls for the analysis of
the full duality transformations group of the Abelian effective
action that, for this purpose, it is convenient to write it as:
\begin{equation}
S_W=\frac{1}{16 \pi}{\rm Im}\int d^4x\; d^2\theta\;
\frac{d\varphi_D}{d\varphi}w^\alpha w_\alpha+\frac{1}{32 \pi
i}\int d^4 x\;d^2\theta\; d^2\bar{\theta}\;(\varphi^\dagger
\varphi_D-\varphi^\dagger_D\varphi) \label{eq:sl2}
\end{equation}
Besides the S-duality transformation
\[
\left(\begin{array}{c}\varphi_D\\\varphi\end{array}\right)\rightarrow
\left(\begin{array}{cc}0&1\\-1&0\end{array}\right)\left(\begin{array}{c}\varphi_D\\\varphi\end{array}\right)
\]
there is also a T-duality transformation,
\[
\left(\begin{array}{c}\varphi_D\\\varphi\end{array}\right)\rightarrow
\left(\begin{array}{cc}1&b\\0&1\end{array}\right)\left(\begin{array}{c}\varphi_D\\
\varphi\end{array}\right)=T(b)\left(\begin{array}{c}\varphi_D\\\varphi\end{array}\right)\quad
, \quad b\in{\mathbb Z}
\]
which is a true symmetry of the effective action. The second term
in (\ref{eq:sl2}) is T-invariant because $b$ is real and the first
term gets shifted by
\[
\frac{b}{16 \pi}{\rm Im}\int d^4\; xd^2\theta\; w^\alpha
w_\alpha=-\frac{b}{16\pi}\int d^4 x\;
F_{\mu\nu}\tilde{F}^{\mu\nu}=0
\]
because there are no instantons in an abelian theory. The
generators of the full group of duality transformations are:
$S=\left(\begin{array}{cc}0&1\\-1&0\end{array}\right),\quad
T=\left(\begin{array}{cc}1&1\\0&1\end{array}\right)$. A generic
element of the $SL(2,{\mathbb Z})$ duality group, ${\cal
D}(m,n,p,q)$, $m,n,p,q\in{\mathbb Z}$, $mp-qn=1$, acts in the form:
\[
\left(\begin{array}{c}\varphi_D^\prime\\\varphi^\prime\end{array}\right)=\left(\begin{array}{cc}m&n\\p&q\end{array}\right)\left(\begin{array}{c}\varphi_D\\\varphi\end{array}\right)
\qquad .
\]
Because of
\[
{\cal D}(n,m,p,q)[{\cal
F}^{\prime\prime}(\varphi)]=\frac{d(m\varphi_D+n\varphi)}{d(p\varphi_D+q\varphi)}=
\frac{m\frac{d\varphi_D}{d\varphi}+n}{p\frac{d\varphi_D}{d\varphi}+q}\qquad
,
\]
the duality action on $\tau(a)$ is a linear fractional
transformation: $\tau^\prime=\frac{m\tau+n}{p\tau+q} .$

The key observation of Seiberg and Witten is that the ${\cal N}=2$
SUSY algebra is anomalous and the renormalized central charge is:
\begin{equation}
Z_R(u)=a(u)n_e+a_D(u)n_m=\left(\begin{array}{cc} n_m&n_e
\end{array}\right)\left(\begin{array}{c}a_D\\a\end{array}\right)\label{eq:qchar}
\end{equation}
(if the fields and the dual fields are re-scaled by $g$ and $g_D$
). $n_e$ and $n_m$ count the number of electrically and
magnetically charged states, whereas $a(u)$ and $a_D(u)$ play the
r$\hat{o}$le of the renormalized electric and magnetic charge. The
rationale behind this crucial hypothesis is electric-magnetic
duality: for electrically charged BPS states it is clear that
${\mathbb Z}_R=n_e a$. The dual formula ${\mathbb Z}_R=n_m a_D$ for
magnetically charged BPS states may be verified through an
argument \'a la Bogomolny. Consider the full high energy theory in
its effective form (\ref{eq:hen}) and examine the bosonic terms in
the Hamiltonian of a magnetic monopole:
\[
E=\frac{1}{4\pi}{\rm Im}\int d^3x \{\tau_{ab}
(\nabla_i\phi^a)(\nabla_i\phi^b)+\frac{1}{2}\tau_{ab}\;B_i^aB_i^b\}\qquad
, \qquad  \tau_{ab}=\frac{\partial^2{\cal
F}}{\partial\phi^a\partial\phi^b}
\]
The duality transformation $\phi^a\to
\phi_{Da}=\frac{\partial{\cal F}}{\partial\phi^a}$ and the
Bogomolny splitting implies:
\begin{eqnarray*}
E&=&\frac{1}{4\pi}{\rm Im}\int d^3
x\{(-\frac{1}{\tau})^{ab}(\nabla_i\phi_D)_a(\nabla_i\phi_D)_b+\frac{1}{2}
\tau_{ab}\;B_i^aB_i^b\}\\ &=&\frac{1}{4\pi}{\rm Im}\{\int d^3x
(-\frac{1}{\tau})^{ab}
[(\nabla_i\phi_D)_a\pm\frac{1}{\sqrt{2}}\tau_{ac}B_i^c]
[(\nabla_i\phi_D)_b\pm\frac{1}{\sqrt{2}}\tau_{bd}B_i^d]\}\\
&=&\sqrt{2}\int\partial_i(B_i^a\phi_{Da})d^3x\geq|\frac{\sqrt{2}}{4\pi}\int_{\partial
{\mathbb R}^3}d^2S_i\;B_i^a\phi_{Da}|=\sqrt{2}|n_ma_D|
\end{eqnarray*}
The lattice (\ref{eq:qchar}) of BPS dyonic states has built in
duality symmetry and the quantum mass of these states is:
$M_D^2=2|Z_R(u)|^2$.

The problem is thus to determine $a(u)$ and $a_D(u)$. In the weak
g coupling limit, when $u\rightarrow\infty$,
\begin{equation}
 a\simeq\sqrt{2u}\ ,\
a_D\simeq\frac{i}{\pi}\sqrt{2u}\;(\ln\frac{2u}{\Lambda^2}+1)\label{eq:per}
\end{equation}
and $u=\infty$ is a branching point -a singularity- of both $a$
and $a_D$. Singularities are characterized by their monodromies:
take $u$ around a counter-clockwise contour of very large radius
in the complex $u$-plane, $u\rightarrow e^{2\pi i}u$. One has
$a\rightarrow -a$ and
\[
a_D\rightarrow -\frac{i}{\pi}\sqrt{2u}\;(\ln\frac{2 e^{2\pi i}
u}{\Lambda^2}+1)=-a_D+2a \qquad .
\]
This can be written in terms of the monodromy matrix:
\[
\left(\begin{array}{c}a_D(u)\\a(u)\end{array}\right)\rightarrow
M_\infty
\left(\begin{array}{c}a_D(u)\\a(u)\end{array}\right)\quad,\quad
M_\infty=PT(-2)=\left(\begin{array}{cc}-1&2\\0&-1\end{array}\right)\quad
,\quad P=\left(\begin{array}{cc}-1&0\\0&-1\end{array}\right)
\]
Multi-valued functions with prescribed singularities and
monodromies around them are unique. Thus, to find $a(u)$ and
$a_D(u)$ there is the need of identifying all the singularities
and their monodromies.

\subsubsection{Strong coupling singularities}
 We start by answering the following question: How many singularities are
 in the renormalized electric and magnetic charges as functions of u?

To identify the branching points in the lattice of quantum charges
when the theory is considered in different points of ${\cal M}$ we
proceed in three steps.
\begin{enumerate}
\item Recall that the prepotential (\ref{eq:pre}) in the weak
coupling regime is invariant under a ${\mathbb Z}_8$ sub-group of
$U(1)_C$. Under this ${\mathbb Z}_8$-symmetry $\phi\rightarrow
e^{i\pi\frac{n}{2}}\phi$ and $\phi^2\rightarrow -\phi^2$ if $n$
odd. $u=\langle {\rm tr}\phi^2\rangle$ breaks this ${\mathbb Z}_8$
invariance further to ${\mathbb Z}_4$ and there is only a ${\mathbb
Z}_4$-symmetry left on a given point of the vacuum moduli space.
Due to this global symmetry $u\to -u$ singularities of ${\cal M}$
should come in pairs: if $u=u_0$ is a singularity also $u=-u_0$
is. The only fixed points of $u\to -u$ are $u=\infty$ and $u=0$.
We know that $u=\infty$ is a singular point of ${\cal M}$. So if
there are two singularities the other must be the fixed point
$u=0$.

If there are only two singularities $u=\infty$ and $u=0$ one concludes by
contour deformation that: $M_0=M_\infty$. Then,
\[
\left(\begin{array}{c}a^\prime_D(0)\\a^\prime(0)\end{array}\right)=
\left(\begin{array}{cc}-1&2\\0&-1\end{array}\right)
\left(\begin{array}{c}a_D(0)\\a(0)\end{array}\right) \qquad
\Rightarrow \qquad a^\prime(0)=-a(0)
\]
and $a^{\prime 2}(0)=a^2(0)$ is not affected by any monodromy.
Hence, $a^2$ would be a good coordinate in all the quantum vacuum
moduli space, which is not; the $\frac{1}{2}\langle\phi^2\rangle = 0$
point does not belong to ${\cal M}_{{\rm quantum}}$ and two
singularities only cannot work.

\item The most important singularity occurs when $a_D$ is zero.
Magnetic monopoles become massless at the (ultra)-strong coupling
regime $u=\Lambda^2$ and therefore,
\[
a_D(u)\simeq c_0(u-\Lambda^2)\qquad , \qquad c_0={\rm constant}
\]
in a neighborhood of this point. In the next sub-section we shall
analyze the effective theory in the monopole patch of ${\cal M}$,
but we advance that
\begin{equation}
a(u)\simeq a_0+\frac{i}{\pi}
c_0(u-\Lambda^2)\ln(u-\Lambda^2)\qquad , \qquad a_0={\rm
constant}\label{monp}
\end{equation}
which is enough to unveil the monodromy matrix around this
singularity:
\[
\left(\begin{array}{c}a_D\\a\end{array}\right)\rightarrow
M_{\Lambda^2}\left(\begin{array}{c}a_D\\a\end{array}\right) \qquad
, \qquad
M_{\Lambda^2}=ST(2)S^{-1}=\left(\begin{array}{cc}1&0\\-2&1\end{array}\right)\quad
.
\]
The monodromy transformation can also be interpreted as changing
the magnetic and electric quantum numbers as
\[
(n_m,n_e)\rightarrow (n_m,n_e) M \qquad .
\]
The state of vanishing mass responsible for a singularity should
be invariant under the monodromy. This is so for the magnetic
monopole:
\[
(1,0)\left(\begin{array}{cc}1&0\\-2&1\end{array}\right)=(1,0)
\]
is a left eigen-vector of $M_{\Lambda^2}$ with unit eigenvalue.
\item  To obtain the monodromy matrix at $u=-\Lambda^2$ one
observes that the counter-clockwise contour around $u=\infty$ can
be deformed into a contour encircling $\Lambda^2$ and a contour
encircling $-\Lambda^2$, both counter-clockwise. The factorization
condition  $M_\infty=M_{\Lambda^2} M_{-\Lambda^2}$ follows, and,
hence
\[
M_{-\Lambda^2}=(TS)T(2)(TS)^{-1}=\left(\begin{array}{cc}-1&2\\-2&3\end{array}\right)
\]
Because $(1,-1)M_{-\Lambda^2}=(1,-1)$, this singularity
corresponds to a dyon becoming massless
\end{enumerate}
In summary, there are three patches in the quantum moduli space,
centered around the three singularities. The appropriate variables
are respectively the fields of a ${\cal N}=2$ supersymmetric
effective abelian gauge theory with three different $U(1)$ abelian
groups: the maximal torus of the gauge group, the maximal torus of
the dual to the gauge group and, the maximal torus of one diagonal
subgroup of the direct product of both.

\subsubsection{The Seiberg-Witten prepotential}

The strategy for finding $a_D(u)$ and $a(u)$ is to consider them
as the two linearly independent solutions of the Schr\"{o}dinger
equation
\[
[-\frac{d^2}{dz^2}+V(z)]\psi(z)=0\quad , V(z)=-\frac{1}{4}[
\frac{1-\lambda_1^2}{(z+1)^2}+\frac{1-\lambda_2^2}{(z-1)^2}-\frac{1-\lambda_1^2-
\lambda_2^2+\lambda_3^2}{(z+1)(z-1)}]\quad , z=\frac{u}{\Lambda^2}
\]
in the complex plane. $V(z)$ is a meromorphic function of $z$ with
second order poles at $-1,1,\infty$ and residues
$-\frac{1}{4}(1-\lambda_1^2), -\frac{1}{4}(1-\lambda_2^2),
-\frac{1}{4}(1-\lambda_3^2)$. Here, $\lambda_1, \lambda_2,
\lambda_3$, are constants -we assume without loss of generality
$\lambda_i\geq 0$- to be fixed later according to the asymptotic
behaviour of $a(z), a_D(z)$. The transformation
\[
\psi(z)=(z+1)^{\frac{1}{2}(1-\lambda_1)}
(z-1)^{\frac{1}{2}(1-\lambda_2)}f(\frac{z+1}{2})
\]
is very useful because f satisfies the hypergeometric  ODE
\begin{equation}
z(1-z)f^{\prime\prime}(z)+[c-(a+b+1)z]f^\prime(z)-abf(z)=0
\label{eq:hyp})
\end{equation}
with
\[
a=\frac{1}{2}(1-\lambda_1-\lambda_2+\lambda_3);\quad
b=\frac{1}{2}(1-\lambda_1-\lambda_2-\lambda_3);\quad
c=1-\lambda_1\quad .
\]
We choose the two independent solutions of the second-order ODE (\ref{eq:hyp})
\[
f_1(z)=(-z)^a {}_1F_2(a,a+1-c,a+1-b;\frac{1}{z})\quad , \quad
f_2(z)=(1-z)^{c-a-b} {}_1F_2(c-a,c-b,c+1-a-b;1-z)
\]
where ${}_1F_2(a,b,c;z)$ is the Gauss hypergeometric function. The
reason for this election is that $f_1$ and $f_2$ has simple
monodromy properties respectively around $z=\infty$ and $z=1$.
Hence, they are good candidates to be identified with $a(z)$ and
$a_D(z)$.

To make the identification precise, we observe that when
$z\rightarrow \infty$ one has
$V(z)\simeq-\frac{1}{4}\,\frac{1-\lambda_3^2}{z^2}$. The two
independent solutions behave asymptotically as
$z^{\frac{1}{2}(1\pm\lambda_3)}$, if $\lambda_3\neq 0$, and
$\sqrt{z}, \sqrt{z}\ln z$, if $\lambda_3=0$. Comparison with (\ref{eq:per}) tells us that 
the latter case is
realized if $a(z)$ is going to be identified with $f_1$. Simili
modo, if $\lambda_3=0$, as $z\rightarrow 1$ one has $ V(z)\simeq
-\frac{1}{4}(\frac{1-\lambda_2^2}{(z-1)^2}-\frac{1-\lambda_1^2-
\lambda_2^2}{2(z-1)})$, where the sub-leading term is kept.

From the logarithmic asymptotics of $a_D=c_0\Lambda^2 (z-1),\
a=a_0+\frac{i}{\pi}c_0\Lambda^2(z-1)\ln (z-1)$ when $z\to 1$ one
sees that $\lambda_2=1$ and
$-\frac{\lambda^2_1}{8}=\frac{ic_0}{\pi a_0}$. The ${\mathbb Z}_2$
symmetry $u\to -u$ on the moduli space implies that, as $z\to -1$
the potential does not have a double pole either, so that
$\lambda_1=1$ also. Therefore,
\[
\lambda_1=\lambda_2=1\quad,\quad \lambda_3=0\quad,\quad
V(z)=-\frac{1}{4}\frac{1}{(z+1)(z-1)}
\]
and $a=b=-\frac{1}{2}, c=0$. The two solutions
\[
a_D(u)=i\frac{u-\Lambda^2}{2\Lambda^2}\,
F(\frac{1}{2},\frac{1}{2},2;\frac{\Lambda^2-u}{u})\quad , \quad
a(u)=\sqrt{\frac{2}{\Lambda^2}}(u+\Lambda^2)^{\frac{1}{2}}\,
F(-\frac{1}{2},\frac{1}{2},1;\frac{2\Lambda^2}{u+\Lambda^2})
\]
have the correct monodromies as well as the correct asymptotics.
One can invert the equation on the right to find $u(a)$. To obtain
$a_D(a)$ one plugs-in this result into $a_D(u)$.
 Integration with
respect to $a$ yield ${\cal F}(a)$ and, hence, the low energy
theory.

In the Seiberg and Witten work, the whole picture is related to an
elliptic curve. A brief summary is as follows: $M_\infty$,$M_{\Lambda^2}$ and 
$M_{-\Lambda^2}$ belong to the
$\Gamma(2)$ sub-group of $SL(2,{\mathbb Z})$ matrices with integer
mod 2 entries. If ${\mathbb H}$ is the half-plane, the quotient
$\frac{{\mathbb H}}{\Gamma(2)}$ is the moduli space of elliptic
curves given by the algebraic equation:
\begin{equation}
y^2=(x-1)(x+1)(x-z)\label{eq:ecu}\qquad .
\end{equation}
The modular forms of weight $\frac{1}{2}$ and level 2 provide a
six-to-one map $\alpha :\frac{{\mathbb H}}{\Gamma(2)}\cong{\cal
M}_q\longrightarrow {\mathbb P}^1$ from this moduli space, isomorphic
to the quantum vacuum moduli space ${\cal M}_q$, to the projective
complex line :
\[
\alpha(\tau(u))=\left(\Theta\left[\begin{array}{c}0\cr
\frac{1}{2}\end{array}\right]
(0,\tau),\Theta\left[\begin{array}{c}\frac{1}{2}\cr 0
\end{array}\right] (0,\tau)\right)\qquad , \qquad z=\frac{\Theta\left[\begin{array}{c}0\cr
\frac{1}{2}\end{array}\right](0,\tau)}{\Theta\left[\begin{array}{c}\frac{1}{2}\cr
0\end{array}\right](0,\tau)}\quad ,
\]
see \cite{Mum}. Here, we refer by $\tau$, not the coupling
constant, but the modular parameter of the elliptic curves
(\ref{eq:ecu}) which have an structure of level 2. Thus, ${\cal
M}_q$ is a 6-fold covering of ${\mathbb P}^1$.

$\tau=\infty$ is a cusp - a fixed point under the action of
$T(1)=\left(\begin{array}{cc}1&1\\0&1\end{array}\right)\in\Gamma(2)$-:
\[
\alpha(\tau +1)=\left(\Theta\left[\begin{array}{c}0\cr
0\end{array}\right]
(0,\tau),e^{i\frac{\pi}{4}}\Theta\left[\begin{array}{c}\frac{1}{2}\cr
0
\end{array}\right] (0,\tau)\right)\qquad ,
\]
and having in mind that $\lim_{{\rm
Im}\tau\to\infty}\Theta\left[\begin{array}{c}0\cr \frac{1}{2}
\end{array}\right] (0,\tau)=1$, $\lim_{{\rm Im}\tau\to\infty}\Theta\left[\begin{array}{c}\frac{1}{2}
\cr 0
\end{array}\right] (0,\tau)=0$, this point corresponds to
$z=\infty$- a cuspidal curve in the family (\ref{eq:ecu})-.

$N=\left(\begin{array}{cc}1&1\\1&0\end{array}\right)$ is other
element in $\Gamma(2)$ which amounts to:
\[
\alpha(-\frac{1}{\tau} \pm
1)=\left((-i\tau)^{\frac{1}{2}}\Theta\left[\begin{array}{c}0\cr
0\end{array}\right] (0,\tau),e^{\pm
i\frac{\pi}{4}}(-i\tau)^{\frac{1}{2}}\Theta\left[\begin{array}{c}0\cr
\frac{1}{2}
\end{array}\right] (0,\tau)\right)\qquad .
\]
$N(\tau=\infty)=\pm 1$ and, therefore, $\tau=\pm 1$ - or $z=\pm
1$- are new cusp points, which also characterize nodal curves in
(\ref{eq:ecu}).

The physical meaning of this construction is related to the
lattice of charges $L(u)={\mathbb Z}a_D(u)\otimes{\mathbb Z}a(u)$. The
$E(u)$ elliptic curve (\ref{eq:ecu}) is nothing but
$E(u)=\frac{{\mathbb C}}{L(u)}$ and singular curves appear at the
singularities of the lattice of BPS states. Moreover, if we denote
by $\gamma_D$ and $\gamma$ the 1-cycles which generates the first
homology group $H_1(E(u),{\mathbb Z})$ of the elliptic curve, and
pick the linear combination
\[
\lambda=\frac{1}{\sqrt{2}\pi}\frac{\sqrt{x-z}}{\sqrt{x^2-1}}.dx=
\frac{1}{\sqrt{2}\pi}\frac{(x-z)dx}{y}
\]
of the two holomorphic differentials on the curve, we have
\[
a_D(u)=\oint_{\gamma_D}\lambda=\frac{\sqrt{2}}{\pi}\int_1^z
\frac{dx\sqrt{x-z}}{\sqrt{x^2-1}}\qquad , \qquad
a(u)=\oint_{\gamma}\lambda=\frac{\sqrt{2}}{\pi}\int_{-1}^1
\frac{dx\sqrt{x-z}}{\sqrt{x^2-1}}\quad .
\]
Derivation with respect to the coordinate in the moduli shows the
effective coupling constant as the modulus of the elliptic curve
$E(u)$:
\[
\frac{da_D}{dz}=\oint_{\gamma_D}\frac{d\lambda}{dz}\quad , \quad
\frac{da}{dz}=\oint_{\gamma}\frac{d\lambda}{dz}\quad ; \quad
\tau(u)=\frac{da_D/dz}{da/dz}=\frac{da_D}{da}\quad .
\]

\subsection{${\cal N}$=2 dual SUSY QED}

\subsubsection{The ultra-strong limit: low energy effective theory}
We now closely focus on the peculiar scaling limit
$a_D(\Lambda^2)=0$, usually called ultra-strong limit.
Magnetically charged states become thus massless and the
magnetically charged massless ${\cal N}=2$  hyper-multiplet,
\[
M(y,\theta)=\phi_m(y)+\sqrt{2}\theta\psi_m(y)+\theta^2F_m(y)\qquad
,\qquad \tilde{M}(y,\theta)=\tilde{\phi}_m(y)+\sqrt{2}\theta
\tilde{\psi}_m(y)+\theta^2\tilde{F}_m(y),
\]
must be \lq\lq integrated in " the low energy effective theory.
The spin of the physical fields, two Weyl spinors $\psi_m$ ,
$\tilde{\psi}_m$ ,and , two complex scalars $\phi_m$ ,
$\tilde{\phi}_m$, is shown next:
\begin{equation}
\begin{array}{ccccc}{\rm Spin}& &0&\frac{1}{2}&0\\ & & &\psi_m& \\ & &\phi_m& &\tilde{\phi}_m\\ & & &\tilde{\psi}_m& \\ & & & & \end{array}\label{diamat}
\end{equation}
All of them are massless and magnetically charged; accordingly,
the associated quanta are light dual electrons, dual positrons,
magnetic monopoles (dual s-electrons) and magnetic anti-monopoles
(dual s-positrons). It is important to note that $\phi_m$ and
$\tilde{\phi}_m$ form one doublet under $SU(2)_I$ whereas $\psi_m$
and $\tilde{\psi}_m$ are singlets.

There is also the ${\cal N}=2
$ chiral super-field
\[
\chi_D(y,\theta)=\varphi_D(y)+\sqrt{2}\tilde{\theta}W_D(y)+\tilde{\theta}^2G_D(y).
\]
which includes all the dual Abelian fields introduced in
sub-Section \S 3.3.3. The spin of the physical fields $v_\mu^D$ ,
$\lambda_D$ , $\psi_D$ , $\phi_D$ is shown next
\[
\begin{array}{ccccc} & & & &{\rm Spin}\\ & &v_\mu^D& &1\\ &\lambda_D& &\psi_D&\frac{1}{2}\\ & &\phi_D& &0\\ & & & & \end{array}
\]
and the corresponding quanta are dual photons , dual photinos ,
dual Higgsinos, and dual complex Higgs particles.

The ${\cal N}=2$ super-symmetric low energy action in this limit
is built from three pieces:

\begin{eqnarray*}
S_M&=&\int d^4x\int d^2\theta\int
d^2\bar{\theta}\left\{\tilde{M}e^{2V_D}\tilde{M}^\dagger+M^\dagger
e^{-2V_D}M \right\}\\ S_W&=&\frac{{\rm Im}}{16\pi}\int d^4x\int
d^2\theta d^2\tilde{\theta}\tau_D(0)\chi_D\chi_D \qquad ; \qquad
S_Y=\sqrt{2}\int d^4x[\int d^2\theta\varphi_D M\tilde{M}+{\rm
h.c.}]
\end{eqnarray*}
$S_M$ describes the $U(1)_D$ gauge dynamics of the light
magnetically charged fields, $S_W$ is the Seiberg-Witten effective
action exactly at the ultra-strong limit, and , $S_Y$ is a Yukawa
type action needed to ensure ${\cal N}=2$ super-symmetry through
the achievement of $SU(2)_I$ invariance in the total effective
action $S_{{\rm eff}}=S_W+S_M+S_Y$, see \cite{Weinberg}.

A better understanding of the low energy effective theory in the
ultra-strong limit requires a closer look to the theory in a
neighborhood of the $u=\Lambda^2$ point in the quantum moduli
space of vacua. One introduces the magnetic coupling $g_D$ as
\[
\tau_D(a_D)=\frac{4\pi i}{g_D^2(a_D)}
\]
because there is no instanton angle in QED and $\Theta_D=0$.
Moreover, from the $\beta$-function of ${\rm (QED)_D}$,
\[
\mu\frac{d g_D}{d\mu}=\frac{g_D^3}{8\pi^2}\qquad ,
\]
and because $\mu\propto a_D=c_0(u-\Lambda^2)$ ( in the monopole
patch ) we obtain
\[
a_D\frac{d\tau_D}{d a_D}=-\frac{i}{\pi}\quad\Rightarrow\quad\tau_D=-\frac{i}{\pi}\ln
a_D \quad .
\]
Now the physical meaning of this limit is clear:
$\lim_{a_D\rightarrow 0}i\tau_D\simeq -\infty$ means that the
ultra-strong limit is the $g_D$ weak coupling limit (the g strong
coupling limit, of course). Note also that from
$\tau_D=\frac{d(-a)}{da_D}$ we can deduce formula (\ref{monp}) in
subsection \S. 3.3.5:
\[
a\simeq a_0+\frac{i}{\pi}a_D \ln
a_D=a_0+\frac{ic_0}{\pi}(u-\Lambda^2)\ln[c_0(u-\Lambda^2)]\quad .
\]

The key observation is that $S_W$ at $a_D=0$ becomes a
Ginzburg-Landau type effective action with all the quantum
fluctuations integrated out. Therefore, the effective theory at
the ultra-strong limit is a classical field theory and, rescuing
the Planck constant, wave-particle duality tells us that the field theoretical coupling 
$g_D$ is related to the magnetic charge of the dual light quanta in the form: $g_d=\hbar \bar{g}_D$. 
Thus, one must replace in $S_W$ $\tau_D(0)$ by
\[
\bar{\tau}_D(0)=\hbar^2\tau_D(0) \qquad ,
\]
which remains finite at the ultra-strong limit $\tau_D\to \infty$ only if $\hbar\to 0$, i.e., the g-strong coupling limit 
is nothing but the classical limit for the magnetically charged quanta !!!.

\subsubsection{Extended states in the ultra-strong limit}
In order to search for solitons in the ultra-strong limit we
perform Berezin integration in the low energy Ginzburg-Landau
action to find:
\[
S_W=\int
d^4x\{-\frac{1}{4}f_{\mu\nu}^Df^{\mu\nu}_D-i\bar{\lambda}_D\bar{\sigma}^\mu\partial_\mu\lambda_D+\frac{D^2}{2}+\nonumber\\
+\partial_\mu\phi^*_D\partial^\mu\phi_D-i\bar{\psi}_D\bar{\sigma}^\mu\partial_\mu\psi_D+F^*_DF_D\}
\]
\begin{eqnarray*}
S_M&=&\int
d^4x\{(\nabla_\mu\phi_m)^*\nabla^\mu\phi_m+\nabla_\mu\tilde{\phi}_m(\nabla^\mu\tilde{\phi}_m)^*+F^*_mF_m+\tilde{F}_m^*\tilde{F}_m+\nonumber\\
&+&\bar{g}_DD(\phi^*_m\phi_m-\tilde{\phi}_m^*\tilde{\phi}_m)-i(\bar{\psi}_m\bar{\sigma}^\mu
\nabla_\mu\psi_m+\bar{\tilde{\psi}}_m\bar{\sigma}^\mu
\nabla_\mu\tilde{\psi}_m)\}
\end{eqnarray*}
\begin{eqnarray*}
S_Y&=&\sqrt{2}\bar{g}_D\int
d^4x[\{F_D\phi_m\tilde{\phi}_m+2\psi_D(\tilde{\psi}_m\phi_m+\psi_m\tilde{\phi}_m)
+\phi_D(\phi_m\tilde{F}_m+\tilde{\phi}_mF_m)+{\rm h.c.}\}\\
&-&\{\phi_m^*\psi_m\lambda_D-\phi_m\bar{\psi}_m\bar{\lambda}_D+\tilde{\phi}_m\bar{\tilde{\psi}}_m\bar{\lambda}_D-\tilde{\phi}^*\psi_m\lambda_D\}].
\end{eqnarray*}
The covariant derivatives on the spinor and scalar fields of the
hypermultiplet are:
\begin{eqnarray*}
\nabla_\mu\phi_m=\partial_\mu\phi_m+i\bar{g}_DA_\mu^D\phi_m,&\;\;\;\;\;\;&\nabla_\mu\psi_m=\partial_\mu\psi_m+i\bar{g}_DA_\mu^D\psi_m\\
\nabla_\mu\tilde{\phi}_m=\partial_\mu\tilde{\phi}_m-i\bar{g}_DA_\mu^D\tilde{\phi}_m&\;\;\;\;\;\;&\nabla_\mu\tilde{\psi}_m=
\partial_\mu\tilde{\psi}_m-i\bar{g}_DA_\mu^D\tilde{\psi}_m \qquad .
\end{eqnarray*}

After performing integration on the auxiliary fields $D$ and
$F_m$, $\tilde{F}_m$ we obtain the bosonic part of the effective
action:
\begin{equation}
S_{eff}^B=\int
d^4x\{-\frac{1}{4}f_{\mu\nu}^Df^{\mu\nu}_D+\partial_\mu\phi_D^*\partial^\mu\phi_D+
(\nabla_\mu S)^\dagger(\nabla^\mu S)
-\frac{\bar{g}_D^2}{2}(4|\phi_D|^2+S^\dagger S)S^\dagger S\}
\label{boseff} \quad .
\end{equation}
We have introduced the $SU(2)_I$ spinor $S$:
\[
S=\left(\begin{array}{cc}\phi_m\\\tilde{\phi}_m^*\end{array}\right)\quad
.
\]
It is obvious that absolute minima of the energy in the effective
theory must satisfy $\phi_D(x)=a_D=0$. Assuming this condition the
energy density for time-independent axially symmetric
configurations can be written à la Bogomolny:
\begin{equation}
{\cal E}_{eff}^B(a_D=0)=\int d^2x\{|f_{12}^D\mp\bar{g}_DS^\dagger
S|^2+|(\nabla_1\pm i \nabla_2)S|^2\}+|\int d^2x
f_{12}^D|\label{bogb}
\end{equation}
Solutions of the first-order equations
\begin{equation}
f_{12}^D=\pm\bar{g}_DS^\dagger S\qquad , \qquad (\nabla_1\pm i
\nabla_2)S=0 \label{first}
\end{equation}
are absolute minima of ${\cal E}_{eff}^B(a_D=0)$. Unfortunately,
the sign combination in (\ref{first}) is such that there are no
regular solutions besides the trivial $S=0, A_i^D=0$ solution: the
system is frustrated.

Nevertheless, an identical analysis in the Poincar\'e disc,
instead ${\mathbb R}^2$, requires to add to the Lagrangian density
the piece: ${\cal L}_R=-\frac{1}{2}RS^\dagger S$, where $R$ is the
constant negative curvature. After some rather trivial re-scaling,
a new system of first-order equations appear:
\begin{equation}
f_{12}^D=\mp\frac{4}{(1-x_1^2-x_2^2)}(1-S^\dagger S)\quad , \quad
(\nabla_1\pm i \nabla_2)S=0  \label{first1}\quad .
\end{equation}
There are vorticial solutions to the new system of first-order
equations (\ref{first1}) of the form:
\[
\tilde{\phi}_m^{(n)}(z)= 0\quad , \quad
\phi_m^{(n)}(z)=\frac{2(1-|z|^2)}{1-f_n^*f_n(z)}.\frac{f_n^{\prime}(z)}{|f_n^{\prime}(z)|}
\quad  , \quad
f_n(z)=\prod_{i=1}^{n}\frac{(z-a_i)}{(1-a_{i}^{*}z)}\quad .
\]
which support a total electric flux of
\[
|\int d^2x f_{12}^D(x_1,x_2)|=\frac{2\pi}{\bar{g}_D}(n-1)
\]
concentrated around the points $z=a_i$ of the complex plane. These
solitonic solutions are nothing but \lq\lq dual" instantons -dual
self-dual configurations- with cylindrical symmetry.

The whole picture of the spectrum of ${\cal N}=2$ $SU(2)$
super-symmetric Yang-mills theory is summarized in the following
boxes:
{\small
\begin{center}
\begin{tabular}{|c||c|}\\ [-0.7cm]\hline  ${\bf weak}\, {\bf coupling}$& ${\bf
strong}\, {\bf coupling}$
\\ \hline light quanta: electrically charged &  light quanta:
magnetically charged  \\ extended states: magnetically charged &
extended states : electrically charged  \\   $S^1$-invariant
instantons & cylindrically-invariant instantons  \\ \hline
\end{tabular}
\end{center}}

\subsubsection{Soft breaking of ${\cal N}$=2 supersymmetry:
monopole condensation} Adding a mass term in the original
non-abelian theory
\[
S_\mu=\mu\int d^4x[\int d^2\theta{\rm tr}\,\Phi^2+{\rm
h.c.}]=-\mu\int d^4x{\rm
tr}(\phi^*\phi+i(\psi\psi+\bar{\psi}\bar{\psi}))
\]
${\cal N}=2$ super-symmetry is \lq\lq softly" broken to ${\cal
N}=1$. This addition affects the low energy effective theory
through the trading of ${\rm tr}\Phi^2$ by an abelian chiral
super-field $U(y,\theta)$ which is a functional of $\varphi_D$ in
the monopole patch,
\[
U(\varphi_D)=U(\phi_D)+\sqrt{2}\theta\psi_DU^\prime(\phi_D)+\theta^2
(U^\prime(\phi_D)F_D-\frac{1}{2}U^{\prime\prime}(\phi_D)\psi_D\psi_D)\qquad
,
\]
supplying -after Berezin integration- a new term to the effective
action:
\[
S_\mu=\mu\int d^4x
\{U^\prime(\phi_D)F_D-\frac{1}{2}U^{\prime\prime}(\phi_D)\psi_D\psi_D+{\rm
h.c.}\}.
\]
 Taking into account only the contribution of the
constant modes, the effective action at $u=\Lambda^2$, reduces to
the effective potential:
\[
V_{eff}=\sqrt{2}\bar{g}_D\varphi_DM\tilde{M}+\mu U(\varphi_D)+{\rm
h.c.}\qquad \qquad .
\]
The expectation values of $M$, $\tilde{M}$ and $\varphi_D$
\begin{eqnarray*}
\langle M(x)\rangle=\langle\phi_m(x)\rangle=m \qquad &,& \qquad
\langle\tilde{M}(x)\rangle=\langle\tilde{\phi}_m(x)\rangle=\tilde{m}
\\ \langle \varphi_D(x)\rangle&=&\langle\phi_D(x)\rangle=a_D
\end{eqnarray*}
are given by the minima of $V_{{\rm eff}}$:
\begin{eqnarray}
\frac{dV_{eff}}{d\varphi_D}=\frac{dV_{eff}}{dM}&=&
\frac{dV_{eff}}{d\tilde{M}}=0 \nonumber \\
\sqrt{2}\bar{g}_Dm\tilde{m}+\mu U^\prime (a_D)&=&0 \qquad , \qquad
a_D\tilde{m}=a_Dm=0 \label{eq:vac} \qquad .
\end{eqnarray}
There is also the constraint $|m|=|\tilde{m}|$ coming from the
integration of the $D$-terms. When $\mu=0$, $m=\tilde{m}=0$,
$a_D=0$ solve (\ref{eq:vac}), and we come back to the quantum
moduli space of the ${\cal N}=2$ theory. If $\mu U^\prime(0)\neq
0$ the vacuum manifold is different:
\begin{equation}
a_D=0 \qquad , \qquad
m\tilde{m}=|m|^2e^{i(\alpha+\tilde{\alpha})}=
-|\frac{\mu}{\bar{g}_D\sqrt{2}}U^\prime(0)|e^{i\beta}
\label{eq:vac1}
\end{equation}
is the solution (\ref{eq:vac}). Only the sum of the phases of $m$
and $\tilde{m}$ is fixed,
$\beta+\frac{\pi}{2}=\alpha+\tilde{\alpha}$, and there is a circle
of gauge-equivalent vacua parametrized by $\alpha-\tilde{\alpha}$.

(Massless) magnetic monopoles, dual s-electrons, condense:
spontaneous symmetry breaking of the dual Abelian group arises
from identifying the vacuum manifold as the orbit of the $U(1)_D$
action. Note that now the quantum moduli space of vacua is formed
by the $u=\pm \Lambda^2$. In each of these two points a $U(1)$
group acts, giving rise to one circle orbit. Expanding around the
expectation values of the $M$ and $\tilde{M}$ fields
\[
M(y,\theta )=m+h_m(y)+\sqrt{2}\theta\psi_m(y)+\theta^2F_m(y)
\]
\[
\tilde{M}(y,\theta
)=\tilde{m}+\tilde{h}_m(y)+\sqrt{2}\theta\tilde{\psi}_m(y)+\theta^2\tilde{F}_m(y)
\]
scalar and vector mass terms arise in the Lagrangian density:
\[
{\cal L}=\bar{g}_D^2|m|^2(|h_m|^2+|\tilde{h}_m|^2+A_\mu^DA_D^\mu)
\qquad .
\]
The electric Higgs mechanism for the fermions, albeit in a more
involved manner also happens.
\subsubsection{Stable semi-local vortices}
Variational calculus on the effective Bosonic action for the
$\phi_d$ field tells us that topological defects can arise as
solutions of the classical field equations such that
$\phi_m(x)\tilde{\phi}_m(x)\neq m\tilde{m}$ in some region of the
plane if and only if $U^{\prime\prime}(0)=0$. But this is indeed
the case because when $u\rightarrow \Lambda^2$,
$U(\varphi_D)=a_0\varphi_D+\Lambda^2$, as one can check from the
Seiberg-Witten solution for the prepotential near $u=\Lambda^2$
and $u=\infty$ .

The effective action being Abelian, we expect to find planar
solitons if $\mu\neq 0$; we look for minima of the bosonic
effective energy per unit of length of time-independent
axi-symmetric configurations:
\begin{eqnarray*}
{\cal E}_{eff}^B(a_D=0)&=&\int
d^2x\{\frac{1}{4}f_{12}^Df_{12}^D+(\nabla_i\phi_m)^*\nabla_i\phi_m+(\nabla_i\tilde{\phi}_m)^*\nabla_i\tilde{\phi}_m+\nonumber\\
&+&|\sqrt{2}\bar{g}_D\phi_m\tilde{\phi}_m+C|^2+\frac{\bar{g}_D^2}{2}(|\phi_m|^2-
|\tilde{\phi}_m|^2)^2\} \qquad ,
\end{eqnarray*}
where $C=C_1+iC_2=\mu U^\prime (0)$ and $i=1,2$.

Following the method of Ref. \cite{Guil} we restrict the search to
the trial surface in the ${\cal N}=2$ $I$-space ${\bf C}^2$
defined by:
\[
\tilde{\phi}_m=-e^{i\beta}\phi_m^* \qquad , \qquad {\rm
tan}\beta=\frac{C_2}{C_1} \quad  .
\]
This condition is compatible with the asymptotic behaviour
\[
\lim_{r\rightarrow\infty}\phi_m(x_1,x_2)=m \qquad , \qquad
\lim_{r\rightarrow\infty}\tilde{\phi}_m(x_1,x_2)=\tilde{m}
\]
guaranteeing finiteness of the \lq\lq string tension" ${\cal
E}_{eff}^B(a_D=0)$. The topological conditions for the existence
of vortices are met, and, introducing the new variables
\[
\phi_m=\frac{1}{2}\phi\ \ \ \ \ f^2=2\sqrt{2}\frac{|C|}{\bar{g}_D}
\]
we end, for a given $\beta$, with the Ginzburg/Landau/Higgs free
energy:
\[
{\cal E}_{eff}^B(a_D=0,\beta)=\int
d^2x\{\frac{1}{2}f_{12}^Df_{12}^D+\frac{1}{2}(\nabla_i\phi)^*\nabla_i\phi+\frac{\bar{g}_D^2}{8}(|\phi|^2-f^2)^2\}
\]
From the Bogomolny splitting
\begin{equation}
\begin{array}{cc}{\cal E}_{eff}^B(a_D=0,\beta)=
\int d^2x\{[F_{12}^D\mp\frac{\bar{g}_D}{2}(f^2-|\phi|^2)]^2+
|(\nabla_1\pm i\nabla_2)\phi|^2\}\pm\\\pm\int
d^2x\{\frac{\bar{g}_D}{2}
(f^2-|\phi|^2)F_{12}^D\pm\frac{i}{2}(\nabla_1\phi^*
\nabla_2\phi-\nabla_2\phi^*\nabla_1\phi)\}\end{array},
\end{equation}
it is seen that the absolute minima satisfy the first order
equations:
\begin{equation}
(\nabla_1\pm i\nabla_2)\phi=0\qquad , \qquad
F_{12}^D=\pm\frac{\bar{g}_D}{2}(f^2-|\phi|^2)
\end{equation}
The string tension is easy to compute for the self-dual vortex
solutions:
\begin{equation}
{\cal E}_{eff}^B(a_D=0)=\frac{\bar{g}_D}{2}f^2|\int d^2xF_{12}^D\,
|\qquad ;
\end{equation}
it is proportional to the electric flux. Due to the topology of
the configuration space of the problem, the ``electric" flux of
the vortex solutions is quantized in ``quanta" of ${\displaystyle
\frac{2\pi}{\bar{g}_D}}$ flux. Therefore, the self-dual solutions
are electric flux lines of energy per length unit ${\cal
E}_{eff}^B[{SD}]=\pi f^2n$ which arise at the critical point
between type I and type II dual superconductivity. The existence
of electric flux filaments warrants the confinement of the
electric charges at this limit of the super-symmetric system.

A subtle question remains: despite favorable topology it is not
clear whether or not these electric vortices are stable because
the amplitude of internal space. To address this question we
define \lq\lq natural " fields:
\[
\phi_\pm=\tilde{\phi}_m\pm e^{i\beta}\tilde{\phi}_m \qquad .
\]
In the new field variables the energy per length unit reads:
\begin{equation}
{\cal E}_{eff}^B(a_D=0)=\frac{1}{2}\int d^2x \left[
f_{12}^Df_{12}^D+|\nabla_i\phi_+|^2+\frac{1}{2}| \nabla_i\phi_-|^2
+\bar{g}_D^2\{\frac{(|\phi_+|^2+|\phi_-|^2-f^2)^2}{4}+|\phi_+|^2\}\right]\label{semi}
\end{equation}
Dropping the last term in (\ref{semi}) the vortices would be in
neutral equilibrium with ${\mathbb CP}^1$-lumps, arising semi-local
topological defects as a continuum interpolation between both
kinds of solitons, see \cite{semi-loc}. The $|\phi_+|^2$-term
means that the evolution from the vortices towards the ${\mathbb
CP}^1$-lumps requires energy; thus, the electric vortices are
stable.

\clearpage
\section{${\cal N}=2$ twisted SUSY Yang-Mills theory}

\subsection{${\cal N}$=2 gauge theory on four-manifolds:
Wick rotation and twisting}

We try now to define a ${\cal N}=2$ super-symmetric gauge theory
on a four-dimensional Riemannian manifold, i.e., we replace ${\mathbb
R}^{1,3}$ by a manifold  $X$ which is locally ${\mathbb R}_E^4$, as
the base space. We would expect that the super-space ${\mathbb
R}^{4|8}$ should also be replaced by a super-manifold $S$ locally
homeomorph to ${\mathbb R}_E^{4|8}$.

The problem is that the fields become sections in associated
bundles to $P({\rm Spin} (4), X)$, e.g., scalars, spinors, vectors,
etcetera. Thus, $X$ must be spin and this requirement breaks
super-symmetry. The only loophole to build a SUSY theory on a 4
manifold locally Euclidean is to perform a process called TWIST. In this Section we shall discuss
the seminal papers by Witten Topological Quantum Field Theory,
\cite{WittenTQFT}, and Monopoles and Four Manifolds, \cite{Witten}, where subtle Donaldson invariants distinguishing between smooth structures in four manifolds are characterized by means of certain expectation values in Topological Quantum Field Theories.

\subsubsection{The twist: mixing of internal and
external symmetries}

We have seen in previous Sections that the group of symmetry of
the ${\cal N}=2$ SUSY gauge theory is:
\[
H=SU(2)_- \otimes SU(2)_+ \otimes SU(2)_R \otimes U(1)_C \qquad .
\]
Here $K=SU(2)_- \otimes SU(2)_+$ is the group of external
``rotations", whereas $U(2)_R=SU(2)_R \otimes U(1)_C$ is the
internal symmetry group. The idea of the twist is to replace the
group $K$ by ${\bar K}={\bar {SU}}(2)_-\otimes SU(2)_+$, where
${\bar {SU}}(2)_-$ is the diagonal sub-group of $SU(2)_-\otimes
SU(2)_R$. The description of how this change affects to the
super-charges, the super-algebra, the super-manifold and the
fields follows.
\begin{enumerate}
\item ${\bf Twisted}$ ${\bf super-charges}$
The \lq\lq quantum" numbers of the SUSY and twisted SUSY charges
with respect to the $K$ and ${\bar K}$ groups are shown in the
next Box. The labels are the representations of the $SU(2)$
sub-groups and the $U(1)_C$ charge ( the super-script).

\[
\begin{array}{|cc|c|cc} \hline && \\[-0.1cm]  {\bf  SUSY}\,
{\bf Charge}
  & & {\bf Twisted}\, {\bf SUSY}\, {\bf Charge} \\ \hline && \\
  [-0.1cm]  Q^I_\alpha \qquad (\frac{1}{2},0,\frac{1}{2})^1
  & & Q^\beta_\alpha \qquad
(\frac{1}{2}\otimes\frac{1}{2},0)^1 \\ \hline && \\[-0.1cm]
  {\bar
Q}_{I{\dot \alpha}}\qquad (0,\frac{1}{2},\frac{1}{2})^{-1} & &
{\bar Q}_{\beta{\dot \alpha}} \qquad
(\frac{1}{2},\frac{1}{2})^{-1} \\
\hline
\end{array}
\]

The explicit expression of the twisted in terms of the non-twisted
SUSY charges can be obtained by means of the analogy with one-half
spin systems.
\begin{itemize}
\item  Understanding the super-charges $Q^I_\alpha$ as the states
of a system of two spin one-half particles, we identify the
following spin arrangements:
\[
Q^1_1\simeq |\uparrow \, \downarrow \rangle \qquad , \qquad
Q^2_1\simeq |\downarrow \, \downarrow \rangle \qquad , \qquad
Q^1_2\simeq |\uparrow \, \uparrow \rangle \qquad , \qquad
Q^2_2\simeq -|\downarrow \, \uparrow \rangle
\]
The interpretation is obvious: the I labels the spin of the first
particle , up or down, and the same job does the $\alpha$ for the
spin of the second particle. We have chosen to identify $I=1$ and
$\alpha=2$ with spin up and a $e^{i\pi}$ phase is assigned to the
$Q^2_2$ state to fit with the fact that ${\rm Tr}Q$ is an scalar.
The decomposition of the tensor product of two $\frac{1}{2}$
representations in terms of irreducible representations,
\[
(\frac{1}{2}\otimes\frac{1}{2},0)=(0,0)\oplus (1,0)
\]
leads us to introduce an scalar super-charge
\[
Q=\frac{1}{2}(Q^1_1+Q^2_2)\simeq \frac{1}{2}(|\uparrow \,
\downarrow \rangle -|\downarrow \, \uparrow \rangle) \qquad ,
\]
with $(0,0)^1$ quantum numbers. There are also three
super-charges,
\begin{eqnarray*}
Q^3=\frac{1}{2}(Q^1_1-Q^2_2)&\simeq & \frac{1}{2}(|\uparrow \,
\downarrow \rangle +|\downarrow \, \uparrow \rangle ) \\
Q^+=Q^1_2\simeq |\uparrow \, \uparrow \rangle  \qquad &,& \qquad
Q^-=Q^2_1\simeq |\downarrow \, \downarrow \rangle \quad ,
\end{eqnarray*}
which carry $(1,0)$ quantum numbers. Defining $Q^1=Q^++Q^-$ ,
$Q^2=i(Q^+-Q^-)$ and using the three complex structures
$\eta^a_{\mu\nu}$ (`t Hooft symbols) of the hyper-Khaler manifold
${\mathbb R}^4$ the $(1,0)^1$ representation is organized as a
self-dual anti-symmetric tensor:
\[
Q_{\mu\nu}=\sum_{a=1}^3\eta^a_{\mu\nu}Q^a \qquad, \qquad
Q_{\mu\nu}=\frac{1}{2}\varepsilon_{\mu\nu\rho\sigma}Q^{\rho\sigma}
\]

\item From the vector of matrices $\sigma^\mu=(i{\bf
1}_2,\sigma_i)$, where $\sigma_i, i=1,2,3$, are the Pauli
matrices, and through the identification,
\[
\bar{Q}_{\beta\dot{\alpha}}=(\sigma^\mu Q_\mu
)_{\beta\dot{\alpha}}\qquad ,
\]
a vector super-charge $Q_\mu$ with quantum numbers
$(\frac{1}{2},\frac{1}{2})^{-1}$ is obtained in terms of the
twisted super-charges ${\bar Q}_{\beta{\dot\alpha}}$:

\begin{eqnarray*}
Q_1=\frac{1}{2}({\bar Q}_{1{\dot 2}}+{\bar Q}_{2{\dot 1}}) \qquad
\qquad &,& \qquad \qquad Q_2=\frac{i}{2}({\bar Q}_{1{\dot
2}}-{\bar Q}_{2{\dot 1}}) \\ Q_3=\frac{1}{2}({\bar Q}_{1{\dot
1}}-{\bar Q}_{2{\dot 2}})\qquad \qquad &,& \qquad \qquad
Q_4=-\frac{i}{2}({\bar Q}_{1{\dot 1}}+{\bar Q}_{2{\dot 2}})
\end{eqnarray*}

\end{itemize}

\item ${\bf Twisted}$ ${\bf SUSY}$ ${\bf algebra}$

The SUSY algebra
\[
\{ Q^I_\alpha,\bar{Q}_{J\dot{\beta}}\}=\delta^I_J
P_{\alpha\dot{\beta}}
\]
becomes in terms of the twisted generators:
\begin{eqnarray}
Q^2=Q_\mu^2=Q_{\mu\nu}^2=0 \qquad &,& \qquad \{Q,Q_\mu\}=P_\mu
\nonumber
\\ \{ Q, Q_{\mu\nu}\}=0 \qquad &,&
\{Q_{\mu\nu},Q_\rho\}=\varepsilon_{\mu\nu\rho\sigma}P^\sigma+
g
_{\mu\rho}P_\nu-g_{\nu\rho}P_\mu \label{eq:tsa}
\end{eqnarray}

\item ${\bf Twisted}$ ${\bf super-manifold}$

The twist prescription also requires the choice of $(x_\mu,{\bar
\theta},\theta_\mu ,{\bar\theta}_{\mu\nu})$ as local coordinates
in the ${\mathbb R}^{4|8}$ super-manifold. Again, the twisted
Grassman variables are built by the same procedure from the
non-twisted Grassman variables, having in mind the behaviour of
both kind of odd coordinates with respect to the groups $K$ and
${\bar K}$ shown in the next Box.

\[
\begin{array}{|cc|c|cc} \hline && \\[-0.1cm]  {\bf  Grassman}\,
{\bf variable}
  & & {\bf Twisted}\, {\bf Grassman}\, {\bf variable} \\ \hline && \\
  [-0.1cm]  \theta^I_\alpha \qquad (\frac{1}{2},0,\frac{1}{2})^1
  & & \theta^{\alpha{\dot \beta}} \qquad
(\frac{1}{2},\frac{1}{2})^1 \\ \hline && \\[-0.1cm]
  {\bar
\theta}_I^{\dot \alpha}\qquad (0,\frac{1}{2},\frac{1}{2})^{-1} & &
{\bar \theta}_{\dot \beta}^{\dot \alpha} \qquad
(0,\frac{1}{2}\otimes \frac{1}{2})^{-1} \\
\hline
\end{array}
\]

Definitions of vector, $\theta_\mu$, scalar, ${\bar\theta}$, and
self-dual, ${\bar\theta}_{\mu\nu}$, Grassman variables immediately
follow:
\begin{itemize}
\item From
$\theta^{\alpha{\dot\beta}}=(\sigma_\mu\theta_\mu)^{\alpha{\dot\beta}}$
we find
\begin{eqnarray*}
\theta_1=\frac{1}{2}(\theta^{1{\dot 2}}+\theta^{2{\dot 1}}) \qquad
\qquad &,& \qquad \qquad \theta_2=\frac{i}{2}(\theta^{1{\dot
2}}-\theta^{2{\dot 1}}) \\ \theta_3=\frac{1}{2}(\theta^{1{\dot
1}}-\theta^{2{\dot 2}})\qquad \qquad &,& \qquad \qquad
\theta_4=-\frac{i}{2}(\theta^{1{\dot 1}}+\theta^{2{\dot 2}})
\end{eqnarray*}
as the components of $\theta_\mu$, which belong to the
$(\frac{1}{2},\frac{1}{2})^1$ representation of $SU(2)_- \otimes
{\bar {SU}}(2)_+\otimes U(1)_C$.
\item The scalar Grassman variable
${\bar\theta}=\frac{1}{2}({\bar\theta}_{\dot 1}^{\dot
1}+{\bar\theta}_{\dot 2}^{\dot 2})$ carries the quantum numbers
$(0,0)^{-1}$ with respect to the same group.
\item The three Grassman variables
\[
{\bar\theta}^1={\bar\theta}_{\dot 2}^{\dot 1}+{\bar\theta}_{\dot
1}^{\dot 2} \quad , \quad {\bar\theta}^2=i({\bar\theta}_{\dot
2}^{\dot 1}-{\bar\theta}_{\dot 1}^{\dot 2}) \quad , \quad
{\bar\theta}^3=\frac{1}{2}({\bar\theta}_{\dot 1}^{\dot
1}-{\bar\theta}_{\dot 2}^{\dot 2})
\]
are assembled in the $(0,1)^{-1}$ self-dual combination
${\bar\theta}_{\mu\nu}=\sum_{a=1}^3\eta^a_{\mu\nu}{\bar\theta}^a$
.
\end{itemize}
Thus, it is easy to check that
\[
Q=\frac{\partial}{\partial{\bar\theta}}\quad , \quad
Q_\mu=\frac{\partial}{\partial\theta_\mu}-i{\bar\theta}\partial_\mu
\quad , \quad
Q_{\mu\nu}=\frac{\partial}{\partial{\bar\theta}_{\mu\nu}}-i(\varepsilon_{\mu\nu\rho\sigma}
\theta^\rho\partial^\sigma+\theta^\mu\partial^\nu-\theta^\nu\partial^\mu
)
\]
satisfy the twisted SUSY algebra (\ref{eq:tsa}), and, therefore,
the transformations of the super-Poincare group
$P(x_\mu,\varepsilon_\mu,{\bar\varepsilon},{\bar\varepsilon}_{\mu\nu})=
e^{i(-x^\mu P_\mu+\varepsilon^\mu
Q_\mu+{\bar\varepsilon}Q+{\bar\varepsilon}^{\mu\nu}Q_{\mu\nu})}$
act infinitesimally on the ${\mathbb R}_E^{4|8}$ as:
\[
x^\mu \, \rightarrow x^\mu-i\varepsilon^\mu{\bar\theta}\quad ,
\quad \theta^\mu \, \rightarrow \theta^\mu+\varepsilon^\mu \quad ,
\quad {\bar\theta}\, \rightarrow {\bar\theta}+{\bar\varepsilon}
\quad , \quad {\bar\theta}^{\mu\nu} \, \rightarrow
{\bar\theta}^{\mu\nu}+{\bar\varepsilon}^{\mu\nu} \quad .
\]

\item ${\bf Twisted}$ ${\bf fields}$

The twist works exactly in the same way for the spinor fields.
Denoting by $\psi_{I\alpha}$ -I=1, the $\psi$ (Higgsino), I=2, the
$\lambda$ (gluino)- the spinor fields of the ${\cal N}=2$ chiral
multiplet, we display the quantum numbers of both the non-twisted
and twisted version in the following Box.
\[
\begin{array}{|cc|c|cc} \hline && \\[-0.1cm]  {\bf  Spinor}\,
{\bf field}
  & & {\bf Twisted}\, {\bf Spinor}\, {\bf field} \\ \hline && \\
  [-0.1cm]  \psi_{I\alpha} \qquad (\frac{1}{2},0,\frac{1}{2})^{-1}
  & & \psi_{\beta\alpha} \qquad
(\frac{1}{2}\otimes\frac{1}{2},0)^{-1} \\ \hline && \\[-0.1cm]
  {\bar
\psi}^I_{\dot \alpha}\qquad (0,\frac{1}{2},\frac{1}{2})^{1} & &
{\bar \psi}^\beta_{\dot \alpha} \qquad
(\frac{1}{2},\frac{1}{2})^{1} \\
\hline
\end{array}
\]

Again the twist process leads to deal with vector, $\psi_\mu$,
scalar, $\eta$, and, self-dual tensor, $\chi_{\mu\nu}$, fields:
\begin{itemize}
\item From
${\bar\psi}^\beta_{{\dot\alpha}}=(\sigma^\mu\psi_\mu)^\beta_{{\dot\alpha}}$
we find
\begin{eqnarray*}
\psi_1=\frac{1}{2}(\psi^{1{\dot 2}}+\psi^{2{\dot 1}}) \qquad
\qquad &,& \qquad \qquad \psi_2=\frac{i}{2}(\psi^{1{\dot
2}}-\psi^{2{\dot 1}}) \\ \psi_3=\frac{1}{2}(\psi^{1{\dot
1}}-\psi^{2{\dot 2}})\qquad \qquad &,& \qquad \qquad
\psi_4=-\frac{i}{2}(\psi^{1{\dot 1}}+\psi^{2{\dot 2}})
\end{eqnarray*}
as the components of field $\psi_\mu$, which belong to the
$(\frac{1}{2},\frac{1}{2})^1$ representation of
$SU(2)_-\otimes{\bar {SU}}(2)_+\otimes  U(1)_C$.
\item The scalar field
$\eta=\frac{1}{2}(\psi_{11}+\psi_{22})$ carries the quantum
numbers $(0,0)^{-1}$ with respect to the same group.
\item The three combinations
\[
\chi^1=\psi_{12}+\psi_{21} \quad , \quad
\chi^2=i(\psi_{12}-\psi_{21}) \quad , \quad
\chi^3=\frac{1}{2}(\psi_{11}-\psi_{22})
\]
are assembled in the $(0,1)^{-1}$ self-dual antisymmetric tensor field
$\chi_{\mu\nu}=\sum_{a=1}^3\eta^a_{\mu\nu}\chi^a$ .
\end{itemize}
The vector field $A_\mu$ trades the quantum numbers
$(\frac{1}{2},\frac{1}{2},0)^0$ by $(\frac{1}{2},\frac{1}{2})^0$
after twisting, keeping its vector nature. The Higgs field $\phi$
and its adjoint $\phi^\dagger$, respectively with quantum numbers
$(0,0,0)^{-2}$ and $(0,0,0)^2$ are twisted to two real scalar
fields, $\lambda$ and $\phi$, labeled by $(0,0)^{-2}$ and
$(0,0)^2$ as irreducible representations of ${\bar
{SU}}(2)_-\otimes U(1)_C$. Finally, a word about the auxiliary
fields that we shall not consider in the sequel. Before twisting,
D and F belong to a vector representation of $SU(2)_R$: the
quantum numbers of $\vec{D}=D\vec{e}_1+{\rm Re}F\vec{e}_2+{\rm
Im}F\vec{e}_3$ are $(0,0,1)^1$. After twisting $\vec{D}$ becomes a
neutral self-dual tensor $G_{\mu\nu}$ with $(1,0)^0$.
\end{enumerate}

A last remark: all the twisted Grassman variables and fields
belong to real representations of the symmetry group. Therefore,
super-symmetry on Riemannian four manifolds is possible for real
fields.

\subsubsection{${\cal N}=2$ super-symmetric Euclidean action: $Q$-cohomology}

We have collected the following data: a compact oriented
4-manifold $X$ endowed with a Riemannian metric $h_{\mu\nu}$ and a
principal fiber bundle $P$ with a simple and compact structural
group $G$. $A_\mu $ belongs to the space of G-connections ${\cal
A}$ in $P$. $G_{\mu\nu},\chi_{\mu\nu}\in \Omega^{2+}(X,{\rm
Lie}G)$ are self-dual two forms, $\psi_\mu\in \Omega^{1}(X,{\rm
Lie}G)$ is a one-form and $\eta ,\lambda ,\phi \in
\Omega^{0}(X,{\rm Lie}G)$ are zero forms.

From these ingredients we build a ${\cal N}=2$ super-symmetric
action as a Berezin integral in the ${\mathbb R}_E^{4|4}$
super-space:
\begin{equation}
S=\frac{1}{4}\int d^4x \; d^4\theta \; \sqrt{{\rm det}h}\;{\rm
tr}\left\{\Phi^2+\Phi_\mu^*\Phi^\mu+\frac{1}{48}\Phi_{\mu\nu}\Phi^{\mu\nu}+
\frac{1}{12}\Phi_{\mu\nu\rho}^*\Phi^{\mu\nu\rho}+\frac{1}{12}\sum_{a=1}^2
\Phi_{\mu\nu\rho\sigma}^a\Phi^{\mu\nu\rho\sigma}_a\right\}
\label{eq:bert} \qquad .
\end{equation}
From the self-dual part of the curvature of the connection
$A_\mu$,
\[
F_{\mu\nu}^+=\frac{1}{2}(F_{\mu\nu}+\frac{1}{2}\varepsilon_{\mu\nu\rho\sigma}F^{\rho\sigma})\quad
,
\]
and the covariant derivatives of the $\phi$ and $\psi_\mu$ fields,
\[
D_\mu\psi_\nu=\nabla_\mu\psi_\nu-\Gamma^\lambda_{\mu\nu}\psi_\lambda\qquad,\qquad
D_\mu\phi=\nabla_\mu\phi \quad ,
\]
where $\Gamma^\lambda_{\mu\nu}$ are the Christoffel symbols of the
metric $h$ and $\nabla_\mu$ is the covariant derivative given  by
the $A_\mu$ connection, we have defined the following ${\cal N}=1$
tensor super-fields:
\begin{eqnarray*}
\Phi(x,\theta_\mu )&=&\phi (x)+\theta_\mu\psi^\mu
(x)+\theta_\mu\theta_\nu F^{\mu\nu}_+ \quad , \quad
\Phi_\mu=iD_\mu\phi(x)-i\theta_\sigma
D_\mu\psi^\sigma+\theta_\mu\theta_\nu\theta_\rho\chi^{\nu\rho}(x)\\
\Phi_{\mu\nu}(x,\theta_\rho)&=&i(\theta_\mu\theta_\nu
+\frac{1}{2}\varepsilon_{\mu\nu\rho\sigma}\theta^\rho\theta^\sigma)[\lambda,\phi]
\quad , \quad
\Phi_{\mu\nu\rho}(x,\theta_\sigma)=-i\varepsilon_{\mu\nu\rho\gamma}
\theta^\gamma D_\delta\psi^\delta
+\theta_\mu\theta_\nu\theta_\rho\; \eta \\
\Phi_{\mu\nu\rho\sigma}^1(x,\theta_\gamma)&=&\varepsilon_{\mu\nu\rho\sigma}(\square\phi-
\frac{i}{2}[\psi_\gamma , \psi^\gamma])+
\theta_\mu\theta_\nu\theta_\rho\theta_\sigma\; \lambda \\
\Phi_{\mu\nu\rho\sigma}^2(x,\theta_\gamma)&=&\varepsilon_{\mu\nu\rho\sigma}\phi-\frac{i}{4}
\theta_\mu\theta_\nu\theta_\rho\theta_\sigma\;([\chi_{\gamma\delta},\chi^{\gamma\delta}]
+\frac{1}{4}[\eta,\eta])
\end{eqnarray*}
All the $\Phi_{\mu\cdots }$ fields have $U(1)_C$ charge equal to
two and the normalization is chosen in such a way that Berezin
integration in (\ref{eq:bert}) gives the famous Witten topological
action:
\begin{eqnarray}
S=\int &d^4x&\;\sqrt{{\rm det}h}\;{\rm
tr}\{\frac{1}{2}F_{\mu\nu}^+ F_+^{\mu\nu}+i\chi^{\mu\nu}
D_\mu\psi_\nu-i\eta D^\mu\psi_\mu-\frac{i}{8}\phi
[\chi_{\mu\nu},\chi^{\mu\nu}]\nonumber \\ &-&\frac{i}{2}\lambda
[\psi_\mu,\psi^\mu ]+\frac{1}{2}\lambda D_\mu
D^\mu\phi-\frac{i}{2}\phi [ \eta,\eta ]
-\frac{1}{8}[\lambda,\phi]^2\} \label{eq:tac}
\end{eqnarray}

To test the invariance of the Euclidean action (\ref{eq:tac})
under twisted super-symmetry we must know how the Q-generator acts
on the fields. This is easily read from the super-fields:
\begin{eqnarray*}
A_\mu (x,{\bar\theta})&=&A_\mu (x)+{\bar\theta}\psi_\mu(x)\qquad ,
\qquad \psi_\mu(x,{\bar\theta})=\psi_\mu(x)+{\bar\theta}D_\mu\phi
(x) \quad , \quad \phi(x,{\bar\theta})=\phi(x) \\\chi_{\mu\nu}
(x,{\bar\theta})&=&\chi_{\mu\nu}(x)+{\bar\theta}F_{\mu\nu}^+(x)\qquad
, \qquad \eta(x,{\bar\theta})=\eta (x)+i{\bar\theta}[\lambda
(x),\phi (x)]\quad , \, \lambda(x,{\bar\theta})=\lambda(x)\, ,
\end{eqnarray*}
of $U(1)_C$ charges 0,1,2,-1,-1,-2. We see that:
\[
\begin{array}{ccccc}
Q
A_\mu(x,{\bar\theta})=\psi_\mu&\qquad&Q\phi(x)=0&\qquad&Q\chi_{\mu\nu}(x,{\bar\theta})=F_{\mu\nu}^+\\
Q\psi_\mu(x,{\bar\theta})=D_\mu\phi&\qquad&Q\eta(x,{\bar\theta})=i[\lambda,\phi]&\qquad&Q\lambda(x)=0
\end{array}\qquad .
\]
From this, one immediately checks that
\[
S=\int \; dx^4 \; \sqrt{{\rm det}h}\; Q\cdot V(x,{\bar\theta})\qquad ,
\]
where
\[
V=\frac{1}{4}{\rm
tr}\;F_{\mu\nu}\chi^{\mu\nu}(x,{\bar\theta})+\frac{1}{2}{\rm
tr}\;\psi_\mu(x,{\bar\theta}) D^\mu \lambda-\frac{1}{4}{\rm
tr}\;\eta(x,{\bar\theta}[\phi,\lambda]\qquad .
\]
Therefore, S is Q-exact and the Witten action is invariant under
Q: $Q\cdot S=0$.

Moreover, the energy-momentum tensor, defined through the
variation of the energy density $S=\int \, dx^4 {\cal L}$ with
respect to the metric tensor,
\[
T_{\mu\nu}=\frac{\delta{\cal L}}{\delta h_{\mu\nu}}\qquad,\qquad
D_\mu T^{\mu\nu}=0
\]
besides of being divergenceless in the covariant sense, is also
Q-exact. $T_{\mu\nu}=Q\Lambda_{\mu\nu}(x,{\bar\theta})$, where
\begin{eqnarray*}
\Lambda_{\mu\nu}(x,{\bar\theta})&=&\frac{1}{2}{\rm
tr}\;(F_{\mu\sigma}\chi_\nu^\sigma(x,{\bar\theta})+F_{\nu\sigma}\chi_\mu^\sigma(x,{\bar\theta})
-\frac{1}{2}h_{\mu\nu}F_{\sigma\rho}\chi^{\sigma\rho}(x,{\bar\theta})+\psi_\mu(x,{\bar\theta})
D_\nu\lambda\\ &+&\psi_\nu(x,{\bar\theta})
D_\mu\lambda-h_{\mu\nu}\psi_\sigma(x,{\bar\theta})
D^\sigma\lambda+\frac{1}{4}h_{\mu\nu}\eta(x,{\bar\theta})[\phi,\lambda])\qquad
.
\end{eqnarray*}
As we shall show, this is a crucial point to establish the
topological meaning of the Witten action (\ref{eq:tac}).

\subsection{Topological field theory}
In this sub-Section we proceed to quantize the classical field
theory described in the sub-Section \S 4.1 . We choose functional
integration as (Feynman) quantization procedure and the main
result is that expectation values of certain operators are ${\bf
C}^\infty$-topological invariants of four-manifolds. Studies of
moduli spaces of self-dual connections in G-bundles on compact
Riemannian manifolds by Atiyah, Hitchin, and Donaldson
unveiled the lack of dipheomormisms between four-manifolds which,
nevertheless, are homeomorphic with respect to each other. This
happened at the beginning of the eighties in the past century and,
slightly later, polynomial invariants were discovered by Donaldson
distinguishing between non-dipheomorphic but homeomorphic
four-manifolds. The idea comes from the analysis of the homology
of the moduli space of self-dual connections. Following Witten, we
shall describe next how the Donaldson invariants arise in the
topological field theory proposed above.

\subsubsection{Functional integral representation of topological invariants }
Let us suppose that there exists an integration measure (whatever
that means )
\[
{\cal D}{\cal F}\simeq {\cal D}A\;{\cal D}\phi\;{\cal
D}\lambda\;{\cal D}\eta\;{\cal D}\psi\;{\cal D}\chi
\]
in the space of all the $(A,\phi,\lambda,\eta,\psi,\chi)$ fields.
The Feynman principle guides us to define the expectation value in
the ground state of a polynomial function of the fields $W$ as the
functional integral:
\[
Z(W)=\int_{{\cal A}/{\cal G}}{\cal D}{\cal F}\exp(-\frac{1}{g^2}
S)\cdot W=<W>  \qquad .
\]
Note that the integration domain is the space of field orbits of
the gauge group; we take quotient by the action of the gauge
group.

Super-symmetry at the quantum level requires:
\[
\exp(\varepsilon Q){\cal D}{\cal F}\exp(-\frac{S}{g^2})\equiv{\cal
D}{\cal F}\exp(-\frac{S}{g^2})\quad\Rightarrow\quad Z_\varepsilon
(W) \neq f(\varepsilon) \, \, ,
\]
i.e., super-symmetry is non-anomalous because both the classical
action and the integration measure are invariant with respect to
the super-symmetry generator. Moreover, the expectation value of
anti-commutators of $Q$ - in the sequel, we consider $Q$ as the
vector field acting on the space of fields through super-Poisson
brackets- with any functional ${\cal O}$ of the fields is zero:
\[
Z_\varepsilon({\cal O})=\int{\cal D}{\cal F}\exp(\varepsilon
Q)\cdot\{\exp(-\frac{S}{g^2})\cdot {\cal O}\}=\int{\cal D}{\cal
F}\exp(-\frac{S}{g^2})({\cal O}+\varepsilon\{Q,{\cal O}\})
\]
\[
\Rightarrow\qquad 0=<\{Q,{\cal O}\}>=\int {\cal D}{\cal
F}\exp(-\frac{S}{e^2})\cdot\{Q,{\cal O}\}\qquad .
\]

The partition function
\begin{equation}
Z=\int_{{\cal A}/{\cal G}}{\cal D}{\cal F}\exp(-\frac{1}{g^2} S) \label{zdon}
\end{equation}
is only built from $Q$-closed states:
$Z_\varepsilon=\exp(\varepsilon Q)Z=Z$. We show next that it is
the simplest topological invariant, by the simple idea of testing
the dependence of $Z$ on the Riemannian metric. Because the
variation of ${\rm exp}[-\frac{S}{g^2}]$ with respect to the
metric is $Q$-exact, assuming that the integration measure is
independent of $h^{\mu\nu}$ (there are no gravitational
anomalies), and taking into account that $<\{Q,{\cal O}\}>=0$ we
see that the partition function $Z$ is a topological invariant
(independent of the metric):
\begin{eqnarray*}
\delta Z&=&\int{\cal D}{\cal
F}\exp[-\frac{S}{g^2}]\cdot(-\frac{\delta S}{g^2})\\
&=&-\frac{1}{g^2}\int{\cal D}{\cal
F}\exp[-\frac{S}{g^2}]\cdot\{Q,\int \, dx^4\, \sqrt{{\det h}}\,
\delta h^{\mu\nu}\Lambda_{\mu\nu}\}=0 \qquad .
\end{eqnarray*}

A semi-classical computation of $Z$ gives an exact answer for the
topological invariant; one easily shows that the partition
function is also independent of the gauge coupling $g$:
\begin{eqnarray*}
\delta Z&=&\delta(-\frac{1}{g^2})\int{\cal D}{\cal
F}\exp[-\frac{S}{g^2}]\cdot S\\ &=&\delta(-\frac{1}{g^2})\int{\cal
D}{\cal F}\exp[-\frac{S}{g^2}]\cdot\{Q, V\}=0 \qquad .
\end{eqnarray*}

\subsubsection{Instantons and the $U(1)_C$ anomaly}
An infinite dimensional generalization of the steepest descent
method for computing integrals of functions of the form $\exp
[\frac{1}{g^2}f(x)]$ where $g^2$ is a very small parameter tells
us that the main contribution to the integral is localized around
the classical minima of $S$. It is well known that the moduli
spaces of anti-self-dual connections,
\begin{equation}
F_{\mu\nu}=-*F_{\mu\nu}\qquad \qquad \Leftarrow\Rightarrow \qquad
\qquad F_{\mu\nu}^+=0  \qquad , \label{eq:self}
\end{equation}
called instantons in physicist's folklore, form such varieties
of absolute minima of Witten's topological action $S$
(\ref{eq:tac}).

Calling $E$ the bundle where the fields of the theory are
sections, the formal dimension of the moduli space ${\cal M}_I$ of
anti-self-dual connections in the associated $G=SU(2)$ $P$-bundle
over $X$ is given by the topological formula
\begin{equation}
d({\cal M}_I)=8 p_1(E)-\frac{3}{2}(\chi(X)+\sigma(X))=8 p_1(E)-3(1+b_+^2)\label{eq:dim}
\end{equation}
where 
\[
p_1(E)=-\frac{1}{8\pi^2}\,\int_X \, {\rm tr} \left( F \wedge F \right)
\]
is the first Pontryagin number of $E$, $F$ is the curvature of the Yang-Mills connection, and $\chi(X)=b^0+b^2+b^4$
and $\sigma(X)=b_+^2-b_-^2$ are respectively the Euler characteristic and the
signature of the 4-manifold $X$.

Formula (\ref{eq:dim}) is derived from deformation theory: if $A$
is a solution of (\ref{eq:self}) $A+\delta A$ is also a solution
of the anti-self-duality equations if and only if:
\begin{equation}
D_\mu\;\delta A_\nu-D_\nu\;\delta
A_\mu+\varepsilon_{\mu\nu\rho\sigma}D^\rho\delta A^\sigma=0
\label{eq:def}
\end{equation}
To count only deformations orthogonal to the orbits of the gauge
group we choose the \lq\lq background\rq\rq gauge condition:
\begin{equation}
D_\mu\; \delta A^\mu=0 \qquad \qquad .\label{eq:defga}
\end{equation}
The dimension $d({\cal M}_I)=n$ of the space of solutions of
(\ref{eq:def})-(\ref{eq:defga}) is formally the dimension of the
moduli space of anti-self-dual connections (the real dimension if
there are no reducible connections).

Looking at formula (\ref{eq:tac}) for Witten's action one
immediately checks that (\ref{eq:def}) is the field equation
coming from variations with respect to $\chi$
\begin{equation}
D_\mu\;\psi_\nu-D_\nu\;\psi_\mu+\varepsilon_{\mu\nu\rho\sigma}D^\rho\psi^\sigma=0
\label{eq:zerm} \quad ,
\end{equation}
while variations with respect to $\eta$ give
\begin{equation}
D_\mu\;\psi^\mu=0 \qquad . \label{eq:zerm1}
\end{equation}
Therefore, the dimension of the space of fermionic zero modes in
the instanton field is equal to the dimension of the moduli space
of anti-self-dual connections. If the gauge group is $SU(2)$ there
are no $\eta$- and $\chi$-zero modes of fluctuation around anti-self-dual instantons. 
An index theorem tells us that, generically,
$n_+-n_-=d({\cal M}_I)$, where $n_+$ is the number of linearly
independent $\psi_\mu$ zero modes ( recall that the $U(1)_C$
charge of $\psi$ is +1) and $n_-$ is the number of linearly
independent $\eta$ and $\chi$ zero modes ( recall that the
$U(1)_C$ carge of $\eta$ and $\chi$ is -1). Physicists read the
index theorem as the anomaly in the $U(1)_C$ symmetry induced by
instantons at the semi-classical level:
\[
d({\cal M}_I)=\Delta U(1)_C=n_+-n_- \qquad ,
\]
i.e., the integration measure ${\cal D}{\cal F}$ is not invariant
under $U(1)_C$ but transform with a weight $-d({\cal M})$.

\subsubsection{k-point correlation functions}
Due to the existence of fermionic zero modes between the
fluctuations around the instanton field and the properties of
Berezin integration measures the partition function - zero-point
correlation function- $Z$ vanishes unless $X\ $,$\ G$ and $\ E\ $
are such that: $d({\cal M}_I)=0$. We address first this case,
assuming that the real dimension of the moduli space is really 0:
there is a finite discrete set of isolated instantons.

Assembling the Bose $\Phi=(A,\phi,\lambda)$ and Fermi
$\Psi=(\eta,\psi,\chi)$ fields of the theory under the labels
$\Phi$ and $\Psi$ the expansion of the action around one of these
instantons up to quadratic order reads:
\[
S_{(2)}=\int_X\sqrt{{\rm det}h}\;(\Phi\Delta_B\Phi+i\Psi D_F\Psi)
\]
Here $\Delta_B$ is a second-order matrix differential operator that
rules the Bosonic small fluctuations around the instanton field.
$D_F$ is a real skew-symmetric matrix first-order operator governing the
Fermionic small deformations of the instanton and supersymmetry
establishes a link between the spectra of these two operators:
\[
iD_F\Psi=\omega\Psi\quad \leftrightarrow\quad
\Delta_B\Psi=\omega^2\Psi \qquad .
\]
The eigenvalues of $D_F$ are purely imaginary and come in pairs.
Therefore, the contribution of one instanton to the partition
function in the weak coupling limit is
\[
\frac{{\rm Pffaf}D_F}{\sqrt{\det
\Delta_B}}=\pm\prod_n\frac{\omega_n}{\sqrt{\omega_n^2}}
\]
because we only need to evaluate the integration of Bosonic and
Fermionic Gaussians of width given by the eigenvalues of
$\Delta_B$ and $D_F$. There is no problem in the regularization of
the infinite product of ratios of eigenvalues but there is no way
in solving the ambiguity in sign. The convention is set in three
steps:
\begin{enumerate}

\item
Choose a sign, e.g. +, for a given instanton $A_\mu^{I_1}$.
\item
Consider a second instanton $A_\mu^{I_2}$ connected to the first
instanton via the family of connections
\[
A_\mu^t=tA_\mu^{I_2}+(1-t)A_\mu^{I_1} \qquad \qquad .
\]
Assign to $A_\mu^{I_2}$ a minus sign if the spectral flow of $D_F$
along the $A_\mu^t$ family crosses an odd number of zeroes.
\item
This process is independent of the path chosen in the space of
connections ${\cal A}/{\cal G}$: this is equivalent to say that
there are no global anomalies and the \lq\lq Pfaffian" line bundle
on ${\cal A}/{\cal G}$ exists and is trivial. Because the moduli
space of (irreducible) anti-self-dual connections is a finite
subset of the space of gauge connections - ${\cal M}\subset {\cal
A}/{\cal G}$- the above statement is tantamount to the
orientability of ${\cal M}$.
\end{enumerate}
Adding the contributions to the partition function we find in the
weak coupling limit the formula:
\begin{equation}
Z\propto \sum_i\frac{{\rm Pfaff}(D^{(i)}_F)}{\sqrt{\det
\Delta^{(i)}_B}}=\sum_i\left(\pm\prod_{n_i}\frac{\omega_{n_i}}{\sqrt{|\omega_{n_i}|^2}}
\right)=\sum_i(-1)^{N_i} \qquad ,\qquad N_i=0,1 \qquad ,
\label{eq:don1}
\end{equation}
where the $N_i$ give the signs as explained above and the discrete index $i$ runs over the number of instantons. The result is
independent of $h_{\mu\nu}$ and $g$: the partition function $Z$ is the first
Donaldson $C^\infty$-topological invariant for four-manifolds.

When $d({\cal M})>0$, the non-vanishing correlation functions are
path integrals of the form:
\begin{equation}
Z({\cal O})=\int{\cal D}{\cal F}\, {\cal O}\cdot
\exp\{-\frac{S}{g^2}\} \qquad , \label{eq:corr1}
\end{equation}
where ${\cal O}$ is any function of the fields such that the
$U(1)_C$ charge of ${\cal O}$ is equal to $d({\cal M})$; ${\cal O}$
supplies the right number of Grassman variables to compensate the
fermionic zero modes in the Berezin measure and obtain a non-zero result.

Is $Z({\cal O})$ a topological invariant? The variation of
(\ref{eq:corr1}) under a change of the metric is:
\begin{eqnarray}
\delta Z({\cal O})&=&\int{\cal D}{\cal F}
\exp(-\frac{S}{g^2})\cdot(-\frac{\delta S}{g^2}{\cal O}
+\delta{\cal O})\nonumber\\ &=&\int{\cal D}{\cal F}
\exp(-\frac{S}{g^2})\cdot(-\frac{1}{2g^2}\{Q,\int\sqrt{{\rm det}h
}\;\delta h^{\mu\nu}\Lambda_{\mu\nu}\}{\cal O}+\delta{\cal O})
\qquad , \label{eq:corr2}
\end{eqnarray}
where $\delta{\cal O}$ is the variation of ${\cal O}$ with respect
to the metric $h_{\mu\nu}$. If $\{Q,{\cal O}\}=0$
\[
\int{\cal D}{\cal F}
\exp(-\frac{S}{g^2})(-\frac{1}{g^2}\{Q,\int\sqrt{{\rm det}h}\, \, \delta
h^{\mu\nu}\Lambda_{\mu\nu}\}{\cal O})
=-\frac{1}{g^2}<\{Q,\int\sqrt{{\rm det}h}\, \, \delta
h^{\mu\nu}\Lambda_{\mu\nu}\}{\cal O}>=0
\]
and the first term in (\ref{eq:corr2}) is zero. Therefore, the
only remaining condition on ${\cal O}$ for topological invariance
of $Z({\cal O})$ is independence of the metric: $\delta{\cal
O}=0$. Thus, we choose ${\cal O}$ such that:
\[
1)\quad\{Q,{\cal O}\}=0 \qquad \qquad 2)\quad {\cal
O}\neq\{Q,\rho\} \qquad \qquad 3)\quad \delta{\cal O}=0 \qquad .
\]
Note that $Z({\cal O})=0$ if ${\cal O}=\{Q,\rho\}$ for any $\rho$
and this justifies the condition 2).

The only field complying with these conditions is the spin zero
scalar field $\phi$: $\phi$ belongs to the kernel of $Q$, does not
depend on the metric, and there is no $\rho$ such that
$\phi=\{Q,\rho\}$. Therefore, ${\cal O}$ must be defined in terms
of the G-invariant polynomials of $\phi$; for $G=SU(2)$ there is
only one independent G-invariant quadratic polynomial:
\[
{\cal O}={\rm tr}\;\phi^2 \qquad \qquad .
\]
To emphasize that it is a local operator that depends on a point P
it is usually written as:
\[
W_0(P)=\frac{1}{2}{\rm tr}\; \phi^2 (P)\qquad .
\]
Because the $U(1)_C$ charge of $W_0(P)$ is 4, when the bundle E
and the 4 manifold X are such that $d({\cal M}_I)=4k$, $k\in{\mathbb
Z}$, the k-point correlation function
\begin{equation}
Z(k)=\int{\cal D}{\cal F}\exp(-\frac{S}{g^2})\prod_{i=1}^k
W_0(P_i)=<W_0(P_1)\cdots W_0(P_k)> \label{eq:kpo}
\end{equation}
is a $C^\infty$-topological invariant of the four manifold $X$, the simplest type of Donaldson polynomial expressed as 
a vacuum expectation value of field operators in Witten's topological quantum field theory.

\subsubsection{Donaldson polynomials}
In particular, $Z(k)$ is independent of the choice of the
$P_1,P_2,\ldots,P_k$  points in X because it is independent of the
metric. To test explicitly the last proposition we compute
\[
\frac{\partial}{\partial x^\mu}W_0={\rm tr}\;\phi
D_\mu\phi=i\{Q,{\rm tr}\;\phi\psi_\mu\}
\]
to find a Q-exact answer. Thus, we write
\[
W_0(P)-W_0(P^\prime)=\int_{P^\prime}^P\frac{\partial W_0}{\partial
x^\mu}d x^\mu=i\{Q,\int_{P^\prime} W_1\} \qquad ,
\]
where $W_1={\rm tr}\; (\phi\psi_\mu)dx^\mu={\rm tr}\;(\phi\wedge
\psi)$ is a one-form on $X$ which gives the derivative of $W_0$ as
a BRST commutator. Therefore,
\[
\langle (W_0(P)-W_0(P^\prime)\rangle \cdot
\prod_jW_0(P_j)>=\langle \{Q,i\int_{P^\prime}^PW_1\cdot
\prod_jW_0(P_j)\rangle =0
\]
and the expectation value at $P$ is equal to the expectation value
at $P^\prime$.

This process can be iterated recursively
\begin{eqnarray*}
0&=&i\{Q,W_0\}\qquad\qquad d W_0=i\{Q,W_1\} \qquad\qquad d
W_1=i\{Q,W_2\} \\d W_2&=&i\{Q,W_3\}\qquad\qquad d W_3=i\{Q,W_4\}
\qquad\qquad  d W_4=0
\end{eqnarray*}
\[
W_2={\rm tr}(\frac{1}{2}\psi\wedge\psi+i\phi\wedge F)\quad 
, \quad \quad  W_3=i\;{\rm tr}\;(\psi\wedge F) \quad , \quad \quad
W_4=-\frac{1}{2}{\rm tr}\;(F\wedge F) \qquad .
\]
Recall that $\phi$, $\psi$ and $F$ are respectively zero, one, and two
forms in $X$.  Note also that the $U(1)_C$ charge of every $W_k$ is
$4-k$.

Given a j-dimensional homology cycle $\gamma$ on $X$ the integral
$I(\gamma)=\int_\gamma W_j$ is BRST invariant:
\[
\{Q, I\}=\int_\gamma \{Q, W_j\}=-i\int_\gamma dW_{j-1}=0
\]
Moreover, $I$ depends only on the homology class of $\gamma$. If
$\gamma=\partial\beta$ is a boundary, Stoke's theorem shows that $I(\gamma)$ is $Q$-exact:
\[
I(\gamma)=\int_\gamma W_j=\int_\beta dW_j=
i\int_\beta\{Q,W_{j+1}\}=i\{Q,\int_\beta W_{j+1}\}\qquad .
\]

The quantum field theory formulas for the Donaldson invariants are
expectation values of products of $I(\gamma)$ integrals.
Suppose that $X$, $G$, and $E$ are such that $d({\cal M})\geq 0$.
Let $\gamma_1, \gamma_2, \cdots , \gamma_r$ be homology cycles on
$X$ of dimension $j_1, j_2, \cdots, j_r$ chosen such that
$d({\cal M})=\sum_{i=1}^r(4-j_r)$. Clearly, $\prod_{i=1}^r W_{j_i}$ is an
operator with $U(1)_C$ charge equal to $d({\cal M})$. The vacuum expectation
value:
\begin{equation}
Z(\gamma_1,\ldots,\gamma_r)=\int{\cal D}{\cal
F}\exp(-\frac{S}{g^2})\prod_{i=1}^r\, I(\gamma_i)=\langle
\int_{\gamma_1} W_{j_1}\int_{\gamma_2} W_{j_2}\cdots
\int_{\gamma_r} W_{j_r}\rangle \label{eq:pol}
\end{equation}
is the TQFT \lq\lq Donaldson polynomial\rq\rq. In the left-hand member of (\ref{eq:pol}) the explicit
dependence in the homology of $X$ is shown:
\[
(\gamma_1, \gamma_2, \cdots , \gamma_r)\in H_{j_1}(X)\times
H_{j_2}(X)\times\cdots\times H_{j_r}(X) \qquad .
\]
The Feynman integral in the right-hand side of (\ref{eq:pol})
reduces to an integration over the Grassman zero modes that span
the cotangent bundle to ${\cal M}$. Therefore, the Donaldson
invariants map the $H_{j_i}(X)$ homology group of $X$ in the
$H^{4-j_i}({\cal M})$ cohomology group of the moduli space of
instantons ${\cal M}$.

\subsection{Twist of the low-energy theory: the Seiberg-Witten equations}

We now perform the twist of the low energy ${\cal N}=2$ SUSY QED described in the Section \S.3.4. The fields forming the ${\cal N}=2$
multiplet, although they are Abelian, present no novelties regarding the process of twisting with respect to the twisting at high energy. We 
only need to define the twist of the matter fields $M$ and $\tilde{M}$ entering in the monopole patch that play such an importantant r$\hat{\rm o}$le in the soft breaking of ${\cal N}=2$ SUSY giving rise to a confinement phase of the system. The magnetically charged matter fields shown in the diamond (\ref{diamat}) are twisted through the transmutation of the $SU(2)_C$ doublet of scalar fields $\phi^1_m=\phi_m$ and $\phi^2_m=\widetilde{\phi}_m$ in a right-handed Weyl spinor: 
\[
\begin{array}{|cc|c|cc} \hline && \\[-0.1cm]  {\bf  Scalar}\,
{\bf fields}
  & & {\bf Twisted}\, {\bf Spinor}\, {\bf field} \\ \hline && \\
  [-0.1cm]  \phi_{m}^I \, , I=1,2 \qquad (0,0,\frac{1}{2})^{0}
  & & S_+=\left(\begin{array}{c} 0^\alpha \\ M^{\alpha}\end{array}\right)=\left(\begin{array}{c} 0\\ 0 \\ \phi^1_m \\ \phi^2_m \end{array}\right)  \qquad
(\frac{1}{2},0)^{0}\\ 
\hline
\end{array}
\]
The $SU(2)_C$-singlet Weyl spinors in the diamond do not change their Weyl spinor character under twist:
\[
\begin{array}{|cc|c|cc} \hline && \\[-0.1cm]  {\bf  Spinor}\,
{\bf fields}
  & & {\bf Twisted}\, {\bf Spinor}\, {\bf fields} \\ \hline && \\
  [-0.1cm]  \psi_{m\alpha}\qquad (\frac{1}{2},0,0)^1& &\left(\begin{array}{c} 0_\alpha \\ \mu_\alpha\end{array}\right)=\left(\begin{array}{c} 0\\ 0\\ \psi_{m1} \\ \psi_{m2}  \end{array}\right)\quad
(\frac{1}{2},0)^1\\ [-0.1cm]\bar{\tilde{\psi}}_{m\dot{\alpha}}\qquad (0,\frac{1}{2},0)^{-1} & &\left(\begin{array}{cc} 0_{\dot\alpha} & \bar{\nu}_{\dot{\alpha}}\end{array}\right)=\left(\begin{array}{cccc} 0 & 0 & - \tilde{\psi}_{m\dot{1}} & -\tilde{\psi}_{m\dot{2}}\end{array}\right)\qquad
(0,\frac{1}{2})^{-1}
  \\ 
\hline
\end{array}
\]
Thus, the twisted spinor fields $M^\alpha$ and $\mu_\alpha$ belong to the space of sections of a Spin-complex bundle over a four-dimensional
Riemannian manifold: 
\[
M_\alpha,\ \mu_\alpha \in \Gamma (S^+\otimes L)\qquad,\qquad
\bar{\nu}_{\dot{\alpha}}\in\Gamma (S^-\otimes L) \, \, .
\]

In order to define the Dirac operator acting on these spinorial sections we start from the Euclidean Clifford algebra
\[
\gamma_a=i\left(\begin{array}{cc}0&\sigma_a\\ \bar{\sigma}_a&0\end{array}\right)\quad,\quad\begin{array}{l}\sigma_a=(\vec{\sigma}, i{\bf
1}_2)\\\bar{\sigma}_a=(-\vec{\sigma},i{\bf
1}_2,)\end{array} \quad , \quad \{\gamma_a,
\gamma_b\}=2\delta_{ab} \quad,\quad\gamma^5=\left(\begin{array}{cc}-{\bf
1}_2&0\\0&{\bf 1}_2\end{array}\right)
\]
where $a,b=1,2,3,4$ and $\vec{\sigma}\equiv (\sigma^1,\sigma^2,\sigma^3)$ are the Pauli matrices. If the equations
\[
\delta_{ab}e^{\mu a} e^{\nu b}=h^{\mu\nu} \quad , \quad h^{\mu\nu}e_{a\mu} e_{b\nu}=\delta_{ab} \, \, \, , \, \, \, \mu,\nu=1,2,3,4
\]
determine the vier-bein $e^{\mu a}$, the \lq\lq square root\rq\rq of the metric $h^{\mu\nu}$ in an oriented Riemannian four-manifold $X$, the Clifford algebra on the curved space $X$ is defined as follows:
\[
\gamma^\mu=e^{\mu a}\gamma_a\, \, \, \Rightarrow \, \, \,  \{\gamma^\mu,\gamma^\nu\}=\gamma^\mu\gamma^\nu+\gamma^\nu\gamma^\mu=2 h^{\mu\nu}\quad .
\]
The Dirac operator $\raisebox{.15ex}{{\large /}}\!\!\!\!\hspace{0.8mm}\partial_A=\gamma^\mu D_\mu$ acts on the space of sections of the  ${\rm Spin}_{\mathbb{C}}$ bundle \newline 
$\Gamma \left((S^+\oplus S^-)\otimes L)\right)$ over $X$. The covariant derivative $D_\mu$ is defined in terms of the spin connection 
$\omega_{\mu ab}$ and the $U(1)$ connection $A_\mu$.
\[
D_\mu=\nabla_\mu+i A_\mu \quad,\quad\nabla_\mu=\partial_\mu+\omega_{\mu a
b}[\gamma_a,\gamma_b] \qquad .
\]

The twisted Seiberg-Witten effective action at low energy involves two contributions:

\begin{enumerate}

\item The action including the twisted matter fields together with their Yukawa couplings to the Abelianized fields in the ${\cal N}=2$
supermultiplet:
\begin{eqnarray}
S_M^{\rm eff}&=&\int_X\sqrt{{\rm det}h}[h^{\mu\nu}D_\mu S^\dagger D_\nu
S+\frac{1}{4}RS^\dagger S-\frac{1}{16}S^\dagger
\Sigma_{\mu\nu}SS^\dagger \Sigma^{\mu\nu}S]\nonumber \\
&+&\int_X\sqrt{{\rm det}h}\{i\phi\lambda S^\dagger
S+\frac{1}{2\sqrt{2}}\chi^{\mu\nu}(S^\dagger\Sigma_{\mu\nu}\mu+\mu^\dagger\Sigma_{\mu\nu}S)\nonumber\\
&-&\frac{i}{2}(\nu^\dagger\, \raisebox{.15ex}{{\large
/}}\!\!\!\!\hspace{1mm}
\partial_A\mu-\mu^\dagger\,\raisebox{.15ex}{{\large
/}}\!\!\!\!\hspace{1mm}
\partial_A\nu)+\frac{1}{2}(S^\dagger\gamma^\mu\psi_\mu
\nu-\nu^\dagger\gamma^\mu\psi_\mu S)\\ &+&\frac{1}{2}\eta(\mu^\dagger
S-S^\dagger\mu)+\frac{i}{4}(\phi \nu^\dagger
\nu-\lambda\mu^\dagger\mu)\} \label{zdonwit1}
\end{eqnarray}
Here $h^{\mu\nu}$ is the metric tensor in $X$, $R$ is the corresponding scalar curvature and $\Sigma^{\mu\nu}=\frac{1}{2}[\gamma^\mu,\gamma^\nu]$.
\item The abelianization\ of the Donaldson action (\ref{eq:tac}) reads
\begin{equation}
S^{\rm eff}=\int d^4x\;\sqrt{{\rm det}h}\;\{\frac{1}{2}f_{\mu\nu}^+ f_+^{\mu\nu}+i\chi^{\mu\nu}
D_\mu\psi_\nu-i\eta D^\mu\psi_\mu +\frac{1}{2}\lambda D_\mu
D^\mu\phi\} \label{eq:tac1}
\end{equation}
where $f_{\mu\nu}^+$ is the self-dual Abelian gauge field tensor and the other fields are the Abelian counterparts of the fields entering 
in the twisted Yang-Mills action (\ref{eq:tac}).
\end{enumerate}
Having in mind that
\[
h^{\mu\nu}D_\mu S^\dagger D_\nu S=\frac{1}{2}D_\mu S^\dagger \{\gamma^\mu,\gamma^\nu\}D_\nu S=\raisebox{.15ex}{{\large /}}\!\!\!\!\hspace{1mm} \partial_A 
S^\dagger\,\raisebox{.15ex}{{\large /}}\!\!\!\!\hspace{1mm}
\partial_A S + D_\mu S^\dagger \Sigma^{\mu\nu} D_\nu S\quad .
\]
we select in the effective action $S^{\rm eff}+S^{\rm eff}_M$ two definite positive terms involving the spinorial field (after twisting) $S$:
\begin{equation}
S^{\rm eff}_{SW} =\int_X\sqrt{{\rm det}h}\, \, \left[
(f_{\mu\nu}^++\frac{i}{2}S^\dagger\Sigma_{\mu\nu}S)
(f^{\mu\nu+}+\frac{i}{2}S^\dagger\Sigma^{\mu\nu}S)
+ \raisebox{.15ex}{{\large /}}\!\!\!\!\hspace{1mm} \partial_A
S^\dagger\,\raisebox{.15ex}{{\large /}}\!\!\!\!\hspace{1mm}
\partial_A S+\cdots \, \, \right] \label{eq:efSWact}
\end{equation}
We observe that the positive perfect square terms in  $S^{\rm eff}_{SW}$ are zero if the following coupled system of non-linear PDE's
\begin{equation}
(1) \, \, \, \, \raisebox{.15ex}{{\large /}}\!\!\!\!\hspace{1mm} \partial_A S=0 \quad \, \, \, , \quad \quad \, \, \, \qquad (2)  \, \, \, \, f_{\mu\nu}^+=-\frac{i}{2}S^\dagger \Sigma_{\mu\nu} S \label{sdseibwit}
\end{equation}
holds. Henceforth, the solutions of the PDE system (\ref{sdseibwit}), referred to as the non-linear Seiberg-Witten equations, are the absolute minima of $S^{\rm eff}_{SW}$. These equations describe the obstruction to self-duality of a $U(1)$-gauge field tensor on a four-manifold due to the anomalous magnetic momentum induced by an harmonic right-handed spinor via the coupling $f_{\mu\nu}^+S^\dagger\Sigma_{\mu\nu}S$. Contrarily to the moduli space of instantons the moduli space of the Seiberg-Witten solutions is \underline{compact} {\footnote{The reason is that there are no solutions of the PDE system (\ref{sdseibwit}) which shrink to zero size  because there are no $L^2$ solutions in ${\mathbb R}^4$.}} and presents no singularities due to reducible connections. Therefore, diffeomorphism invariants of four-manifolds are more accessible through the topological invariants of the moduli spaces of Seiberg-Witten solutions than via the topological structures of moduli spaces of instantons, see \cite{Witten}, \cite{Taubes}.

The dimension of the moduli space of Seiberg-Witten solutions is envisaged from the linearization of the (\ref{sdseibwit}) system of PDE equations. Up to  first-order in the perturbations $A_\mu^{\rm SW}(x)+t a_\mu(x)$, $S_{\rm SW}(x)+s \Psi(X)$ around a given SW solution $(A_\mu^{\rm SW}, S_{\rm SW})$, we find the linearized Seiberg-Witten equations
\begin{eqnarray}
\gamma_\mu \left(\nabla_\mu + i a_\mu(x)\right) S_{\rm SW}(x)&+&\gamma_\mu \left(\nabla_\mu+i A_\mu^{\rm SW}(x)\right)\Psi(x)=0 \label{swdirper}\\  \nabla_\mu a_\nu(x) - \nabla_\nu a_\mu(x)&+& \frac{\sqrt{\vert g\vert}}{2}\varepsilon_{\mu\nu\rho\delta}\left(\nabla_\rho a_\delta(x) - \nabla_\delta a_\rho(x)\right)\nonumber=\\ &=& \frac{i}{4}\left(\Psi^\dagger(x) [\gamma_\mu,\gamma_\nu]S_{\rm SW}(x)+S_{\rm SW}^\dagger(x) [\gamma_\mu,\gamma_\nu]\Psi(x)\right) \, \, \label{swvorper}
\end{eqnarray}
where $\vert g \vert$ is the determinant of the tensor metric in $X$ and $\nabla_\mu$ denotes the covariant derivative acting on spinorial and/or tensorial sections. Setting, e.g., the background gauge $\left(\nabla_\mu+ i A_\mu^{\rm SW}\right) a_\mu=0$ to avoid pure gauge
perturbations one may identify the dimension of the moduli space of Seiberg-Witten solutions by an index theorem argument. The number of zero modes arising as normalizable solutions $(a_\mu,\Psi)$ of equation (\ref{swdirper}) is essentially captured by {\footnote{In fact the line bundle $L$ is accounted for as a real bundle in such a way that we count twice the index of the Dirac operator.}} the index of the Dirac operator acting on sections of the ${\rm Spin}^C$ bundle $S_+\otimes L$: $c_1(L)^2- \frac{\sigma(X)}{4}$, i.e., it is determined in terms of the first Chern class of the complex line bundle $L$ and the signature of the four manifold $X$. Simili modo the number of zero modes coming from the solutions of 
(\ref{swvorper}) is established from the index of the Hodge operator $d+d^*$: $-\chi(X)/2-\sigma(X)/2$ where now the Euler characteristic
$\chi(X)$ enters. All together the dimension of the moduli space of solutions of the Seiberg-Witten equations is found to be:
\begin{equation}
d({\cal M}_{\rm SW})=c_1(L)^2-\frac{2 \chi(X)+3\sigma(X)}{4} \, \, \ . \label{dmodssw} 
\end{equation}

\subsection{Kronheimer-Mrowka basic classes and Seiberg-Witten invariants}

We now briefly describe the differential invariants of smooth four manifolds derived from the topological structures arising in
the moduli space of solutions of the Seiberg-Witten equations. A good physics flavoured treatment of this topic is offered in Labastida Lectures, see \cite{Labastida}. In Reference \cite{Atiyah} a deep and extremely condensed explanation can be found about how the basic classes of Kronheimer and Mrowka, \cite{Kronheimer}, prompt zero dimensional moduli spaces of Seiberg-Witten solutions, which in turn
provide a very effective procedure of computation of the Seiberg-Witten invariants. Following Reference \cite{Witten} we asign to an homology
two-cycle in the four manifold $X$, $\gamma\in H_2(X)$, Donaldson polynomials of the form, recall the formula (\ref{eq:pol}),
\[
p_{s,r}(\gamma , X)=\int {\cal D}{\cal F} \, e^{-S/g^2} \, \prod_{j=1}^r W_0(P_j)\left(\int_\gamma \, W_2 \right)^s =\langle \prod_{j=1}^r W_0(P_j)\left(\int_\gamma \, W_2 \right)^s \rangle \quad ,
\]
which are non null only if the dimension of the moduli space of anti-self-dual instantons is such that : $d({\cal M})=4r+2s$, equivalent to
$b_2^+={\rm odd}$.

Suppose that the four manifold $X$ is such that $p_{s,r+2}(\gamma, X)=4 p_{s,r}(\gamma, X)$. Manifolds enjoying this property are called of simple type. The primitive Donaldson polynomials arising in this situation
\[
q_s(\gamma,X)=\left\{ \begin{array}{c} \, p_{s,0}(\gamma, X) \, \, \, {\rm if} \, \, \, s=1+b_2^+ \, \, {\rm mod}\, 2 \\ \\ p_{s,1}(\gamma, X) \, \, \,  {\rm if}  \, \, \, s=b_2^+ \, \, {\rm mod}\, 2\end{array}\right.
\]
are assembled in a generating function: $q(\gamma, X)=\sum_{s=0}^\infty \, \frac{1}{s!}\, q_s(\gamma, X)$. The Kronheimer-Mrowka formula
for this generating function is:
\begin{equation}
q_s(\gamma, X)= {\rm exp}[\frac{\gamma \cdot \gamma}{2}]\sum_{I=1}^N \, a_I \, {\rm exp}[\kappa_I \cdot \gamma] \label{eq:KroMro} \quad .
\end{equation}
In this expansion $a_1, a_2, \cdots, a_N$ are non null rational numbers which define the Seiberg-Witten invariants. $\kappa_I\in H^2(X,\mathbb{Z})$ are the Kronheimer-Mrowka \underline{basic} classes. A 2D cohomology class, e.g. $\kappa_I$, defines a line bundle over a 2D surface through the first Chern class, i.e. $c_1(L_I^2)=\kappa_I$ such that KM classes are related to the square of the line bundle $L_I$: $c_1(L_I)= c_1(L_I^2/2$. The basic classes are characterized by the property: $c_1^2(L_I^2)=\kappa_I^2 =2 \chi(X)+3 \sigma(X)$, which is thus an integer number $\kappa_I^2 =5 b_2^+-b_2^-$.

Evaluation of the $a_I$ coefficients was achieved by Witten in Reference \cite{Witten}. The pillars of his calculation were settle down in 
the effective twisted Abelian ${\cal N}=2$ SUSY gauge theory governed by the action $S^{\rm eff}_M+S^{\rm eff}=S^{\rm eff}_{\rm SW}$, see (\ref{zdonwit1}), (\ref{eq:tac1}) and (\ref{eq:efSWact}). Instead of looking at vacuum expectation values of the microscopic ${\cal N}=2$ twisted SUSY non-Abelian gauge theory on $X$ in the high energy regime computations are performed in the infrared (strong coupling) domain where the macroscopic Abelian effective theory emerges. Recall that the partition function $Z$ is independent of the coupling constant $g$. Thus, identical expectation values can be equivalently calculated either at weak $g$-coupling, where the action (\ref{eq:tac}) and the partition function (\ref{zdon}) are the main instruments, or, at strong $g$-coupling, where computations are based on the action (\ref{eq:tac1}) and the partition function (\ref{zdonwit1}) in the monopole patch of the moduli space of vacua {\footnote{A completely equivalent treatment at strong $g$-coupling is possible near the $u=-\Lambda^2$ singularity where massless dyons replace massless monopoles as new particles in the spectrum.}}.The steepest descent approximation applied to the Feynman path integral in this patch shows that it is
localized near the saddle points of $S^{\rm eff}_{\rm SW}$, i.e., the solutions of the Seiberg-Witten equations (\ref{sdseibwit}), rather 
than around anti-self-dual instantons. Because formula (\ref{dmodssw}) tells us that
\[
d({\cal M}_{\rm SW})=\frac{1}{4}\left(c_1^2(L^2)-2\chi(X)-3\sigma(X)\right) 
\]
the Kronheimer-Mrowka basic classes are those for which the dimension of the moduli space of solutions of the Seiberg-Witten equations is zero: $d({\cal M}_{\rm SW})=0$. Therefore, there is a finite number of points $\nu$ in ${\cal M}_{\rm SW}$ and the Seiberg-Witten
partition function collects the number of SW-solutions weighted with their signs, a formula completely analogous to (\ref{eq:don1}):
\[
Z_{\rm SW} \propto n_L=\sum_{i=1}^\nu \, \varepsilon_i = \sum_{i=1}^\nu \, (-1)^{N_i} \, \, \, , \, \, \, N_i=0,1 \quad .
\]
Again, $N_i=0,1$ counts the number of zeroes ${\rm mod}\, 2$ crossed by the spectrum of $D_F^t$ when the fermionic fluctuation operator $D_F$ varies through a family of fields joining two solutions of the Seiberg-Witten equations (recall the analysis just before (\ref{eq:don1})).
The generating function of the Donaldson polynomials of a four manifild $X$ of simple type is then obtained by summing all the contributions of this kind for the $N$ basic classes:
\begin{equation}
q(\gamma, X)= 2^{(1+d(X))} {\rm exp}[\frac{\gamma\cdot\gamma}{2}]\, \sum_{I=1}^N \, n_{L_I} \, \cdot {\rm exp}[c_1(L_I^2)\cdot \gamma] \label{eq:swiqft} \quad.
\end{equation}
The factor of $2$ is due to the fact that, even though instanton moduli spaces are invariant with respect to the center of $SU(2)$ , the Donaldson invariants are defined without dividing by two. The critical exponent $d(X)=\frac{1}{4}\left(7 \chi(X)+11 \sigma(X)\right)$ is a $c$-renormalization factor which appear when one compares the expectation values of Donaldson polynomials computed in the microscopic $SU(2$ non-Abelian theory with the outcome in the effective Abelian theory of massless monopoles, see \cite{Witten}-\cite{WittenA}. To set the values of the coefficients $7/4$, $11/4$ the argument runs as follows: there is no perfect duality invariance when one consider the theory in any pont $u$ of the vacuum moduli space. On a curved 4-manifold gravitational anomalies in the Fermionic fields integration measure of the form:
\begin{eqnarray}
&& d\mu^F={\rm exp}\left[b(u) \chi(X)+c(u)\sigma(X)\right]d\mu^F_D  \, , 
\, \, \,  \chi(X)=\frac{1}{24\pi^2}\, \int_X \, R \wedge \widetilde{R} \, , \, \, \, \sigma(X)=\frac{1}{24\pi^2}\, \int_X \, R \wedge R 
\nonumber \\ && R=R_{\mu \, \, \, \nu \alpha}^{\, \, \, \alpha} dx^\mu\wedge dx^\nu \, \, , \quad \quad  \widetilde{R}=\frac{\sqrt{\vert g\vert}}{2}\varepsilon_{\rho\sigma}^{\, \, \, \, \, \, \alpha\beta}R_{\alpha \, \, \, \beta \gamma}^{\, \, \, \gamma} dx^\rho\wedge dx^\sigma \, \, \label{gravan}
\end{eqnarray}
arise. Here $R$ and $\widetilde{R}$ are respectively the Ricci tensor and its dual of the Riemannian manifold $X$. $\chi(X)$, the Euler number, and the signature, $\sigma(X)$, are the only observables of dimension four which are topological invariants and may arise in the topological, twisted, ${\cal N}=2$ Supersymmetric Gauge Theory. Thus a factor ${\rm exp}\left[b(u) \chi(X)+c(u)\sigma(X)\right]$ must be included in the twisted action. Witten's cunning strategy was to identify in the weak coupling limit $b(u\to \infty)$ as $b=\frac{7}{4}\log 2$ and $c(u\to\infty)$ as $c=\frac{11}{4}\log 2$ to fit with the Donaldson invariants of known four manifolds as $K3$ surfaces and/or manifolds of
simple type. Because topological invariance these values must be constant all over the moduli space and the coefficient $2^{d(X)}$ arise this way also in the strong coupling $u=\pm\Lambda^2$ regimes. It is astonishing how an extremely subtle renormalization coefficient
due to the physics of gravitational, and perhaps modular, anomalies in ${\cal N}=2$ SUSY gauge theory may be derived from purely mathematical information about the differential structures of some specific four manifolds.

\subsection{Hidden physics behind the low-energy twisted Seiberg-Witten action}

Understanding of its physical meaning suggests to scrutinize the two-component theory, see Reference \cite{Veltman}, which is akin to the second-order dual QED governed by the twisted SW action on the Euclidean ${\mathbb R}^4$ space-time. The action is:
\begin{eqnarray}
S_E^{\rm SW} &=& \int \, dx^4 \, \left\{ \frac{1}{4}\left(f_{\mu\nu}f_{\mu\nu}+\frac{1}{2}\varepsilon_{\mu\nu\rho\sigma}f_{\mu\nu}f_{\rho\sigma}\right)+\frac{1}{2} D_\mu S_+^\dagger D_\mu S_++D_\mu S_+^\dagger \Sigma_{\mu\nu}D_\nu S_+\right\}\nonumber\\ &+& \int \, dx^4 \, \left\{-i\frac{\lambda}{2}S_+^\dagger \Sigma_{\mu\nu}S_+\cdot f_{\mu\nu}^++\frac{\lambda^2}{8}S_+^\dagger \Sigma_{\mu\nu}S_+\cdot S^\dagger \Sigma_{\mu\nu}S_+\right\} \, \, .
\label{thirquarsw}
\end{eqnarray}
It is convenient to write explicitly the quantities involving the two-component spinor $S$:
\[
S_+^\dagger (x)=\left( 0 \, \, \,  0 \, \, \, \phi_1^*(x) \, \, \phi_2^*(x)\right) \, \, \, , \, \, \, D_\mu=\frac{\partial}{\partial x_\mu}+ i g A_\mu(x) \, \, \, , \,   \, \, S_+(x)=\left(\begin{array}{c} 0 \\ 0 \\ \phi_1(x) \\ \phi_2(x)\end{array}\right)
\]
\[
\Sigma_{\mu\nu}=\frac{1}{2}[\gamma_\mu,\gamma_\nu]=\frac{1}{2}\left(\begin{array}{cc} \sigma_\mu\bar{\sigma}_\nu-\sigma_\nu\bar{\sigma}_\mu
& 0 \\ 0 & \bar{\sigma}_\mu\sigma_\nu-\bar{\sigma}_\nu\sigma_\mu \end{array}\right)
\]
where $g=\frac{4\pi}{e}$ is the magnetic charge dual to the electric charge $e$ and $\lambda$ is a non-dimensional coupling which sets the strength of the anomalous magnetic momentum of the spinorial particle.

Besides the conventional photon propagator due to the Maxwell term in the action (\ref{thirquarsw}) there is a propagator of a spin one-half particle of the form:
\begin{equation}
P^{-1}=\Big(p_\mu\left[\delta_{\mu\nu}{\bf 1}_2+\bar{\sigma}_\mu\sigma_\nu-\bar{\sigma}_\nu\sigma_\mu\right]p_\nu\Big)^{-1} \label{fprop}\, \, .
\end{equation}
Note that 
\[
{\rm det}P=p_\mu p_\mu-\frac{1}{12}\varepsilon_{\mu\nu\rho\sigma}p_\mu p_\nu p_\rho p_\sigma-\frac{1}{4}\sum_{a=1}^3 \, \eta^a_{\mu\nu}p_\mu p_\nu\cdot \eta^a_{\rho\sigma}p_\rho p_\sigma  \, \, ,
\]
where  $\eta^a_{\mu\nu}$ are the  three complex structures in $R^4$ or 't Hooft symbols, and the $P$-matrix is invertible. There is also
a (ineffective) toological term in the SW action: the Abelian second Chern class. More interesting; six types of electro-magnetic couplings 
give rise to trivalent or fourvalent vertices:
\begin{eqnarray}
&& \hspace{-3cm}({\rm 1})  -i \frac{g}{2} A_\mu \left(S_+^\dagger\partial_\mu S_+-\partial_\mu S_+^\dagger S_+ \right)\hspace{1.5cm}\quad , \quad ({\rm 2}) \, \,\, \, \frac{g^2}{2}A_\mu A_\mu S_+^\dagger S_+  \label{u1couplings}  \\
&& \hspace{-3cm}({\rm 3}) -i g \Big(A_\mu S_+^\dagger \Sigma_{\mu\nu} \partial_\nu S_+ -\partial_\mu S_+^\dagger \Sigma_{\mu\nu} A_\nu S_+ \Big) \quad , \quad ({\rm 4}) \, \, -g^2 A_\mu S_+^\dagger \Sigma_{\mu\nu} A_\nu S_+ \label{emcouplings}\\
&& \hspace{-3cm} ({\rm 5}) -i\frac{\lambda}{2}S_+^\dagger \Sigma_{\mu\nu}S_+\cdot f_{\mu\nu}^+ \hspace{3.5cm}\quad , \quad ({\rm 6}) \, \, \, \, \, \frac{\lambda^2}{8}S_+^\dagger \Sigma_{\mu\nu}S_+\cdot S^\dagger \Sigma_{\mu\nu}S_+ \, \, .\label{magcouplings}
\end{eqnarray}
These strange Feynman rules are akin to those emerging in Veltman Two component theory and electron magnetic moment, see \cite{Veltman}
The Veltman second-order QED Feynman rules come from the Lagrangian obtained from the QED Lagrangian through multiplication of the electron spinor field by the Dirac operator:
\[
\psi \, \longrightarrow \, \left(-\gamma_\mu D_\mu+m\right)\psi 
\]
and afterwards projecting to two-component (e.g. only positron) spinors. The twisted Seiberg-Witten action is a dual version of this gauge theory. There are more vertices than in Veltman theory where only the vertices of type (1), (2), and (3) arise. The dual electron propagator
becomes also more complex in the Seiberg-Witten framework because the anomalous dual magnetic moment enter. Nevertheless, the main virtue, the separation between electric and magnetic couplings, is shared by the dual Seiberg-Witten twisted QED and Veltman second-order QED, whereas both Lagrangians are non-hermitian.

If $g=1=\lambda$ the twisted dual SW effective action can be written in the form
\begin{equation}
=\int \, dx^4 \, \Big[ \frac{1}{2}\left(f_{\mu\nu}^++\frac{i}{2}S_+^\dagger \Sigma_{\mu\nu} S_+\right)\left(f_{\mu\nu}^++\frac{i}{2}S_+^\dagger \Sigma_{\mu\nu} S_+\right)+\left(\gamma_\mu D_\mu S_+\right)^\dagger \gamma_\nu D_\nu S_+\Big] \quad , \label{seibwit}
\end{equation}
plus \lq\lq topological\rq\rq terms. 

Solutions of the first-order Seiberg-Witten PDE system (\ref{sdseibwit}) in ${\mathbb R}^4$
\begin{equation}
f^+_{\mu\nu}(x)=-\frac{i}{2}S^\dagger(x)\Sigma_{\mu\nu}S(x) \quad \quad , \quad \quad
\gamma_\mu D_\mu S(x)=0 \label{seibwitt}
\end{equation}
are thus absolute minima of $S_E^{\rm W}$. We know that there are no $L^2$-integrable solutions of of this PDE system in 
Euclidean 4D space. Nevertheless, it is possible to obtain non-$L^2$ solutions by means of several dimensional reduction
procedures that help to grasp the physical meaning of the Seiberg-Witten equations.

\subsection{Low dimensional Freund and Seiberg-Witten solutions}

We devote this subsection to analyze dimensionally reduced solutions to the Seiberg-Witten equations \cite{Freund}-\cite{Adam} in order to elucidate the topological nature of the non perfect square terms in the action (\ref{seibwit}) as well as the hidden physics in the twisted  effective dual second-order QED arising at the ultrastrong coupling. 

\subsubsection{Three-dimensional Freund solutions}
We write the Freund equations, the Seiberg-Witten equations with  a flip in the spinor term sign, in components:
%\clearpage
\begin{eqnarray}
&& f_{12}^+(x)=f_{34}^+(x)=-\frac{1}{2}\Big(\vert \phi_1(x)\vert^2-\vert\phi_2(x)\vert^2\Big) \label{fsw1}\\
&& f_{31}^+(x)=f_{24}^+(x)=\frac{i}{2}\Big( \phi_1^*(x)\phi_2(x)-\phi_2^*(x)\phi_1(x)\Big) \label{fsw2} \\
&& f_{23}^+(x)=f_{14}^+(x)=-\frac{1}{2}\Big( \phi_1^*(x)\phi_2(x)+\phi_2^*(x)\phi_1(x)\Big) \label{fsw3}
\end{eqnarray}
This PDE system is nothing but the equation in the left of formula (\ref{seibwitt}) with a plus sign in the right-hand fellow and must be solved together with the two-component Dirac equation:
\begin{equation}
\left(\begin{array}{cc} i D_3-D_4 & iD_1+D_2 \\ i D_1-D_2 & -i D_3-D_4 \end{array}\right)\cdot\left(\begin{array}{c} \phi_1(x) \\ \phi_2(x)\end{array}\right)=\left(\begin{array}{c} 0 \\ 0\end{array}\right) \label{dfsb} \quad .
\end{equation}
First, we address dimensional reduction to ${\mathbb R}^3$, i.e. spinors and gauge potentials are $x_4$-independent and choose $A_4=0$. One can check that the Dirac monopole singular potential, in the axial gauge $A_3(x_1,x_2,x_3)=0$,
\begin{equation}
A_1(x_1,x_2,x_3)=-\frac{x_2}{2 r(r-x_3)} \, \, \, \, , \, \, \, \,  A_2(x_1,x_2,x_3)=\frac{x_1}{2 r(r-x_3)} \label{dirmonpot}
\end{equation}
together with the harmonic monopole spinors
\begin{equation}
\phi_1(x_1,x_2,x_3)=\frac{x_1-i x_2}{2r\sqrt{r(r-x_3)}} \quad , \quad \phi_2(x_1,x_2,x_3)=\frac{1}{2r}\cdot\sqrt{\frac{r-x_3}{r}} \label{harmonspin}
\end{equation}
satisfy the 3D Freund equations:
\begin{eqnarray*}
&& f_{12}^+(x_1,x_2,x_3)=-\frac{x_3}{r^3}= -\frac{1}{2}\Big(\vert \phi_1(x)\vert^2-\vert\phi_2(x)\vert^2\Big) \\
&& f_{31}^+(x_1,x_2,x_3)=-\frac{x_2}{4r^3}=\frac{i}{2}\Big( \phi_1^*(x)\phi_2(x)-\phi_2^*(x)\phi_1(x)\Big) \\
&& f_{23}^+(x_1,x_2,x_3)=-\frac{x_1}{4r^3}=-\frac{1}{2}\Big( \phi_1^*(x)\phi_2(x)+\phi_2^*(x)\phi_1(x)\Big) \quad .
\end{eqnarray*}
A little more work is needed to show that the spinor (\ref{harmonspin}) satisfies the Dirac equation (\ref{dfsb}), reduced to three dimensions, in the monopole background. The physical meaning of this Freund solution is clear. The gauge potential is singular at $r=x_3$
positive half-axis where the famous Dirac string is located. The harmonic spinors are even more singular at the Dirac string location because they have branching points along that curve. Nevertheless, there are other quantities which are less singular, e.g., the dual
magnetic (electric) field only has a pole at the origin:
\begin{eqnarray*}
 b_1(x_1,x_2,x_3)&=&f_{23}(x_1,x_2,x_3)=-\frac{x_1}{2 r^3} \, \, \, , \, \, \, b_2(x_1,x_2,x_3)=f_{31}(x_1,x_2,x_3)=-\frac{x_2}{2 r^3}
\\ b_3(x_1,x_2,x_3)&=&f_{12}(x_1,x_2,x_3)=-\frac{x_3}{2 r^3} \quad .
\end{eqnarray*}
The \lq\lq electric\rq\rq flux through an sphere of radius $R$ centered at the origin is:
\[
\Phi=\int_{S^2_R} \, \left( b_1(x_1,x_2,x_3)dx_2\wedge dx_3+ b_2(x_1,x_2,x_3)dx_3\wedge dx_1+ b_3(x_1,x_2,x_3)dx_1\wedge dx_2\right)=2\pi \, \, , 
\]
i.e., the monopolar electric charge carried by the Freund solution is the minimal compatible with the Dirac quantization condition.

\subsubsection{Planar Seiberg-Witten solutions}
Finally, we consider the true Seiberg-Witten equations
\begin{eqnarray}
&& f_{12}^+(x)=f_{34}^+(x)=\frac{1}{2}\Big(\vert \phi_1(x)\vert^2-\vert\phi_2(x)\vert^2\Big) \label{ssw1}\\
&& f_{31}^+(x)=f_{24}^+(x)=-\frac{i}{2}\Big( \phi_1^*(x)\phi_2(x)-\phi_2^*(x)\phi_1(x)\Big) \label{ssw2} \\
&& f_{23}^+(x)=f_{14}^+(x)=\frac{1}{2}\Big( \phi_1^*(x)\phi_2(x)+\phi_2^*(x)\phi_1(x)\Big) \label{ssw3}
\end{eqnarray}
together with the Dirac equation (\ref{dfsb}):
\begin{eqnarray}
 ( D_3+i D_4)\phi_1(x)+(D_1-i D_2)\phi_2(x) &=& 0 \label{dir1}\\ (D_1+i D_2)\phi_1(x) -( D_3-i D_4)\phi_2(x) &=& 0 \label{dir2} \, \, .
\end{eqnarray}
Multiplying (\ref{dir1}) by $D_1+i D_2$, (\ref{dir2}) by $D_3+i D_4$, subtracting the second equation from the first, and using (\ref{ssw2})-(\ref{ssw3}) we find:
\begin{equation}
-2\vert \phi_1 \vert^2(x)\phi_2(x)+\bar{\partial}_{A_{12}}\partial_{A_{12}}\phi_2(x)+\bar{\partial}_{A_{34}}\partial_{A_{34}}\phi_2(x)=0\, \, , \label{obst}
\end{equation}
where we have defined:
\[
\bar{\partial}_{A_{12}}\partial_{A_{12}}=(D_1+iD_2)(D_1-iD_2) \quad , \quad \bar{\partial}_{A_{34}}\partial_{A_{34}}=(D_3+iD_4)(D_3-iD_4) \, \, .
\]
Multiplication of (\ref{obst}) by $\phi_2^*(x)$, and integration over all $\mathbb{R}^4$ leads to the identity
\begin{equation}
\int_{\mathbb{R}^4}\, d^4x \, \left\{\vert\phi_1\vert^2(x)\cdot\vert\phi_2\vert^2(x)+\vert \partial_{A_{12}}\phi_2(x)\vert^2 + \vert \partial_{A_{34}}\phi_2(x)\vert^2 \right\}=0 \label{intobst} \, \, ,
\end{equation}
after a partial integration, to be satisfied by the Seiberg-Witten solutions.

There are two possibilities:

\begin{itemize}

\item - A. $\phi_2=0$. On configurations with \lq\lq electric\rq\rq spin $1/2$, $\sigma_3\left(\begin{array}{c} \phi_1 \\ 0 \end{array}\right)=\left(\begin{array}{c} \phi_1 \\ 0 \end{array}\right)$, we consider the planar ansatz:
\[
\phi_1(x)=\phi_1(x_1,x_2)\, , \, \, \, \phi_2=A_3=A_4=0 \, , \, \, \, \, A_1(x)=A_1(x_1,x_2) \, \, \, \, \, A_2(x)=A_2(x_1,x_2) \, .
\]
The Seiberg-Witten equations (\ref{ssw1}-\ref{ssw2}-\ref{ssw3}-\ref{dir1}-\ref{dir2}) reduce to the first-order PDE system:
\begin{equation}
f_{12}(x_1,x_2)=\frac{1}{2}\vert \phi_1(x_1,x_2)\vert^2 \quad , \quad (D_1+iD_2)\phi_1(x_1,x_2)=0 \, \, \, .\label{sdcs}
\end{equation}
The solution of the covariant analyticity condition in (\ref{sdcs}) is:
\[
A_{\bar{z}}=\frac{1}{2}(A_1+iA_2)=-i\partial_{\bar{z}} \log \phi_1 = \frac{1}{2}(\partial_1+i\partial_2)\log \phi_1 \, \, ,
\]
which, because $f_{12}=-2 i (\partial_z A_{\bar{z}}-\partial_{\bar{z}}A_z)$, converts the vortex equation  (\ref{sdcs}-left) into the
Liouville equation:
\begin{equation}
2 \partial_{\bar{z}}\partial_z \log \phi_1+ \phi_1^*\phi_1=0 \, \, \, . \label{liouvv}
\end{equation}
The general solution of (\ref{liouvv}) such that $\lim_{x_1^2+x_2^2\to\infty} \phi_1(x_1,x_2)=0,$, guaranteeing finite energy density in
the $x_1$-$x_2$-plane, is:

\begin{equation}
\phi_1^{(k)}(z)=\frac{2 f^\prime(z)V^2(z)}{\vert V(z)\vert^2+ \vert f(z)V(z)\vert^2} \quad , \quad z=x_1+i x_2 \label{gensolliou}
\end{equation}
where the following choice of the $V$ and $f$ functions as
\[
V(z)=\prod_{j=1}^k (z-z^{(j)}) \, \, \quad , k\in{\mathbb N}^* \, \, \, , \quad \quad f(z)=f_0+\sum_{j=1}^k \frac{c_k}{z-z^{(j)}}
\]
gives rise to non-singular solutions supporting quantized electric flux: 
\[
\Phi_E= \int\int \, dx_1 dx_2 \, f_{12}(x_1,x_2)=\frac{1}{2}\int\int \, dx_1 dx_2 \, \vert \phi_1^{(k)}(z)\vert^2=\frac{2\pi}{g}k
\]
where the g-coupling constant has been re-surfaced. 

Observe that $k$ is positive and the flux is located around $z^{(j)}$, the zeroes of $\phi_1(z)$. It spreads out from these zeroes, however, with $\vert c_j\vert $, the length scale of the solution, which is a free parameter due to the breaking by the flux tube of the scale invariance of the theory. There is also freedom in choosing ${\rm arg}c_j$ because the $U(1)_d$ symmetry and the moduli space of solutions is $\mathbb{C}^{2k}$; the moduli space parameters are the centers of the solitons $z^{(j)}$ and the modulus and phase of $c_j$ determining the scale and phase of each individual soliton.

\item -B. $\phi_1=0$, $\vert\partial_{A_{12}}\phi_2\vert=\vert\partial_{A_{34}}\phi_2\vert=0$. 

On configurations with \lq\lq electric\rq\rq spin $-1/2$, $\sigma_3\left(\begin{array}{c} 0 \\\phi_2 \end{array}\right)=-\left(\begin{array}{c} 0 \\ \phi_2 \end{array}\right)$, the appropriate planar ansatz reads:
\[
\phi_2(x)=\phi_2(x_3,x_4)\, , \, \, \, \phi_1=A_1=A_2=0 \, , \, \, \, \, A_3(x)=A_3(x_3,x_4) \, \, \, \, \, A_4(x)=A_4(x_3,x_4) \, .
\]
The Seiberg-Witten equations (\ref{ssw1}-\ref{ssw2}-\ref{ssw3}-\ref{dir1}-\ref{dir2}) reduce to the first-order PDE system:
\begin{equation}
f_{34}(x_3,x_4)=-\frac{1}{2}\vert \phi_2(x_3,x_4)\vert^2 \quad , \quad (D_3-iD_4)\phi_2(x_3,x_4)=0 \, \, \, .\label{sdcsb}
\end{equation}
The solution of the covariant anti-analyticity condition in (\ref{sdcsb}) is:
\[
A_{w}=\frac{1}{2}(A_3-iA_4)=i\partial_{w} \log \phi_2 = \frac{1}{2}(\partial_3-i\partial_4)\log \phi_2 \, \, ,
\]
which, because $f_{34}=-2 i (\partial_w A_{\bar{w}}-\partial_{\bar{w}}A_w)$, converts the vortex equation  (\ref{sdcsb}-left) into the
Liouville equation:
\begin{equation}
2 \partial_{\bar{w}}\partial_w \log \phi_2- \phi_2^*\phi_2=0 \, \, \, . \label{liouvvb}
\end{equation}
\end{itemize}
The general solution of (\ref{liouvv}) such that $\lim_{x_3^2+x_4^2\to\infty} \phi_2(x_3,x_4)=0,$, guaranteeing finite energy density in
the $x_3$-$x_4$-plane, is:

\begin{equation}
\phi_2^{(k)}(\bar{w})=\frac{2 f^\prime(\bar{w})V^2(\bar{w})}{\vert V(\bar{w})\vert^2+ \vert f(\bar{w})V(\bar{w})\vert^2} \quad , \quad \bar{w}=x_3-i x_4 \label{gensollioub}
\end{equation}
where the following choice of the $V$ and $f$ functions as
\[
V(\bar{w})=\prod_{j=1}^k (\bar{w}-\bar{w}^{(j)}) \, \, \quad , k\in{\mathbb N}^* \, \, \, , \quad \quad f(\bar{w})=f_0+\sum_{j=1}^k \frac{c_k}{\bar{w}-\bar{w}^{(j)}}
\]
gives rise to non-singular solutions supporting quantized \lq\lq magnetic\rq\rq flux: 
\[
\Phi_M= \int\int \, dx_3 dx_4 \, f_{34}(x_3,x_4)=\frac{1}{2}\int\int \, dx_3 dx_4 \, \vert \phi_2^{(k)}(\bar{w})\vert^2=-\frac{2\pi}{g}k
\]
The moduli space of these Type B planar solutions of the Seiberg-Witten equations present analogous features to those of the Type A moduli space described above. 

There are also pairs of Type A-Type B planar solutions with the same characteristics- obeying similar ansatzes-in the $x_2$-$x_3$-/ $x_4$-$x_1$-planes as well as in the $x_3$-$x_1$-/$x_2$-$x_4$-planes, to wit:

\begin{itemize}

\item  -A. $\left(\begin{array}{c}\, \, \, \psi_2 \\ - \psi_2\end{array}\right)=0$. On configurations such that $\sigma_1 \Big[\frac{1}{\sqrt{2}}\left(\begin{array}{c} \psi_1 \\ \psi_1 \end{array}\right)\Big]=\frac{1}{\sqrt{2}}\left(\begin{array}{c} \psi_1 \\ \psi_1 \end{array}\right)$, we consider the planar ansatz:
\[
\psi_1(x)=\psi_1(x_2,x_3)\, , \, \, \, \psi_2=A_1=A_4=0 \, , \, \, \, \, A_2(x)=A_2(x_2,x_3) \, \, \, \, \, A_3(x)=A_2(x_2,x_3) \, .
\]
Because $\phi_1=\phi_2=\frac{1}{\sqrt{2}}\psi_1$ the Seiberg-Witten equations (\ref{ssw1}-\ref{ssw2}-\ref{ssw3}-\ref{dir1}-\ref{dir2}) reduce to the first-order PDE system:
\begin{equation}
f_{23}(x_2,x_3)=\frac{1}{2}\vert \psi_1(x_2,x_3)\vert^2 \quad , \quad (D_2+iD_3)\psi_1(x_2,x_3)=0 \, \, \, ,\label{sdcs1}
\end{equation}
allowing for an identical moduli space of solutions as (\ref{sdcs}).

\item  -B. $\left(\begin{array}{c}\, \, \, \psi_1 \\ \psi_1\end{array}\right)=0$. On configurations such that $\sigma_1 \Big[\frac{1}{\sqrt{2}}\left(\begin{array}{c} \psi_2 \\ -\psi_2 \end{array}\right)\Big]=-\frac{1}{\sqrt{2}}\left(\begin{array}{c} \psi_2 \\ -\psi_2 \end{array}\right)$, we consider the planar ansatz:
\[
\psi_2(x)=\psi_2(x_1,x_4)\, , \, \, \, \psi_1=A_2=A_3=0 \, , \, \, \, \, A_1(x)=A_1(x_1,x_4) \, \, \, \, \, A_4(x)=A_2(x_1,x_4) \, .
\]
Because $\phi_1=-\phi_2=\frac{1}{\sqrt{2}}\psi_2$ the Seiberg-Witten equations (\ref{ssw1}-\ref{ssw2}-\ref{ssw3}-\ref{dir1}-\ref{dir2}) reduce to the first-order PDE system:
\begin{equation}
f_{14}(x_1,x_4)=-\frac{1}{2}\vert \psi_2(x_1,x_4)\vert^2 \quad , \quad (D_1-iD_4)\psi_2(x_1,x_4)=0 \, \, \, , \label{sdcsb1}
\end{equation}
giving rise to the same moduli space of solutions as (\ref{sdcsb}).

\item  -A. $\left(\begin{array}{c}\, \, \, \psi_2 \\ - i \psi_2\end{array}\right)=0$. On configurations such that $\sigma_2 \Big[\frac{1}{\sqrt{2}}\left(\begin{array}{c} \psi_1 \\ i \psi_1 \end{array}\right)\Big]=\frac{1}{\sqrt{2}}\left(\begin{array}{c} \psi_1 \\ i \psi_1 \end{array}\right)$, we consider the planar ansatz:
\[
\psi_1(x)=\psi_1(x_1,x_3)\, , \, \, \, \psi_2=A_2=A_4=0 \, , \, \, \, \, A_1(x)=A_2(x_1,x_3) \, \, \, \, \, A_3(x)=A_2(x_1,x_3) \, .
\]
Because $\phi_1=-i \phi_2=\frac{1}{\sqrt{2}}\psi_1$ the Seiberg-Witten equations (\ref{ssw1}-\ref{ssw2}-\ref{ssw3}-\ref{dir1}-\ref{dir2}) reduce to the first-order PDE system:
\begin{equation}
f_{31}(x_2,x_3)=\frac{1}{2}\vert \psi_1(x_1,x_3)\vert^2 \quad , \quad (D_1+iD_3)\psi_1(x_1,x_3)=0 \, \, \, ,\label{sdcs2}
\end{equation}
allowing for an identical moduli space of solutions as (\ref{sdcs}).

\item  -B. $\left(\begin{array}{c}\, \, \, \psi_1 \\ i \psi_1\end{array}\right)=0$. On configurations such that $\sigma_2 \Big[\frac{1}{\sqrt{2}}\left(\begin{array}{c} \psi_2 \\ -i \psi_2 \end{array}\right)\Big]=-\frac{1}{\sqrt{2}}\left(\begin{array}{c} \psi_2 \\ -i \psi_2 \end{array}\right)$, we consider the planar ansatz:
\[
\psi_2(x)=\psi_2(x_2,x_4)\, , \, \, \, \psi_1=A_1=A_3=0 \, , \, \, \, \, A_2(x)=A_1(x_2,x_4) \, \, \, \, \, A_4(x)=A_2(x_2,x_4) \, .
\]
Because $\phi_1=-i\phi_2=\frac{1}{\sqrt{2}}\psi_2$ the Seiberg-Witten equations (\ref{ssw1}-\ref{ssw2}-\ref{ssw3}-\ref{dir1}-\ref{dir2}) reduce to the first-order PDE system:
\begin{equation}
f_{24}(x_1,x_4)=-\frac{1}{2}\vert \psi_2(x_2,x_4)\vert^2 \quad , \quad (D_2-iD_4)\psi_2(x_1,x_4)=0 \, \, \, , \label{sdcsb2}
\end{equation}
giving rise to the same moduli space of solutions as (\ref{sdcsb}).

\end{itemize}

Finally, we mention that adding a mass term for the spinors that explicitly breaks the scale invariance and spontaneously breaks the 
$U(1)_D$-gauge symmetry the left equation in (\ref{sdcs}) is perturbed to
\begin{equation}
f_{12}(x_1,x_2)=\frac{1}{2}\Big(\vert \phi_1(x_1,x_2)\vert^2 -1\Big) \, \, \, ,\label{sdcsv}
\end{equation} 
and becomes the self-dual or BPS vortex equation, see \cite{Jaffe}. Thus, the solutions are the celebrated BPS or self-dual vortices
and the real dimension of the moduli space, now $\mathbb{C}^k$, diminishes to $2k$ responding to the freedom of motion of the centers of the quantized \lq\lq electric\rq\rq flux tubes. Of course there are analogous electric flux tubes in the $x_1$-$x_3$- and $x_2$-$x_3$-planes. In the planes containing $x_4$, however, the flux is \lq\lq magnetic\rq\rq but the interpretation of these tubular solutions is doubtful because they should be properly recognized as \lq\lq instantons\rq\rq when $x_4$ refers to \lq\lq Euclidean\rq\rq time.

\section{A very brief epilogue}

This essay has been elaborated by expanding the lecture notes of a short Course that I taught in the Workshop on Geometry and Physics 
in Miraflores de la Sierra, Spain, September 2001, to a mixed audience of Geometers and Theoretical Physicists. The main theme was the description of the impact in Geometry of special concepts and techniques developed in non-perturbative Quantum Field Theory. This 
subject achieved impressive success during the fourth quarter of the last Century establishing deep links between Quantum Physics and 
Algebraic Topology and Geometry, in contrast with Classical Physics more tied to Differential Topology and Geometry. The presentation 
here is strongly inclined towards the side of Physics.

\end{document}